# Strain fields in twisted bilayer graphene

Nathanael P. Kazmierczak,[1,2,†] Madeline Van Winkle,[1,†] Colin Ophus,[3] Karen C. Bustillo,[3] Stephen Carr,[4,5] Hamish G. Brown,[3] Jim Ciston,[3] Takashi Taniguchi,[6] Kenji Watanabe,[7] D. Kwabena Bediako[1,8,9]*

[1] *Department of Chemistry, University of California, Berkeley, CA 94720, USA*

[2] *Department of Chemistry and Biochemistry, Calvin University, Grand Rapids, MI 49546, USA*

[3] *National Center for Electron Microscopy, Molecular Foundry, Lawrence Berkeley National Laboratory, Berkeley, CA 94720, USA*

[4] *Department of Physics, Brown University, Providence, Rhode Island 02912, USA*

[5] *Brown Theoretical Physics Center, Brown University, Providence, Rhode Island 02912, USA.*

[6] *International Center for Materials Nanoarchitectonics, National Institute for Materials Science, 1-1 Namiki, Tsukuba 305-0044, Japan*

[7] *Research Center for Functional Materials, National Institute for Materials Science, 1-1 Namiki, Tsukuba 305-0044, Japan*

[8] *Chemical Sciences Division, Lawrence Berkeley National Laboratory, Berkeley, CA 94720, USA*

[9] *Azrieli Global Scholar, Canadian Institute for Advanced Research (CIFAR), Toronto, Ontario M5G 1M1, Canada*

* Correspondence to: bediako@berkeley.edu

† These authors contributed equally to this work

*Abstract*

**Van der Waals heteroepitaxy allows deterministic control over lattice mismatch or azimuthal orientation between atomic layers to produce long wavelength superlattices. The resulting electronic phases depend critically on the superlattice periodicity as well as localized structural deformations that introduce disorder and strain. Here, we introduce Bragg interferometry, based on four-dimensional scanning transmission electron microscopy, to capture atomic displacement fields in twisted bilayer graphene with twist angles < 2°. Nanoscale spatial fluctuations in twist angle and uniaxial heterostrain are statistically evaluated, revealing the prevalence of short-range disorder in this class of materials. By quantitatively mapping strain tensor fields we uncover two distinct regimes of structural relaxation—in contrast to previous models depicting a single continuous process—and we disentangle the electronic contributions of the rotation modes that comprise this relaxation. Further, we find that applied heterostrain accumulates anisotropically in saddle point regions to generate distinctive striped shear strain phases. Our results thus establish the reconstruction mechanics underpinning the twist angle dependent electronic behaviour of twisted bilayer graphene, and provide a new framework for directly visualizing structural relaxation, disorder, and strain in any moiré material.**

Stacking two-dimensional (2D) van der Waals (vdW) bilayers with a slight offset in lattice periodicity—due to dissimilar lattice constants and/or rotational misalignment—produces a moiré superlattice with a periodicity that is inversely related to the magnitude of interlayer mismatch[1,2]. The moiré pattern superimposes a nanoscale periodic potential on the vdW material and can dramatically alter the electronic band structure of the system[2,3]. As such, moiré materials assembled from graphene, hexagonal boron nitride (hBN), and transition metal dichalcogenides (TMDCs) have proven to be versatile platforms for designing electronic band structures. Secondary Dirac points and Hofstadter's butterfly[4–6] have been observed in graphene/hBN superlattices, and twisted bilayer graphene (TBG) displays a host of correlated electronic phases—including unconventional superconductivity[7–9], correlated insulating behaviour[10], and magnetism[9,11]—that are associated with the formation of ultraflat electronic bands[12] near an interlayer 'magic angle' (MA) of 1.1°. TMDC heterobilayers can stabilize excited states like moiré excitons[13–15], and twisted $WSe_2$ exhibits correlated electronic states over a range of twist angles[16]. Other multilayer graphene moirés also show correlated electronic behaviour[17–19] and moiré-hosted magnetic skyrmions may be realized in magnetic heterobilayers[20]. While valued for their tunability, the flat bands and correlated electronic states created by moiré superlattices are also fragile and can be manipulated or suppressed by small structural deformations. One of the most consequential structural modifications is an intrinsic intralayer atomic lattice reconstruction process[21–29]. In TBG, this reconstruction introduces intralayer strain[23,25,30] and frustrates flat band formation at other theoretically predicted magic angles[12] below 1.1°. Scanning tunnelling measurements[31] and theoretical calculations[32,33] have also provided evidence that symmetry breaking due to extrinsic uniaxial heterostrain, where one layer is stretched relative to the other, may strongly alter the observed electronic phases. In addition, spatial variations in twist angle, an unconventional type of disorder that is unique to moiré materials, strongly impacts the observed electronic phases in MA-TBG[34,35].

Although visualizing the structure and strain fields of moiré materials is paramount to understanding and controlling emergent electronic phases, directly and quantitatively mapping the reconstruction mechanics in these systems has been elusive and the complete strain tensor fields have not been measured. One significant obstacle is the presence of hBN multilayers, typically used in device fabrication[7–11]. These several-layers thick hBN crystals present major complications for imaging complex, multilayer structures and encapsulated heterostructures. A variety of microscopy techniques have been used to image reconstructed moiré bilayers[21,28,29,31,36], but these require the bilayer to be exposed or fully suspended. Conventional dark-field transmission electron microscopy (DF-TEM) can indirectly probe reconstruction in encapsulated samples[27], but resolution is limited at larger twist angles, such as the MA regime for TBG, and key structural details cannot be directly extracted from these measurements. Strain measurements in reconstructed moiré materials have been restricted to determinations of 1D strain[30] and uniaxial heterostrain[31], but 2D strain tensors have thus far only been measured in lateral heterostructures[37].

Here we use four-dimensional scanning transmission electron microscopy (4D-STEM)[38–40] to directly visualize the deformations underlying reconstruction at the nanometre scale and to precisely measure localized strain in TBG. To accomplish this, we develop a new diffraction-based method, which we term Bragg interferometry, that allows high-resolution mapping of the structure and complete 2D strain tensors of TBG, despite the presence of hBN. We present displacement field maps at and around the MA that now make it possible to quantify short-range disorder in the form of both twist angle and heterostrain fluctuations on the length scale of individual stacking domains. In contrast to earlier models depicting a single continuous process, strain field mapping unveils a previously undetected evolution of reconstruction mechanics in TBG as a function of twist angle, and *ab initio* calculations show the influence of individual rotational deformation processes on band structure. Our results also demonstrate the strong interplay between reconstruction strain fields and uniaxial heterostrain in MA-TBG.

**Bragg interferometry and visualization of displacement maps**

We fabricate TBG samples using the common 'tear and stack' method (see Methods and Supplementary Fig. 1)[41], introducing a twist angle of $0.1° < \theta_m < 1.6°$ between the graphene lattices. Fig. 1a shows a schematic of the 4D-STEM experiment, where a focused electron beam is rastered through an hBN/TBG heterostructure and the diffracted electron signal is collected at each probe position[38]. The overlapping Bragg disks for each graphene layer, offset by $\theta_m$, are discernible in Fig. 1a. The total intensity $I_j$ in the overlapping region of the *j*th interfering Bragg disk pair corresponding to a graphene reciprocal lattice vector, **g**, is given by:

$$I_j = A_j \cos^2(\pi \boldsymbol{g}_j \cdot \boldsymbol{u}) \qquad (1)$$

Here $\boldsymbol{u} = \langle u_x, u_y \rangle$ is the local displacement vector from an atom in the first graphene layer to the nearest atom in the same sub-lattice in the second graphene layer (Fig. 1a) and $A_j$ is a scaling factor representing the average number of pixel counts at maximum diffraction intensity (see Methods). Crucially, the intensities in the overlap regions of the Bragg disk pairs—isolated using a virtual aperture during analysis of the 4D-STEM dataset—are observed to vary across the sample, corresponding to different local stacking arrangements. Using equation (1), the spatial arrangement of atomic stacking regions in the TBG layers can be determined by measuring $I_j$ for all ⟨1100⟩ and ⟨2110⟩ reflections and fitting a local **u** assignment for each pixel in the 2D real-space scan (Fig. 1b). Figs. 1c–f show maps of the local displacement vectors for representative TBG samples with $\theta_m$ ranging from 0.16° to 1.37°, where each pixel encodes information about the local displacement vector according to the half-hexagon displacement legend in Fig. 1b. Thus, the hue and value at each pixel indicate displacement vector direction and magnitude, respectively. By using information from Bragg overlap regions of all twelve Bragg disk pairs simultaneously, the displacement vector field provides a more comprehensive picture of the TBG structure compared to DF-TEM images (Supplementary Fig. 1b,c), allowing clear and quantitative visualization of the reconstructed moiré superlattice over many moiré wavelengths even at and above the MA. Additional displacement field maps are provided in Supplementary Fig. 2.

By registering the centroids of each AA region (see Methods), these displacement vector maps enable a direct geometric analysis of the local variations in twist angle and heterostrain, $\varepsilon_H$ (estimated using a previous model[31]), at the resolution of individual AB/BA domains. Figs. 2a,b exemplify the results of these analyses for the displacement map shown in Fig. 1f. Additional $\theta_m$ and $\varepsilon_H$ maps are shown in Supplementary Fig. 3. Mapping local $\theta_m$ for four ostensibly uniform samples near the MA, we find standard deviations in $\theta_m$ to be approximately constant around 0.03º (Fig. 2c). Likewise, mapping $\varepsilon_H$ for the same samples reveals average $\varepsilon_H$ around 0.2% and standard deviations between 0.06% and 0.09% (Fig. 2d). We note that the $\theta_m$ and $\varepsilon_H$ values obtained from propagating uncertainties in AA centroid registration are 0.01º and 0.026%, respectively (see Supplementary Information), indicating these disorder distributions cannot be explained by measurement error. Since the band structure is highly sensitive to supercell size and geometry[34,35], the local spatial fluctuations and distributions in both $\theta_m$ and $\varepsilon_H$ that are resolved within these apparently homogeneous 100 nm × 100 nm regions may provide a gauge of the intrinsic short-range structural disorder to be expected from MA-TBG. The physical mechanism of the disorder remains unclear and will be the subject of future work. Additionally, analysing these displacement fields, we present measurements of the geometric properties of AA and saddle point (SP) stacking regions (see Methods) as displayed in Fig. 2e. These data provide qualitative validation of trends previously predicted from multiscale modeling[25,26]. However, our measurements show larger AA region diameters and thinner SP widths than those predicted from previous simulations, providing new experimental insights for future modelling efforts.

**Strain field mapping**

Traditionally, 4D-STEM strain mapping has obtained strain values from the changes in Bragg disk positions across the dataset, thereby detecting local changes in the lattice constants[37,42]. This technique presents significant challenges for TBG samples because overlapping Bragg disks preclude accurate position registration at low signal-to-noise ratios. However, since strain is the gradient of a displacement field[43,44], maps like Figs. 1c–f allow us to determine the complete 2D strain tensor describing all directions of in-plane deformation in TBG at each pixel as a function of $\theta_m$ (see Methods and Supplementary Figs. 4–6). Consequently, we can measure both interlayer azimuthal rotation and intralayer deformation mechanics.

The interlayer component is the total 'fixed-body' rotation[43] field $\theta_T$, from which the local reconstruction rotation field ($\theta_R$) can be determined by removing $\theta_m$ (that is, $\theta_T = \theta_R + \theta_m$). The maximum shear (also known as principal shear) field, $\gamma_{max}$, provides the maximum amount of intralayer 'engineering' shear strain in any direction experienced by the material[43,44]. Neither $\theta_R$ nor $\gamma_{max}$ require definition of a local tensor coordinate system, allowing isotropic, quantitative visualization of strain across many datasets.

Figs. 3a–d show maps of $\theta_R$ and $\gamma_{max}$ for two exemplary values of $\theta_m$ (additional maps at other $\theta_m$ are provided in Supplementary Fig. 7). Figs. 3a,b provide the first direct experimental evidence for a reconstruction mechanism predicted by theoretical studies[22,23,25,26] and suggested by indirect electron diffraction data[27]: both maps display significant positive $\theta_R$ in AA regions and negative $\theta_R$ in AB/BA domains. Positive $\theta_R^{AA}$ signifies rotation in the direction of $\theta_m$, which shrinks the area of the higher energy AA region. Negative $\theta_R^{AB}$ shows that reconstruction is counteracting $\theta_m$ to bring the AB domains closer to commensurate, low-energy Bernal-stacking. These effects of $\theta_m$ on rotational reconstruction can be clearly visualized in a sample possessing $\theta_m$ varying rapidly from 1.3º to 0.6º owing to a nearby tear in one of the graphene layers (Fig. 3e). In Fig. 3e, because of the variation in $\theta_m$ over the field of view, we plot $\theta_T$ instead of $\theta_R$. Data from a similar region are shown in Supplementary Fig. 8.

Fig. 3f summarizes $\theta_T^{AA}$ and $\theta_R^{AA}$ as a function of $\theta_m$ based on twenty twist angle-homogenous images and two additional datasets over regions with a nearby tear. The two types of datasets show excellent agreement, with greater precision from the homogenous maps. As $\theta_m$ nears zero, $\theta_R^{AA}$ approaches a limiting value of approximately 1.2º. For $\theta_m < 0.5°$, reconstruction keeps $\theta_T^{AA}$ approximately constant. Extrapolation of $\theta_R^{AA}$ to large $\theta_m$ suggests an onset of significant reconstruction begins below $\theta_m \sim 2°$.

These rotational mechanics provide a basis to understand the intralayer shear strain produced by reconstruction. Maps of $\gamma_{max}$ (Fig. 3c, d) show that intralayer strain is localized in the SP at both values of $\theta_m$, with peak values of $\gamma_{max}$ exceeding 0.8% in both cases. The changing direction of the principal strain[44] axes reveals that reconstruction does not generate global strain (Supplementary Fig. 9). For both $\theta_m = 0.26°$ and $\theta_m = 1.03°$, despite the large rotational reconstruction taking place (Fig. 3a,b), $\gamma_{max}$ decreases rapidly upon approaching the core of the AA region. This is due to the bivariate Gaussian radial profile of AA reconstruction[26,21] (Supplementary Fig. 10): near the centre of the AA region, the approximately constant $\theta_R^{AA}$ produces no intralayer strain. For $\theta_m = 0.26°$ (Fig. 3c), AB domains exhibit no intralayer strain over an extended region, which is again consistent with the constant $\theta_R^{AB}$ observed over the same area (Fig. 3a). By contrast, $\theta_R^{AB}$ in MA-TBG changes more rapidly through space (Fig. 3b), as extended Bernal domains have not formed. Consequently, intralayer strain in MA-TBG appears less localized than strain at smaller twists, relative to the moiré unit cell size. While some regions of Fig. 3d show nearly six-fold symmetric SP strain, other regions display more striped features, an observation we shall return to later.

In addition to $\gamma_{max}$, SP strain can also be understood in terms of simple shear strain ($s_{yx} = \partial u_y/\partial x$ and $s_{xy} = \partial u_x/\partial y$). The quantity $s_{yx}$, used in previous one-dimensional strain analysis of shear soliton walls[30], considers the displacement change parallel to a soliton wall (i.e., misfit dislocation)[21,45]. Based on $s_{yx}$ alone, Fig. 3g shows that intralayer shear strain would appear to be minimal in MA-TBG. However, both $s_{xy}$ and $s_{yx}$ are directly obtained from our 2D strain measurements. As $\theta_m$ decreases through 1.1º, $s_{xy}$ is larger and increases more rapidly than $s_{yx}$ until a maximum around $\theta_m = 0.8º$, after which an inversion in $s_{yx}$ and $s_{xy}$ occurs at ~0.5º. This plot of $\gamma_{max}(\theta_m)$ shows that the average intralayer shear strain loading in MA-TBG is substantially greater than suggested by $s_{yx}$ or $s_{xy}$ alone, and comparable to that at smaller $\theta_m$, with a limiting mean $\gamma_{max}$ of ~0.8%.

The crossing of the $s_{yx}$ and $s_{xy}$ magnitudes in SP regions at $\theta_m = 0.5º$ arises because simple shear strain combines intralayer pure shear with interlayer fixed-body rotation (see Methods)[43], and is therefore helpfully explained by evaluating the interlayer reconstruction rotation. In Fig. 3h $\theta_R^{SP}$ and $\theta_R^{AB}$ are plotted as a function of $\theta_m$. We find that $\theta_R^{SP}$ undergoes a sign change from negative to positive as $\theta_m$ decreases, consistent with the changing relative magnitudes of $s_{yx}$ and $s_{xy}$ in Fig. 3g. These $\theta_R^{SP}$ data imply SP expansion in MA-TBG and shrinkage at very small $\theta_m$. Although $\theta_R^{AB} < 0$ over the entire range to counteract $\theta_m$, Fig. 3h also reveals that $\theta_R^{AB}$ varies non-monotonically with $\theta_m$, reaching its minimum value of $\theta_R^{AB} = -0.3°$ at $\theta_m \sim 0.8º$. Fully commensurate AB stacking is achieved when $\theta_R^{AB} = -\theta_m$, a condition met for $\theta_m < 0.2°$. Notably, the shear-induced displacement on the AB boundary grows as $\theta_m$ decreases, despite this diminution in the magnitude of $\theta_R^{AB}$ (Fig. 3h inset). Moreover, for $\theta_m < 0.5°$, the displacement effect of $\theta_R^{AB}$ accelerates and in the limit of $\theta_m < 0.2°$, we calculate an induced displacement on each side of the AB boundary of about one-half the C–C bond length (see Methods). This displacement change is sufficient to explain the formation of thin shear solitons in their entirety, indicating that AB counter-rotation is a mechanism for generating soliton walls.

**Dual regimes of reconstruction mechanics in TBG**

To explore how these observed trends directly impact the structure of TBG, we extract the area percentage of AA, AB, and SP stacking from our displacement maps (see Methods, Supplementary Fig. 17) and plot these as a function of $\theta_m$ in Fig. 4a. Interestingly, we again find two regimes with a trend change near $\theta_m = 0.5°$, the point at which $\theta_R^{SP}$ changes sign (Fig. 3h). For $\theta_m > 0.5°$, AA region fractions shrink steadily as $\theta_m$ decreases, a process driven by increasing $\theta_R^{AA}$. In this regime, as $\theta_m$ decreases, AB and SP region areas both steadily increase, consistent with a dominant reconstruction process that does not distinguish between these stacking orders. In contrast, AB and SP areas diverge for $\theta_m < 0.5°$. AB domains increase rapidly in size to dominate the material, while SP regions decrease in relative area fraction (despite increasing in absolute width as shown in Fig. 2e) as they form true shear soliton walls bordering the AB domains in the small angle limit. Even though AA regions have approximately constant radius in this regime (Fig. 2e), they continue to decrease in area fraction because of the expanding moiré unit cell. Inclusion of intermediate stacking order categories in the analysis leads to the same conclusions (see Supplementary Material and Supplementary Fig. 17).

These two regimes of reconstruction can be further understood by analysing the reconstruction mechanics of TBG entirely through maps of simple shear strain (Figs. 4b,c) using local axis rotations (see Methods), providing a new conceptual picture of TBG reconstruction as an interplay between AA and AB/BA rotation (Fig. 4d). Again, AA reconstruction exerts the main effect for $\theta_m > 0.5°$ (Fig. 4b). In this regime, the dominant simple shear is perpendicular to the SP region path between closely spaced AA regions ($s_{xy} > s_{yx}$ in Fig. 3g), driven by positive $\theta_R^{AA}$. While AB counter-rotation does occur, the induced displacement change is minimal because the moiré wavelength is small (Fig. 3h, inset). Further, the fixed-body rotation produced by AA simple shear is expected to be negative (Fig. 4d, right), explaining the observed negative SP local rotation for $\theta_m > 0.5°$ (Fig. 3h). For $\theta_m < 0.5°$, AB reconstruction dominates. Because the AA rotation field decays quickly away from the AA core (Fig. 3a) and only a small $\theta_R^{AB}$ is required to counteract the small $\theta_m$, AB counter-rotation alone serves to maintain true soliton walls in this regime (Fig. 4c). Adjoining AB–BA domains rotating in the same direction (with negative $\theta_R^{AB}$) generate dominant simple shear parallel to the soliton wall (Fig. 4c), demonstrating the case where $s_{yx} > s_{xy}$ in Fig. 3g and $\theta_R^{SP}$ is expected to be positive (Fig. 3h and Fig. 4d, left). When AA and AB simple shear forces are balanced in the SPs, $\theta_R^{SP}$ passes through 0° because the SP region experiences a pure shear force (Fig. 4d, centre). This occurs near $\theta_m = 0.5°$, the critical angle separating the two regimes.

**Predicted electronic modifications by local rotations**

The two-regimes model of Fig. 4 provides a useful framework for examining the perturbation of the electronic structure by reconstruction and, in particular, the destruction of ultraflat bands at smaller angles. We now examine the separate effect of the two relaxation modes on the electronic structure of TBG in Figure 5. These computations are enabled by high-quality *ab-initio* electronic tight-binding models for TBG[46] (see Methods) and a simple parameterized atomic reconstruction model, which is constructed based on our experimental strain maps, that allows selective implementation of $\theta_R^{AA}$- or $\theta_R^{AB}$-dominated reconstruction (see Supplementary Information). We consider three values of $\theta_m$: 0.35º, 0.5º, and 1.15º. $\theta_m$ = 0.5º and 0.35º approximate the second and third magic angles predicted for a rigid (no reconstruction) TBG moiré[12]. Prior to reconstruction, their band structures admit a large number of tangled bands near the Fermi energy (Fig. 5b,c). Focusing first on the 0.35º case (Fig. 5b), application of $\theta_R^{AB}$ rotation alone removes the large number of low-energy bands and frames the lowest four by two pairs of neighbouring bands on each side, but in the process the extreme flatness is lost. On the other hand, we find that including only $\theta_R^{AA}$ retains band flatness, but does not remove as many of the low-energy bands. After including both rotations, we observe emergence of a set of four nearly flat bands and two pairs of parabolic bands that touch the flattened bands at the Γ point, somewhat reminiscent of the band structure of MA-TBG[12,27], albeit more dispersive in nature. At 0.5º (Fig. 5c) the results of either $\theta_R^{AA}$ or $\theta_R^{AB}$ alone initially appear quite similar, though $\theta_R^{AB}$ noticeably produces more dispersive bands while $\theta_R^{AA}$ alone preserves some flatness. At 1.15º (Fig. 5d), $\theta_R^{AA}$ alone more closely replicates the flat band structure of the full reconstruction and opens gaps at the Γ point both above and below the flat band.

It is impossible to exactly ascribe features of the doubly-rotated band structure to individual rotation modes, but the trends observed in Figures 5b–d imply that $\theta_R^{AA}$ helps define the flat low-energy modes, while $\theta_R^{AB}$ ensures only four such bands exist at low energy. This interpretation is also consistent with the geometric effects of the rotations. At smaller angles (e.g. $\theta_m = 0.35°$ or 0.5º), positive rotation near the AA stacking spots (Figs. 3a,e)—where the flat band wave-functions predominantly localise[10]—makes the atomic geometry in these regions more similar to that seen in MA-TBG (i.e. $\theta_T \approx 1.2° > \theta_m$ as shown in Fig. 3f). On the other hand, negative rotation in AB/BA regions (Figs. 3a,g) makes those domains look less like TBG and more like Bernal stacked bilayer graphene, encouraging a more dispersive band structure and in turn reducing the number of bands near the Fermi energy.

These qualitative observations of electronic modifications (Fig. 5b–d) arise because the band structure of TBG is predominantly described by variation in interlayer electronic tunnelling over the moiré superlattice[12], which is highly sensitive to atomic reconstruction[47,48]. It is therefore useful to quantitatively assess the relative electronic importance of the isolated rotation modes by directly comparing the interlayer tunnelling functions under different rotation assumptions. In Fig. 5e, we find that the relative importance of $\theta_R^{AA}$ and $\theta_R^{AB}$ for electronic interlayer tunnelling indeed changes with $\theta_m$. At $\theta_m = 1.15°$, sole application of $\theta_R^{AA}$ yields better agreement in the calculated interlayer tunnelling with the fully reconstructed structure than pure $\theta_R^{AB}$. For $\theta_m < 0.5°$, the converse is true, and at $\theta_m = 0.5°$, the influence of both rotations is almost balanced. This quantitative result agrees with qualitative comparisons between the full electronic interlayer tunnelling functions (Supplementary Figs. 18–26). Thus, the relative contributions of the separate relaxation modes to the fully relaxed electronic structure (Fig. 5e) agree with our two-regimes concept (Fig. 4d), providing the fundamental connection between the reconstruction modified electronic band structures (Fig. 5b–d) and our two-regimes model developed from strain field mapping (Fig. 4d).

We also note that our full-rotation model provides bands that are in good agreement with those obtained by realistic finite-element simulations, which relax the atomic structure self-consistently[22,24,26], but cannot interrogate the impact of individual rotation mechanics as permitted by our model.

**Effect of heterostrain on reconstruction strain fields**

Finally, we consider the influence of uniaxial heterostrain ($\varepsilon_H$) in modifying these intrinsic reconstruction strain fields. Returning to the region in the vicinity of a tear (Fig. 3e), we estimate $\varepsilon_H$ over the field of view (Fig. 6a), revealing regions with nearly identical $\theta_m$ near the MA, but possessing $\varepsilon_H$ varying between 0.1 and 1%. Fig. 6b shows that MA-TBG regions with minimal $\varepsilon_H$ (box 1) exhibit a fully six-fold symmetric strain pattern with localized, isolated pockets of shear strain on each individual SP region. By contrast, regions with large $\varepsilon_H$ (box 2) display striking striped features in $\gamma_{max}$. Additionally, SP shear strain fields are magnified both in value and in extent in heterostrained regions (see also Supplementary Fig. 11), suggesting that the extra strain loading from heterostrain localizes in the SP regions. Fig. 6c captures this heterostrain-induced modification in a sample at $\theta_m = 0.63º$ where the regions are more zig-zag in nature and the unstrained AB domains are consequently offset away from the shortened SP region angles. Prompted by these experimental observations, Fig. 6d displays the results of finite-element relaxation of heterostrained TBG (see Supplementary Information). Simulations show excellent agreement with the experimentally extracted strain distributions and help to explain the formation of these quasi one-dimensional (1D) strain features on geometric grounds. By changing the moiré cell geometry through the superimposition of two moiré patterns, heterostrain decreases the angle between at least two pairs of SP regions, mandating a more rapid change in displacement. This 'displacement pinching' effect implies the need for a connected shear strain field in the decreased SP angle area in order to maintain reconstruction. Rather than shrinking or bending to avoid contact, the SP strain fields remain approximately constant width under heterostrain, and therefore blend near the shortened SP region angles to break rotational symmetry and form striped regions. The tendency for TBG to generate this strain field rather than lessening the degree of reconstruction points to the importance of stacking energy over intralayer strain energy for driving reconstruction mechanics. This model also explains the observation of pronounced 1D striped regions in MA-TBG by comparison to TBG at smaller twists (Fig. 3c and Supplementary Fig. 27).

**Discussion**

Our 4D-STEM Bragg interferometry methodology and analysis make it possible to image moiré superlattices in MA-TBG notwithstanding the real space colocalization of hBN multilayers. The visualization of domain stacking distributions at the level of individual AB/BA domains enables evaluation of the intrinsic superlattice disorder[35] in TBG. In previous squid-on-tip studies of twist angle disorder[34], electronic effects from "long-range" variations in $\theta_m$ were considered as though, locally, each moiré superlattice represented an ideal twisted bilayer at a given $\theta_m$, with an angle that varies in space. Accordingly, since different electronic states dominate at different twist-angles, on the scale of micrometres samples would possess patches of different electronic states, complicating transport measurements. In contrast, the "short-range" $\theta_m$ and $\varepsilon_H$ disorder visualized here would cause fundamentally different effects[35]: the local ideal band structure is modified owing to spatial fluctuations in $\theta_m$ from one AB domain to another and regions with the same effective $\theta_m$ could also present different electronic behaviour due to disorder in $\varepsilon_H$. A combination of both these effects—microscale variations in $\theta_m$ causing patches of different phases and local nanoscale fluctuations in $\theta_m$ and $\varepsilon_H$ causing significant modifications to the band structure itself—may help explain the large variation in the observed low-temperature phases in TBG at or below the magic-angle as well as the preponderance of intertwined correlated states.

2D strain field mapping unveils a rich landscape of structural mechanics in TBG. Intralayer shear strain due to reconstruction is largely concentrated in SP regions, and is found to be substantial near the MA. We find two regimes of reconstruction in TBG involving a competition between AA and AB/BA local rotations that are balanced near a moiré angle of 0.5º. We show that this competition manifests directly in the electronic interlayer coupling that governs the band structure. Modelling the effect of individual reconstruction rotations reveals the general band structure features that are promoted by isolated relaxation modes. The greater influence of AB counter-rotation at small angles compared to the dominance of AA rotation at 1.1º helps explain why flat bands are disrupted by reconstruction at smaller angles, whereas they persist at 1.1º. These results also provide hints at the possibility of a new type of band structure engineering through local physical or chemical perturbations of isolated rotations. Additionally, MA-TBG has recently been found to possess either intrinsic nematic order or a strong nematic susceptibility[31] that appears triggered by heterostrain. Our strain field maps and displacement pinching model show how mesoscale heterostrain in TBG is translated into localized, symmetry-breaking nanoscale features through the anisotropic amplification and deformation of SP regions into 1D strain-structures.

The new Bragg interferometry method introduced here is also applicable to non-moiré heterostructures with colocalized reciprocal lattice vectors (such as intercalation compounds and misfit compounds) and may be performed in a manner compatible with *in situ* mechanical straining, now permitting such manipulations to be visualized directly and quantitatively. While our methodology only considers displacements and strain in the lateral plane, emerging holography techniques may provide a route to obtaining complementary z-axis information[49]. The investigation of wide-ranging moiré materials by this methodology will elucidate the complex interplay between intrinsic reconstruction strain, extrinsic uniaxial strain, and the diverse array of physical phases, including correlated electronic states.

**Acknowledgements**

We are grateful to Philip Kim for discussions. The major experimental work is supported by the Office of Naval Research Young Investigator Program under Award No. N00014-19-1-2199. MV acknowledges support from an NSF GRFP award and UC Berkeley Chancellor's Fellowship. Work at the Molecular Foundry was supported by the Office of Science, Office of Basic Energy Sciences, of the U.S. Department of Energy under Contract No. DE-AC02-05CH1123. CO acknowledges support of the Department of Energy Early Career Research Award program. S.C. acknowledges support from NSF grant No. OIA-1921199. JC and HGB acknowledge support from the Presidential Early Career Award for Scientists and Engineers (PECASE) through the U.S. Department of Energy. DKB acknowledges support from the Rose Hills Foundation through the Rose Hills Innovator Program. K.W. and T.T. acknowledge support from the Elemental Strategy Initiative conducted by the MEXT, Japan, Grant Number JPMXP0112101001, JSPS KAKENHI Grant Number JP20H00354 and the CREST (JPMJCR15F3), JST.

**Author Contributions**

NPK, MV, CO, KCB, HB, and DKB conceived the study. MV designed and fabricated the samples. MV, KCB, and JC acquired the 4D-STEM data. NPK, CO, and HB created the data analysis code. SC carried out the band structure calculations and finite-element modelling. TT and KW provided the bulk hBN crystals. NPK and MV processed the data. NPK, MV, and DKB analysed the data and wrote the manuscript. All authors contributed to the overall scientific interpretation and edited the manuscript.

**Competing Interests**

The authors declare no competing interests.

**Author Information**

The authors declare no competing financial interests. Correspondence and requests for materials should be addressed to D.K.B. (e-mail: bediako@berkeley.edu).

**References**

(1) Abbas, G. et al. Recent Advances in Twisted Structures of Flatland Materials and Crafting Moiré Superlattices. *Adv. Func. Mater.* **30**, 2000878 (2020).

(2) Yankowitz, M., Ma, Q., Jarillo-Herrero, P. & LeRoy, B.J. van der Waals heterostructures combining graphene and hexagonal boron nitride. *Nat. Rev. Phys.* **1**, 112 – 125 (2019).

(3) Balents, L., Dean, C. R., Efetov, D. K. & Young, A. F. Superconductivity and strong correlations in moiré flat bands. *Nat. Phys.* **16,** 725–733 (2020).

(4) Yankowitz, M. et al. Emergence of superlattice Dirac points in graphene on hexagonal boron nitride. *Nat. Phys.* **8**, 382–386 (2012).

(5) Panomarenko, L.A. et al. Cloning of Dirac fermions in graphene superlattices. *Nature* **497**, 594–597 (2013).

(6) Hunt, B. et al. Massive Dirac Fermions and Hofstadter Butterfly in a van der Waals Heterostructure. *Science* **340**, 427–1430 (2013).

(7) Cao, Y. et al. Unconventional superconductivity in magic-angle graphene superlattices. *Nature* **556**, 43–50 (2018).

(8) Yankowitz, M. et al. Tuning superconductivity in twisted bilayer graphene. *Science* **363**, 1059–1064 (2019).

(9) Lu, X. et al. Superconductors, orbital magnets and correlated states in magic-angle bilayer graphene. *Nature* **574**, 653–657 (2019).

(10) Cao, Y. et al. Correlated Insulator Behaviour at Half-Filling in Magic-Angle Graphene Superlattices. *Nature* **556**, 80–84 (2018).

(11) Sharpe, A. L. et al. Emergent ferromagnetism near three-quarters filling in twisted bilayer graphene. *Science* **365**, 605–608 (2019).

(12) Bistritzer, R. & MacDonald, A. H. Moiré bands in twisted double-layer graphene. *Proc. Natl. Acad. Sci. USA* **108,** 12233–12237 (2011).

(13) Seyler, K. L. et al. Signatures of moiré-trapped valley excitons in $MoSe_2/WSe_2$ heterobilayers. *Nature* **567,** 66–70 (2019).

(14) Tran, K. et al. Evidence for moiré excitons in van der Waals heterostructures. *Nature* **567**, 71–75 (2019).

(15) Jin, C. et al. Observation of moiré excitons in $WSe_2/WS_2$ heterostructure superlattices. *Nature* **567,** 76–80 (2019).

(16) Wang, L. et al. Correlated electronic phases in twisted bilayer transition metal dichalcogenides. *Nat. Mater.* **19**, 861–866 (2020).

(17) Tsai, K.-T. et al. Correlated Superconducting and Insulating States in Twisted Trilayer Graphene Moire of Moire Superlattices. arXiv:1912.03375 (2019).

(18) Liu, X. et al. Tunable spin-polarized correlated states in twisted double bilayer graphene. *Nature* **583,** 221–225 (2020).

(19) Chen, G. et al. Signatures of tunable superconductivity in a trilayer graphene moiré superlattice. *Nature* **572,** 215–219 (2019).

(20) Hejazi, K., Luo, Z.-X. & Balents, L. Heterobilayer moiré magnets: moiré skyrmions, commensurate–incommensurate transition and more. Preprint at https://arXiv.org/abs/2009.00860 (2020).

(21) Woods, C. R. et al. Commensurate–incommensurate transition in graphene on hexagonal boron nitride. *Nat. Phys.* **10,** 451–456 (2014).

(22) van Wijk, M. M., Schuring, A., Katsnelson, M. I. & Fasolino, A. Relaxation of moiré patterns for slightly misaligned identical lattices: graphene on graphite. *2D Mater.* **2,** 034010 (2015).

(23) Dai, S., Xiang, Y. & Srolovitz, D.J. Twisted Bilayer Graphene: Moiré with a Twist. *Nano Lett.* **16**, 5923–5927 (2016).

(24) Nam, N. N. T. & Koshino, M. Lattice relaxation and energy band modulation in twisted bilayer graphene. *Phys. Rev. B* **96,** 075311 (2017).

(25) Jain, S.K., Juričić, V. & Barkema, G.T. Structure of twisted and buckled bilayer graphene. *2D Mater*. **4**, 015018 (2016).

(26) Zhang, K. & Tadmor, E. B. Structural and electron diffraction scaling of twisted graphene bilayers. *J. Mech. Phys. Solids* **112,** 225–238 (2018).

(27) Yoo, H. et al. Atomic and electronic reconstruction at the van der Waals interface in twisted bilayer graphene. *Nat. Mater.* **18,** 448–453 (2019).

(28) Rosenberger, M.R. et al. Twist angle-dependent atomic reconstruction and moiré patterns in transition metal dichalcogenide heterostructures. *ACS Nano* **14**, 4550–4558 (2020).

(29) Weston, A. et al. Atomic reconstruction in twisted bilayers of transition metal dichalcogenides. *Nat. Nanotechnol.* **15**, 592–597 (2020).

(30) Alden, J. S. et al. Strain solitons and topological defects in bilayer graphene. *Proc. Natl. Acad. Sci. USA* **110,** 11256–11260 (2013).

(31) Kerelsky, A. et al. Maximized electron interactions at the magic angle in twisted bilayer graphene. *Nature* **572,** 95–100 (2019).

(32) Huder, L. et al. Electronic Spectrum of Twisted Graphene Layers under Heterostrain. *Phys. Rev. Lett.* **120,** 156405 (2018).

(33) Bi, Z., Yuan, N. F. Q. & Fu, L. Designing flat bands by strain. *Phys. Rev. B* **100,** 035448 (2019).

(34) Uri, A. et al. Mapping the twist-angle disorder and Landau levels in magic-angle graphene. *Nature* **581,** 47–52 (2020).

(35) Wilson, J.H., Fu, Y., Das Sarma, S. & Pixley, J.H. Disorder in twisted bilayer graphene. *Phys. Rev. Research* **2**, 023325 (2020).

(36) McGilly, L.J. et al. Visualization of moiré superlattices. *Nat. Nanotechnol.* **15**, 580–584 (2020).

(37) Han, Y. et al. Strain Mapping of Two-Dimensional Heterostructures with Subpicometer Precision. *Nano Lett.* **18**, 3746–3751 (2018).

(38) Yang, H. et al. 4D STEM: High efficiency phase contrast imaging using a fast pixelated detector. *J. Phys. Conf. Ser.* **644,** 012032 (2015).

(39) Jiang, Y. et al. Electron ptychography of 2D materials to deep sub-ångström resolution. *Nature* **559,** 343–349 (2018).

(40) Ophus, C. Four-dimensional scanning transmission electron microscopy (4D-STEM): from scanning nanodiffraction to ptychography and beyond. *Microsc. Microanal.* **25,** 563–582 (2019).

(41) Kim, K. et al. van der Waals Heterostructures with High Accuracy Rotational Alignment. *Nano Lett.* **16,** 1989–1995 (2016).

(42) Ozdol, V.B. et al. Strain mapping at nanometer resolution using advanced nano-beam electron diffraction. *Appl. Phys. Lett.* **106**, 253107 (2015).

(43) Kelly, P. *Mechanics Lecture Notes*. (University of Auckland, Auckland, NZ, 2013).

(44) McGinty, B. *Continuum Mechanics* (2012). At https://www.continuummechanics.org

(45) Butz, B. et al. Dislocations in bilayer graphene. *Nature* **505,** 533–537 (2014).

(46) Fang, S. & Kaxiras, E. Electronic Structure Theory of Weakly Interacting Bilayers. *Phys. Rev. B* **93**, 235153 (2016).

(47) Carr, S., Fang, S., Zhu, Z. & Kaxiras, E. Exact continuum model for low energy electronic states of twisted bilayer graphene. *Phys. Rev. Research* **1**, 013001 (2019).

(48) Guinea, F., & Walet, N. R. Continuum models for twisted bilayer graphene: Effect of lattice deformation and hopping parameters. *Phys. Rev. B* **99**, 205134 (2019).

(49) Latychevskaia, T. et al. Holographic reconstruction of interlayer distance of bilayer two-dimensional crystal samples from their convergent beam electron diffraction patterns. *Ultramicroscopy* **219**, 113021 (2020).

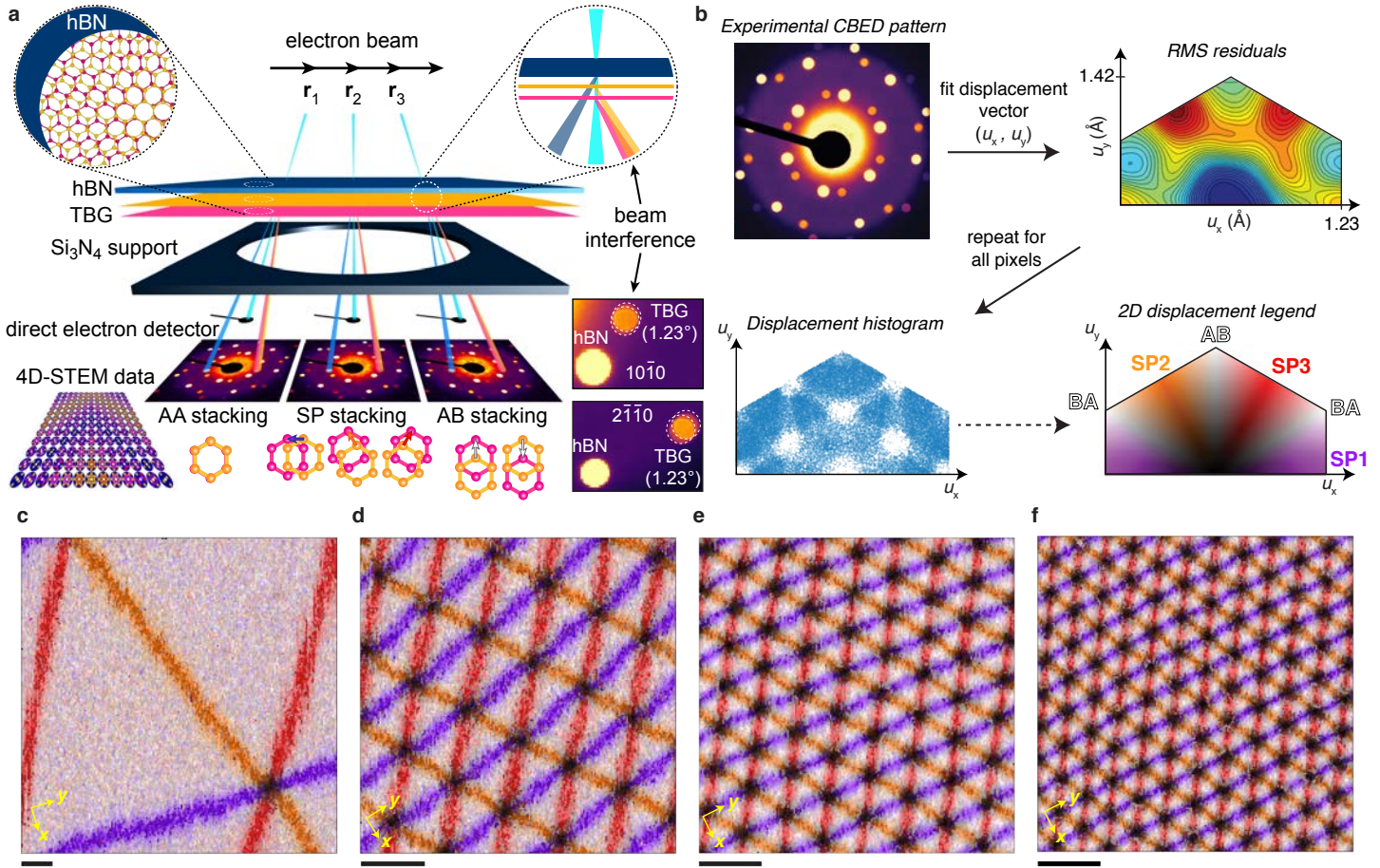

**Figure 1. 4D-STEM Bragg interferometry of TBG. a**, Schematic of 4D-STEM of an hBN/TBG heterostructure, showing Bragg disks of azimuthally misaligned layers for three common TBG stacking orders. Bottom: Common stacking order types with the corresponding displacement vectors depicted with arrows. **b**, Schematic of routine for fitting Bragg disk intensities, $I_j$, to local displacement vectors $\boldsymbol{u}$. Bottom right: Two-dimensional hue–value colourisation scheme used to produce displacement maps from the fitted displacement vectors. Displacement vectors in **a** correspond to $\boldsymbol{u} = (u_x, u_y)$ displacement points in the half-hexagon and are coloured accordingly. **c–f**, Displacement field map for TBG at $\theta_m = 0.16°$ (**c**), 0.63° (**d**), 1.03° (**e**), and 1.37° (**f**). Scale bars: 20 nm.

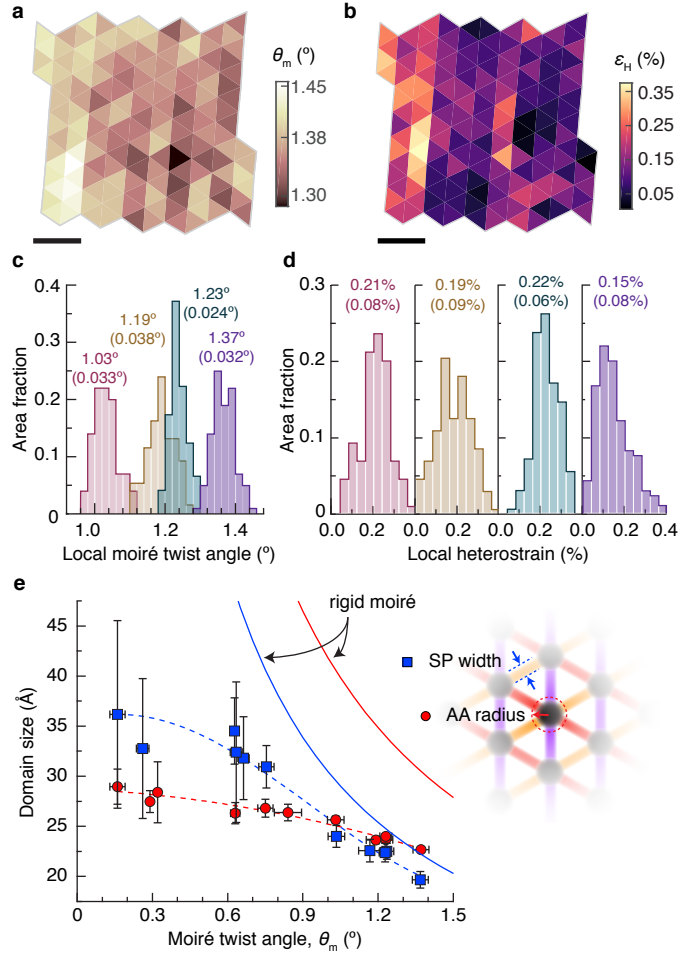

**Figure 2. Short range disorder and geometry analysis of TBG. a,b**, Maps of local twist angle, $\theta_m$ (**a**) and uniaxial heterostrain, $\varepsilon_H$ (**b**) determined from AA-triangulated moiré domains over a region with an average twist angle of 1.37° (from displacement map shown in Figure 1f). Scale bars: 20 nm. **c,d**, Intrinsic local twist angle (**c**) and heterostrain (**d**) disorder for four 100 nm × 100 nm datasets of samples around the magic angle. Mean values are noted with standard deviations in parentheses. **e**, Domain size variation as a function of $\theta_m$ for measured samples (markers) and simulated rigid moiré superlattice (solid lines). Vertical axis error bars represent 95% confidence intervals in domain size and horizontal axis error bars represent standard deviations of $\theta_m$. Dashed lines are polynomial fits to the experimental data that are drawn as visual guides to the overall trends.

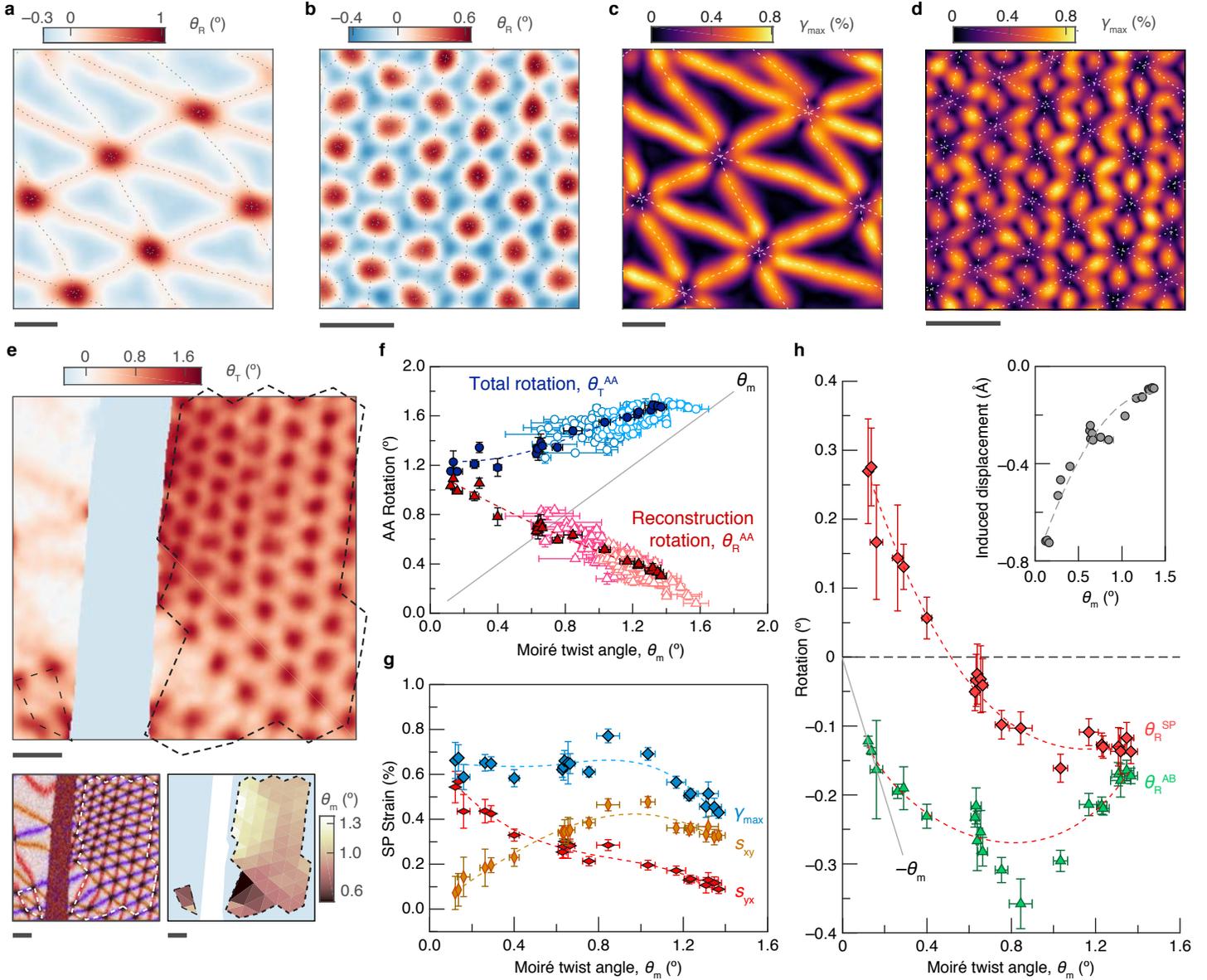

**Figure 3. Strain mapping of TBG. a–d**, $\theta_R$ (**a**, **b**) and $\gamma_{max}$ (**c**, **d**) maps for $\theta_m = 0.26°$ (**a**, **c**) and $\theta_m = 1.03°$ (**b**, **d**). Overlaid dashed lines depict moiré unit cell geometry from displacement maps. $\theta_R$ maps display combined reconstruction rotation of both layers at each pixel and $\gamma_{max}$ maps represent the average strain per graphene layer at each pixel. **e**, $\theta_T$ for a TBG region in the vicinity of a tear in one of the graphene layers. Insets show maps of the displacement field (left) and moiré angle (right). **f**, Total (blue, purple circles) and reconstruction (red, pink triangles) rotation in AA domains as a function of $\theta_m$. Filled markers indicate average values obtained from a complete 4D-STEM dataset over a homogenous twist angle region with a field of view ≥ 50 nm × 50 nm, with vertical-axis error bars depicting 95% confidence intervals. Open markers indicate individual AA domains from two datasets that possess rapidly-changing $\theta_m$ due to a nearby tear (**e** and Supplementary Fig. 8), with error bars from standard deviation of pixels. The solid line represents the moiré rotation. **g**, Three metrics for shear strain in SP domains ($\gamma_{max}$, $s_{xy}$, and $s_{yx}$) as a function of $\theta_m$. **h**, AB and SP local rotation as a function of $\theta_m$ showing crossover in SP reconstruction rotation near $\theta_m = 0.5°$ and AB commensurability criterion $\theta_R^{AB} = -\theta_m$ (grey line) for $\theta_m < 0.2°$. Inset: reconstruction-induced displacement on the boundary of two counter-rotating AB domains (see Supplementary Information). In **g** and **h**, all horizontal-axis error bars depict standard deviations of moiré angles and vertical-axis error bars depict 95% confidence intervals. All scale bars: 20 nm. Dashed lines are polynomial fits to the experimental data that are drawn as visual guides to the overall trends.

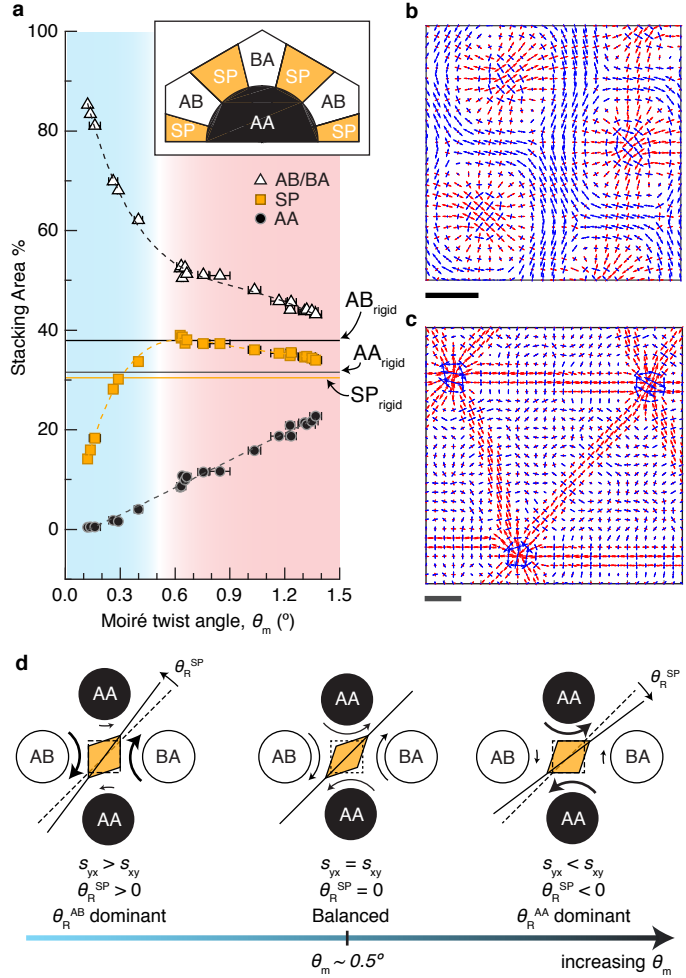

**Figure 4. Regimes of reconstruction in TBG. a**, Variation of relative stacking order areas with twist angle. Solid horizontal lines show the constant stacking area in the case of a rigid moiré (no reconstruction), for comparison. See Methods for stacking order assignment criteria. **b**, **c**, Simple shear decompositions for TBG at $\theta_m = 1.03°$ (**b**) and $\theta_m = 0.14°$ (**c**). Red and blue arrows give the directions and relative magnitudes of the two simple shear components (see Methods for details). Scale bars: 5 nm (**b**) and 20 nm (**c**). **d**, Schematic of the AA- and AB-dominated reconstruction regimes for TBG, explaining the observed changes in simple shear and SP reconstruction rotation.

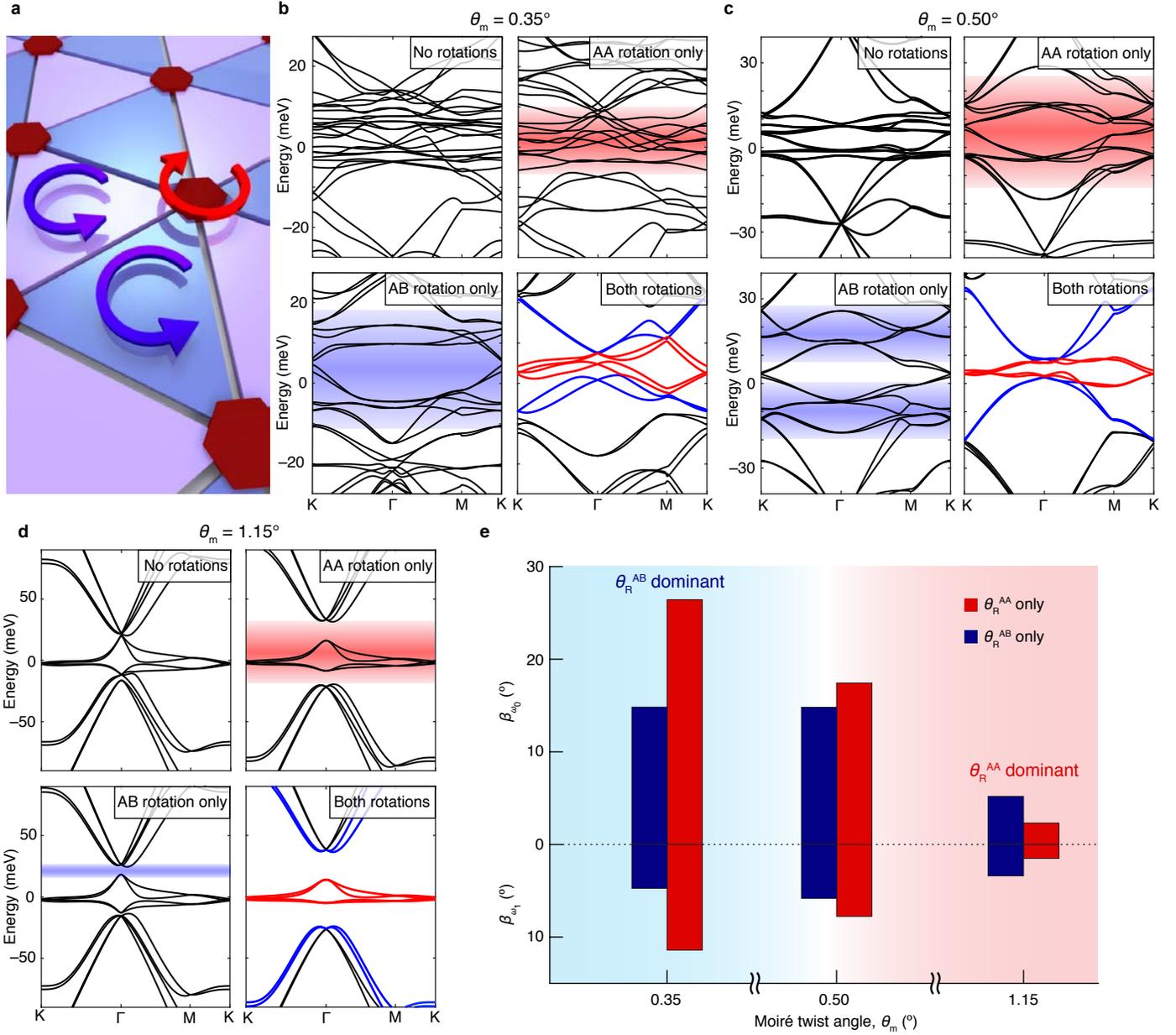

**Figure 5. Effects of isolated relaxation modes on band structure. a**, Schematic of the TBG moiré superlattice, with the AA stacking regions shown as red hexagons, the AB (BA) stacking regions as blue (purple) triangles, and the two rotation modes caused by atomic relaxation: $\theta_R^{AA}$ (red arrow) and $\theta_R^{AB}$ (blue arrows). **b–d**, Calculated band structures for TBG at $\theta_m$ = 0.35° (**b**), 0.5° (**c**), and 1.15° (**d**) under various rotation assumptions. The band structure effects caused by AA (AB) rotation are highlighted in red (blue). **e**, The similarity in the *ab-initio* calculated electronic interlayer scattering between the application of singular rotations (either $\theta_R^{AA}$ or $\theta_R^{AB}$) and the full reconstruction (both $\theta_R^{AA}$ and $\theta_R^{AB}$) is given by a generalized 'angle', $\beta$: a smaller angle indicates better agreement with full reconstruction. This similarity is assessed for interlayer scattering between similar ($\beta_{\omega_0}$) and dissimilar ($\beta_{\omega_1}$) orbitals for both $\theta_R^{AA}$ only (red) and $\theta_R^{AB}$ only (blue) relaxation assumptions.

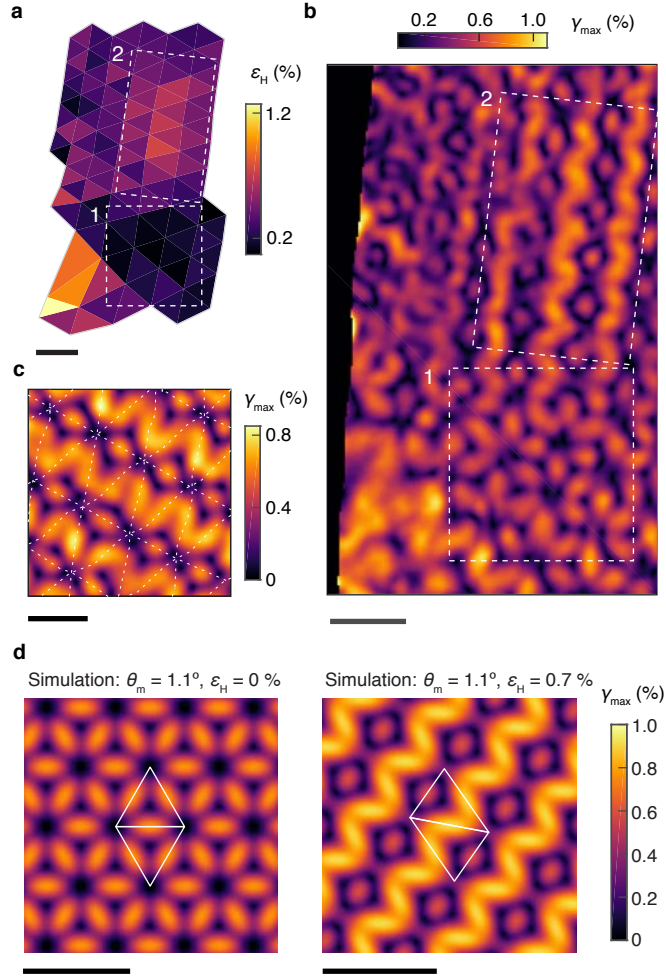

**Figure 6. Visualizing the effect of uniaxial heterostrain on TBG reconstruction. a**, Map of heterostrain, $\varepsilon_H$, determined from AA triangulation (see Methods) over the sample shown in Figure 3e. Boxes 1 and 2 highlight two areas with similar $\theta_m \sim 1.1°$ but possessing significantly different amounts of $\varepsilon_H$. **b**, Map of $\gamma_{max}$ (average per layer) over the region in a showing six-fold symmetric SP strain patterns in box 1 (minimal $\varepsilon_H$) and striped strain features in box 2 ($\varepsilon_H \sim 0.7\%$). **c**, Map of $\gamma_{max}$ (average per layer) over a homogenously heterostrained ($\varepsilon_H \sim 0.45\%$) sample with $\theta_m = 0.63°$ showing pronounced zig-zag features. Overlaid dashed lines depict the moiré pattern based on the displacement field maps. **d**, Calculations from a finite-element relaxation method using parameters extracted from density functional theory (see Methods) for $\theta_m = 1.1°$ without (left) and with (right) uniaxial heterostrain ($\varepsilon_H = 0.7\%$). Solid lines depict the moiré supercell. All scale bars: 20 nm.

## Methods

**Sample preparation**

TBG samples were fabricated using the common 'tear and stack' technique[41,50]. Briefly, monolayer graphene (from Kish graphite, Graphene Supermarket) and ~5 nm thick hBN were mechanically exfoliated onto $SiO_2$/Si substrates and selected using optical microscopy and atomic force microscopy. A poly bisphenol-A-carbonate (PC)/polydimethylsiloxane (PDMS) stamp was used to pick up the hBN. The hBN was then engaged with half of a monolayer graphene crystal and the edge of the hBN was used to tear the graphene in half. The substrate was then rotated by $\theta_m$ prior to picking up the remaining half of the graphene monolayer. During this stacking process, the hBN and graphene lattices were deliberately misaligned >10° using the straight edges of the crystal layers as guides to prevent overlap of the hBN and graphene diffraction disks during 4D-STEM. Finally, the hBN/TBG stack was released from the polymer stamp onto a 50 nm thick amorphous $Si_3N_4$ membrane with 2 µm holes for imaging (Supplementary Fig. 1a). Atomic force microscope images (Supplementary Fig. 1b) show that the stack slightly bends over the edges of the holes in the support and the sample is nearly flat over the majority of the region of interest.

**Electron microscopy measurements**

Electron microscopy was performed at the National Center for Electron Microscopy facility in the Molecular Foundry at Lawrence Berkeley National Laboratory. Low magnification dark-field TEM images were acquired using a Gatan UltraScan camera on a Thermo Fisher Scientific Titan-class microscope operated at 60 kV. Three frames each with an acquisition time of 5 s were summed to produce each dark-field image. These dark-field images were used as a reference for selecting regions of interest for 4D-STEM (Supplementary Fig. 1c).

4D-STEM data sets were acquired using a Gatan K3 direct detection camera located at the end of a Gatan Continuum Imaging Filter on the TEAM I microscope (an aberration-corrected Thermo Fisher Scientific Titan 80 – 300) operated in EFSTEM mode at 80 kV with a 10 eV energy filter centred around the zero-loss peak. In general, two sets of acquisition conditions were used, involving convergence semi-angles of 1.71 mrad (condition A) and 3 mrad (condition B), both of which allowed sufficient signal to noise and avoided overlap between the hBN and graphene diffraction disks. The beam current was 62 – 65 pA and 68 pA for conditions A and B, respectively. By fitting the centre lobe of the STEM probes in real space using a 2D Gaussian function, we measured the full-width at half-maximum (FWHM) values to be 1.3 nm (A) and 0.8 nm (B). Diffraction patterns were collected using a step size of 0.5 nm with 100 × 100 to 300 × 300 scan positions covering an area of 50 nm × 50 nm to 150 nm × 150 nm. The K3 camera was used in full frame electron counting mode with a binning of 4 and an EFSTEM camera length of 800 mm. Each diffraction pattern had an exposure time of 10 – 13 ms, which is a sum of multiple counted frames.

**Displacement Fitting Overview**

A multi-step procedure is followed to convert a 4D-STEM dataset into a TBG displacement map (see Supplementary Information for more details). First, the background scattering is fit and removed from each diffraction pattern so as not to bias the convergent beam electron diffraction (CBED) disk intensities. Next, the overlap regions between each pair of first-order TBG CBED disks are manually defined and all pixel intensity values in each region are summed, converting each diffraction pattern into a twelve-component vector. A displacement vector is then calculated via nonlinear regression, where equation (1) in the text is the fitting function. Equation (1) can be derived from the weak phase approximation (see Supplementary Information). Though the prefactor coefficients $A_j$ are not known *a priori*, they too are fit via the multi-step nonlinear regression procedure discussed in the Supplementary Information. Repeating the fit for each real-space pixel (i.e., each individual diffraction pattern) produces the displacement maps shown in Figs. 1c–f and Supplementary Fig. 2. Note that this fitting process produces some bias due to finite probe width, which is later removed by a filter; see Supplementary Information Section 3 for details.

**Strain Mapping**

To obtain the strain field, we differentiate the displacement field data according to infinitesimal strain theory[43,44,51]. The displacement fitting procedure produces vectors contained entirely within the half-hexagon fitting region, which displays discontinuities at the edges. Before differentiation can occur, the displacement data must therefore be phase-unwrapped to both eliminate the 180° ambiguity[52] and also establish a continuous vector field between adjacent moiré domains. This multidimensional phase-unwrapping problem presents a problem for standard algorithms, which typically can handle either 180° ambiguity or multidimensional unwrapping but not both. To overcome this problem, we developed a geometry-based approach. AA, AB, and SP regions are algorithmically detected from their characteristic displacement vectors, and the stacking order change from crossing each SP region is stored. An initial reference displacement is assigned to a starting AB domain centroid, and then a reference displacement vector is assigned to each neighbouring AB centroid by finding the vectors that satisfy the SP stacking order change criteria. Each AB centroid is assigned recursively. Next, each individual real space pixel is assigned to an AB centroid via the geometry domain registration, so each pixel has a reference vector indicating the approximate region where its phase-unwrapped displacement vector should fall. The precise vector is obtained by choosing the new displacement vector that (a) produces an equivalent interferometry pattern to that of the original displacement vector in the half-hexagon fitting region, and (b) is as close as possible to the reference vector. The unwrapped displacement field is then refined via a 3 × 3 moving window that interchanges displacement vectors to maximize local continuity. After the moving window is applied ten times, remaining discontinuities at the AA and SP boundaries are eliminated. Note that the phase unwrapping process does not change the model fit to the virtual dark field images because each unwrapped displacement vector predicts the same interferometry pattern as the original vector. The unwrapping process converts a single half-hexagon fitting region into a series of continuously connected fitting regions amenable to differentiation, as shown in Supplementary Fig. 14. The unwrapped displacement field is denoised (Supplementary Fig. 15) by Total Generalized Variation (TGV)[53,54] and differentiated to produce the strain maps. We note that the calculated strain and rotation values can change by around ±10% depending on the exact TGV filter settings used, implying some systematic uncertainty in the exact magnitude of the reconstruction strain. However, filter settings are kept consistent so that the twist angle trends are not impacted and the good agreement between FEM simulations and experiment provides support for the filter settings used.

We draw a distinction between the interlayer displacement field $u_{inter}(x,y)$ as opposed to the single-layer or intralayer displacement field $u_{intra}(x,y)$. The experimentally measurable quantity is $u_{inter}(x,y)$, which determines the lattice plane offset giving rise to the interferometry signal. However, to obtain strain quantities that relate to the deformation experienced by a single layer of graphene, it is useful to consider a reference state in which both layers have AA-commensurate stacking, reaching their final positions by some equal-and-opposite combination of rotation and deformation. Because rotations are ≤ 2°, the small angle approximation implies that $u_{intra}(x,y) = u_{inter}(x,y)/2$. In its present form, Bragg interferometry does not give the means of separately resolving strain fields in the top and bottom layers of graphene, so therefore $u_{intra}(x,y)$ is the best quantity to use to determine the average deformation of a single layer of graphene.

A vector-valued displacement field $u(x,y)$ has four associated derivatives, arising from the gradients of the scalar-valued $x$ and $y$ displacements. The elements of the strain tensor can then be calculated from the displacement field derivatives as follows:

$$\epsilon_{xx} = \frac{\partial u_{intra,x}}{\partial x} = \frac{1}{2}\left(\frac{\partial u_{inter,x}}{\partial x}\right) \tag{2}$$

$$s_{xy} = \frac{\partial u_{intra,x}}{\partial y} = \frac{1}{2}\left(\frac{\partial u_{inter,x}}{\partial y} + \theta_m\right) \tag{3}$$

$$s_{yx} = \frac{\partial u_{intra,y}}{\partial x} = \frac{1}{2}\left(\frac{\partial u_{inter,y}}{\partial x} - \theta_m\right) \tag{4}$$

$$\epsilon_{yy} = \frac{\partial u_{intra,y}}{\partial y} = \frac{1}{2}\left(\frac{\partial u_{inter,y}}{\partial y}\right) \tag{5}$$

$$\epsilon_{xy} = \frac{1}{2}(s_{xy} + s_{yx}) \tag{6}$$

The strain tensor is formally composed of the normal strains $\epsilon_{xx}$ and $\epsilon_{yy}$ for the diagonal elements along with the tensorial pure shear strain $\epsilon_{xy}$ on the off-diagonal elements. The terms $s_{xy}$ and $s_{yx}$ are referred to as the simple shear strains[43]. These terms contain information about both strain and fixed-body rotation. To analyse the simple shear quantities arising from reconstruction, we first remove the moiré rotation $\theta_m$ so that only reconstruction rotation will be included. $\theta_m$ is estimated from the moiré superlattice geometry via triangulation (Fig. 2a and Supplementary Fig. 3). For the twist-angle-homogenous datasets, we use the average $\theta_m$ value across the field of view in the strain equations. This procedure has no effect on the strain tensor itself, as the $\theta_m$ terms cancel out in the sum for calculating $\epsilon_{xy}$ (equation (6)). Note that $\epsilon_{xy}$ is the correct term to use for the strain tensor to perform correct tensor rotations, but is not directly comparable in magnitude to the normal strains $\epsilon_{xx}$ and $\epsilon_{yy}$. It is therefore useful to define the 'engineering' pure shear strain as follows[43,44,51]:

$$\gamma_{xy} = s_{xy} + s_{yx} = 2\epsilon_{xy} \tag{7}$$

The quantity $\gamma_{xy}$ exerts the same magnitude of deformation per unit strain as $\epsilon_{xx}$ and $\epsilon_{yy}$.

The quantities $\epsilon_{xx}$, $\epsilon_{yy}$, $s_{xy}$, $s_{yx}$, and $\epsilon_{xy}$ (and therefore $\gamma_{xy}$) are each dependent on the choice of coordinate axes in which to visualize the strain tensor (Supplementary Figs. 4–6). This makes it challenging to compare strains between different SP regions in one image. To overcome this, it is useful to employ the principal strain equations[44]:

$$\varepsilon_{max} = \frac{\epsilon_{xx} + \epsilon_{yy}}{2} + \sqrt{\left(\frac{\epsilon_{xx} - \epsilon_{yy}}{2}\right)^2 + (\epsilon_{xy})^2} \tag{8}$$

$$\varepsilon_{min} = \frac{\epsilon_{xx} + \epsilon_{yy}}{2} - \sqrt{\left(\frac{\epsilon_{xx} - \epsilon_{yy}}{2}\right)^2 + (\epsilon_{xy})^2} \tag{9}$$

These equations correspond to local rotations of the tensor coordinate system in order to express the strain at each pixel entirely in terms of normal strain. The tensor rotation angle is known as the principal angle,

$\theta_P$ [44]:

$$\tan(2\theta_P) = \frac{2\epsilon_{xy}}{\epsilon_{xx} - \epsilon_{yy}} \qquad (10)$$

The principal angle describes the orientation of the rotated x-axis (that is, the orientation of $\varepsilon_{max}$) relative to the starting coordinate system of the tensor. By convention, we choose the direction perpendicular to the purple SP region (SP 1) as the x-axis starting coordinate system. The direction of maximum shear is located 45° counter-clockwise from $\theta_P$. Thus, for a SP region undergoing shear strain due to either AA or AB rotation, $\theta_P$ should be offset by 45° from the direction of the SP region. We find this is indeed the case, confirming that the strain in SP regions is predominantly characterized by shearing (Supplementary Fig. 9).

The maximum shear strain (also known as principal shear strain), $\gamma_{max}$, occurs at a 45° angle from the $\varepsilon_{max}$ coordinate axis at each pixel, and is given by the difference in principal strains[44]:

$$\gamma_{max} = \varepsilon_{max} - \varepsilon_{min} \qquad (11)$$

Because the intralayer strain in TBG arises almost entirely from shearing processes, $\gamma_{max}$ is a natural quantity to summarize the strain mechanics of a sample in one image (Figure 3c–d, Figure 6b–c and Supplementary Figs. 7, 8, 11, 16). As noted in the main text, $\gamma_{max}$ does not require definition of a local tensor coordinate system, unlike the elements of the strain tensor defined in Eqs. 2–7.

In the evaluation of the magnitude of simple strains $s_{xy}$ and $s_{yx}$ as a function of $\theta_m$ in Fig. 3g, for each displacement field map at a particular $\theta_m$, the simple shear strains have their axes rotated three times for maximum compatibility with each of the three SPs, and then the $s_{xy}$ and $s_{yx}$ values from these three tensor rotations are averaged to plot the result vs $\theta_m$.

All equations thus far have considered the average intralayer strain experienced when a single layer of graphene deforms due to reconstruction. When analysing rotation, however, it is more natural to consider the effect of both layers simultaneously to obtain the relative rotational misalignment. The relationship between simple shear and interlayer reconstruction rotation, $\theta_R$, is therefore given by

$$\theta_R = s_{yx} - s_{xy} \qquad (12)$$

The fixed-body rotation equation normally has a factor of one-half; however, here we have multiplied by two to emphasise that we have gone from *intralayer* quantities $s_{xy}$ and $s_{yx}$ to the *interlayer* quantity $\theta_R$ (Figs. 3a,b). The total rotation, including the moiré rotation, can be expressed from the interlayer displacement field directly:

$$\theta_T = \frac{1}{2}\left(\frac{\partial u_{inter,y}}{\partial x} - \frac{\partial u_{inter,x}}{\partial y}\right) \qquad (13)$$

We employ this expression when the moiré angle is changing rapidly over the field of view (Fig. 3e and

Supplementary Fig. 8). The moiré angle and the reconstruction angle are related by
$$\theta_T = \theta_m + \theta_R \tag{14}$$

**Simple Shear Decomposition**

Simple shear strain from AA and AB/BA rotation constitutes the dominant strain mechanic in TBG reconstruction, incorporating both pure strain and fixed-body rotation as discussed previously. While the maximum shear strain $\gamma_{max}$ is commonly used to isotropically visualize the strain component, we seek a similar metric for visualization of both strain and rotation effects found in simple shear. We construct a simple shear decomposition to show the magnitude and direction of simple shear, by analogy to the principal strains technique for normal strain. In principal strain, the coordinate system is rotated to diagonalize the strain tensor, thereby completely eliminating shear strain[44]:

$$\begin{bmatrix} \varepsilon_{max} & 0 \\ 0 & \varepsilon_{min} \end{bmatrix} = \begin{bmatrix} \cos(\theta_P) & \sin(\theta_P) \\ -\sin(\theta_P) & \cos(\theta_P) \end{bmatrix} \begin{bmatrix} \epsilon_{xx} & \epsilon_{xy} \\ \epsilon_{xy} & \epsilon_{yy} \end{bmatrix} \begin{bmatrix} \cos(\theta_P) & -\sin(\theta_P) \\ \sin(\theta_P) & \cos(\theta_P) \end{bmatrix} = \mathbf{Q^T E Q} \tag{15}$$

Here $\theta_P$ is the principal angle, defining the rotated coordinate system. Analogously, we could seek to "off-diagonalize" the "strain-rotation" tensor to obtain a simple shear strain description:

$$\begin{bmatrix} 0 & s_1' \\ s_2' & 0 \end{bmatrix} = \begin{bmatrix} \cos(\theta_s) & \sin(\theta_s) \\ -\sin(\theta_s) & \cos(\theta_s) \end{bmatrix} \begin{bmatrix} \frac{\partial u_{intra,x}}{\partial x} & \frac{\partial u_{intra,x}}{\partial y} \\ \frac{\partial u_{intra,y}}{\partial x} & \frac{\partial u_{intra,y}}{\partial y} \end{bmatrix} \begin{bmatrix} \cos(\theta_s) & -\sin(\theta_s) \\ \sin(\theta_s) & \cos(\theta_s) \end{bmatrix} \tag{16}$$

This equation has one free variable ($\theta_s$, the simple shear angle) but two variables on the diagonal to eliminate (call them $\varepsilon_{xx}'$ and $\varepsilon_{yy}'$). Thus, the equation will in general not have an exact solution, but we can solve for $\theta_s$ in the least-squares sense to minimize $\varepsilon_{xx}'^2 + \varepsilon_{yy}'^2$. By performing this least-squares regression for each pixel, we obtain the best possible simple shear representation of the strain field, which we refer to as the simple shear decomposition. $\theta_s$ obtained in this way has a 90° phase ambiguity, which can make visualization challenging. To obtain components of a continuous simple shear vector field, we examine the rotated tensor value $s_2'$ for both $\theta_s$ and $\theta_s + 90°$, and choose the simple shear angle which maximizes the signed value of $s_2'$. When plotted as a vector field quiver plot with two-headed arrows, the components $s_1'$ and $s_2'$ take on the natural interpretation of simple shear strain produced by AA and AB/BA reconstruction rotation (Figure 4b, c).

**Rotational Calibration and Sample Drift**

The displacement vectors obtained from the fitting process are initially aligned to the diffraction pattern. To perform strain mapping via displacement field differentiation, however, it is essential that the displacement vectors be aligned to the real space scan direction of the image. This requires knowing the rotational calibration between the diffraction pattern and the scan direction. To accomplish this, we obtained defocused images of Au nanoparticles[55], determining the diffraction pattern is rotated 19.0° degrees clockwise from the STEM image. Alternatively, a self-consistency approach can be used where the coordinate axes are rotated so that purple soliton walls are parallel to the y-axis, the expected orientation on account of their displacement vectors (Fig. 1c–f). We found that the two methods typically agreed within a few degrees, and the choice between them did not introduce substantive changes in the strain maps.

A distorted moiré image alone is insufficient evidence to conclude the presence and magnitude of heterostrain, as sample drift could induce similar distortions. To estimate the amount of sample drift present, we collected replicate images at different STEM scan angles of two twist-angle-homogeneous regions that exhibited heterostrain in the dark-field TEM images (Supplementary Fig. 16). Because sample drift is typically determined by the orientation of sample holder and not the STEM scan direction, the distortions produced by two subsequent scans should be different for different scan directions relative to the true moiré geometry[56]. We compute the average angles between different SP regions to quantify the change in unit cell distortion. For both pairs of images, the change in angle with STEM scan direction is no greater than 2°, while the difference between the smallest and largest SP region angles is greater than 20° (Supplementary Figs. 16 and Supplementary Table 1). Furthermore, the 1D shear strain features discussed in the text (Fig. 6b,c) rotate consistently with the STEM scan direction (Supplementary Fig. 16). We conclude that the moiré superlattice distortions seen in our images can be reliably attributed to heterostrain. This conclusion is further corroborated by (1) conventional dark-field TEM images of the 4D-STEM scan areas (Supplementary Fig. 1) that also reveal these same distortions, and (2) the strong variations observed within individual scans, particularly near a tear in one graphene layer (Fig. 6b,c and Supplementary Fig. 8).

**Geometry and Strain Trends**

AA region radii (Fig. 2e) are calculated by curve-fitting the displacement amplitude to a two-dimensional Gaussian function with equal variances and no correlation. Pixels with strong SP character are removed from the fit so as not to bias the background AB/BA displacement amplitude. The reported radii are for the circular level curve of the Gaussian at a displacement amplitude[21] of 0.71 Å. SP region widths are calculated on the basis of the displacement vector angle with the origin before phase unwrapping. Each pixel is assigned an angle score between 0 (displacement angle equivalent to precise AB/BA stacking) and 1 (angle equivalent to precise SP stacking). The angle scores are interpolated perpendicular to the boundary of the SP region, and the angle score threshold of 0.5 is used to determine the width of the SP region. Both AA and SP geometry fits are performed without TGV filtering.

Simple shear strain trends (Fig. 3g) in $s_{xy}$ and $s_{yx}$ were obtained as averages over all three SP directions. For each SP direction, the $s_{xy}$ and $s_{yx}$ values were computed in the right-handed tensor coordinate system with x-axis perpendicular to the SP and y-axis parallel to the SP direction.

Stacking area trends (Fig. 4a) are determined by partitioning the displacement vectors into three stacking order categories (AA, SP, and AB/BA) as depicted in the inset of Figure 4a. All displacement vectors with amplitude less than 0.71 Å are assigned to AA stacking[21], while the remaining vectors are assigned to whichever pure stacking order is closer in displacement space (AB/BA or SP) leading to maps shown in the first row of Supplementary Fig. 17a–d. To avoid the influence of outliers, this calculation is performed on the TGV filtered data. A second stacking area analysis was also conducted using a five-category partition (see Supplementary Figure 17).

All strain and rotation trends are obtained through an ROI-based approach. Based on the geometry registration obtained during the phase-unwrapping process, masks are built selecting all pixels within a given distance of the registration position. For AA regions, all pixels within 1 nm of the AA centre are included. For SP regions, all pixels within 1 nm of the line down the centre of the region are included, excepting a mask of variable size that prevents the AA region from being used. For AB/BA domains, first the AA and SP regions are removed with wide masks, and then the remaining area is used as the AB region. Therefore, transitional pixels between two domains are not included in these calculations. Within each masked region, all pixels are averaged to produce the calculated value for that specific domain, and then all domains are averaged together to produce the single reported value for the dataset.

### *Ab-initio* Electronic Structure Models

We used a first principles multiscale approach for the analysis of the interlayer electronic tunnelling functions between the twisted graphene layers and the electronic band structure calculations. An interlayer coupling function for carbon atoms, derived from density functional theory[46], is applied to a low-energy continuum model which can exactly reproduce tight binding results from twisted and atomically relaxed supercells[47,48]. In generating our continuum models, the tunnelling between orbitals of opposite layers is Fourier transformed to assess all possible Umklapp scattering processes introduced by the twist angle. Additional details pertaining to calculation of electronic tunnelling functions, band structures, and finite element simulations are provided in the Supplementary Information.

### Computational Implementation

4D-STEM image processing and analysis was conducted in MATLAB (version ≥ R2016b) on a personal computer. Total generalized variation (TGV) denoising was conducted according to a published algorithm[54,55]. All other code for analysis of 4D-STEM data in this project was custom-written by the authors. Additional details on 4D-STEM data analysis are provided in the Supplementary Information.

**Code Availability**

The computer code used for the dataset processing and strain analysis has been made publicly available at https://github.com/bediakolab/StrainFieldsInTwistedBilayerGraphene.

**References**

(50) Cao, Y. *et al.* Superlattice-Induced Insulating States and Valley-Protected Orbits in Twisted Bilayer Graphene. *Phys. Rev. Lett.* **117,** 116804 (2016).

(51) Boresi, A. P. & Schmidt, R. J. *Advanced Mechanics of Materials*. 55–72 (John Wiley & Sons, Inc., 2003).

(52) Metcalf, T. R. Resolving the 180-degree ambiguity in vector magnetic field measurements: the 'minimum' energy solution. *Solar Phys.* **155**, 235–242 (1994).

(53) Lu, W. et al. Implementation of higher-order variational models made easy for image processing. *Math. Method. Appl. Sci.* **39**, 4208–4233 (2016).

(54) Duan, J. et al. An edge-weighted second order variational model for image decomposition. *Digit. Signal Process.* **49**, 162–181 (2016).

(55) Savitzky, B. et al. py4DSTEM: a software package for multimodal analysis of four-dimensional scanning transmission electron microscopy datasets. Preprint at https://arxiv.org/abs/2003.09523 (2020).

(56) Ophus, C., Ciston, J. & Nelson, C. T. Correcting nonlinear drift distortion of scanning probe and scanning transmission electron microscopies from image pairs with orthogonal scan directions. *Ultramicroscopy* **162**, 1–9 (2016).

*Supplemental Information*

# Strain fields in twisted bilayer graphene

Nathanael P. Kazmierczak,[1,2†] Madeline Van Winkle,[1†] Colin Ophus,[3] Karen C. Bustillo,[3] Stephen Carr,[4,5] Hamish G. Brown,[3] Jim Ciston,[3] Takashi Taniguchi,[6] Kenji Watanabe,[7] D. Kwabena Bediako[1,8,9]*

[1] *Department of Chemistry, University of California, Berkeley, CA 94720, USA*
[2] *Department of Chemistry and Biochemistry, Calvin University, Grand Rapids, MI 49546, USA*
[3] *National Center for Electron Microscopy, Molecular Foundry, Lawrence Berkeley National Laboratory, Berkeley, CA 94720, USA*
[4] *Department of Physics, Brown University, Providence, Rhode Island 02912, USA*
[5] *Brown Theoretical Physics Center, Brown University, Providence, Rhode Island 02912, USA.*
[6] *International Center for Materials Nanoarchitectonics, National Institute for Materials Science, 1-1 Namiki, Tsukuba 305-0044, Japan*
[7] *Research Center for Functional Materials, National Institute for Materials Science, 1-1 Namiki, Tsukuba 305-0044, Japan*
[8] *Chemical Sciences Division, Lawrence Berkeley National Laboratory, Berkeley, CA 94720, USA*
[9] *Azrieli Global Scholar, Canadian Institute for Advanced Research (CIFAR), Toronto, Ontario M5G 1M1, Canada*
* Correspondence to: bediako@berkeley.edu
† These authors contributed equally to this work

**Table of Contents**

## 1. Derivation of the Fitting Function

To derive Eq. 1 in the text, we assume a bilayer graphene structure where each monolayer has a projected electrostatic potential given by $V(\mathbf{r})$. The layers are locally displaced from each other by an amount $\mathbf{u}$, a two-dimensional vector quantity which varies across the specimen. Since the bilayer thickness (< 1 nm) is much smaller than the depth of field of the electron probe (>100 nm for an 80 keV probe with 1.71–3.00 mrad convergence semi-angle) and the electron illumination wave function $\Psi_{\text{illum}}(\mathbf{r})$ will only be weakly scattered by the two layers of graphene, we may express the scattered electron probe wavefunction $\Psi(\mathbf{r})$ to a very good approximation using the weak phase object approximation[1]:

$$\Psi(\mathbf{r}) = e^{i\sigma V(\mathbf{r}-\mathbf{u}/2)+i\sigma V(\mathbf{r}+\mathbf{u}/2)}\Psi_{\text{illum}}(\mathbf{r}) \approx \left(1 + i\sigma V(\mathbf{r}-\mathbf{u}/2) + i\sigma V(\mathbf{r}+\mathbf{u}/2)\right)\Psi_{\text{illum}}(\mathbf{r}) \qquad (S1)$$

We assume an aberration-free focused electron probe which has the reciprocal space form of the aperture function, $\widehat{\Psi}_{\text{illum}}(\mathbf{k}) = \hat{A}(\mathbf{k})$, which is a top-hat function equal to 1 for $|\mathbf{k}| < \mathbf{k}_{max}$ and 0 otherwise. Here $\mathbf{k}_{max}$ is the size of the probe-forming aperture in units of inverse length. In the diffraction plane the measured scattered electron probe intensity is given by the modulus squared Fourier transform of Eq. S1,

$$|\Psi(k)|^2 = \left|\left(\delta(k) + i\sigma\sum_g V_g \delta(\mathbf{k}-\mathbf{g})\,e^{i\pi \mathbf{g}\cdot \mathbf{u}} + i\sigma\sum_g V_g \delta(\mathbf{k}-\mathbf{g})\,e^{-i\pi \mathbf{g}\cdot \mathbf{u}}\right) \otimes \hat{A}(k)\right|^2$$

$$= \hat{A}(\mathbf{k}) + 4\sigma^2 \sum_g \hat{A}(\mathbf{k}-\mathbf{g})\,\cos^2(\pi \mathbf{g}\cdot \mathbf{u}) \qquad (S2)$$

Here δ is the Dirac delta function and ⊗ represents convolution. It has been assumed that the probe forming aperture radius, $\mathbf{k}_{max}$, is less than half the size of the separation of the Fourier coefficients of the monolayer $V_g$, that is, there is no overlapping of different Bragg disks from the same layer in the diffraction pattern. The local displacement of the monolayers can be measured through the $cos^2(\pi \mathbf{g} \cdot \mathbf{u})$ term.

## 2. Displacement Fitting Details

To remove the unwanted background scattering, the averaged diffraction pattern is fit to a Lorentzian function after masking off all convergent beam electron diffraction (CBED) disks and the beamstop. The residuals of the background fit are interpolated radially through the CBED disks from the centre of the diffraction pattern, and both the fit and the interpolated residuals are subtracted from each diffraction pattern. This correction removes unwanted scattering contributions from the CBED intensities. Next, the overlap regions between each pair of TBG CBED disks are manually defined, and all pixel intensity values in each region are summed. Because we use all first order TBG Bragg reflections in the fitting procedure, this converts each diffraction pattern into a twelve-component vector characterizing the local interferometry pattern. Shifts in the CBED disk positions contribute no information in this method. Therefore, the manually defined summation region may be centred on the reciprocal lattice vector from either graphene layer, as long as the summation region lies entirely within the CBED overlap region. In general, we place the summation region on the centre of the CBED overlap region, corresponding to the average of the reciprocal lattice vectors from each layer. This procedure allows visualization of the virtual dark-field images from each TBG reflection pair (Supplementary Fig. 1c). Note that this technique is superior to sequentially acquiring twelve separate conventional dark-field (DF) images for each reflection, as it eliminates systematic/correlated errors due to sample drift in-between acquisitions, does not incidentally incorporate signals from adjacent hBN diffraction disks owing to the typical sizes of selected area diffraction apertures, and enables measurement of a wide range of twist angles without changing microscope parameters.

The twelve "virtual" dark-field datasets are then converted into a single displacement map by nonlinear regression (Supplementary Fig. 12). The fitting function for any given pixel is given by Eq. 1 in the main text, where the $I_j$ values for each of the twelve reflections ($j \in \{\langle 1100 \rangle, \langle 2110 \rangle\}$) are the response variables and the two-component displacement vector $\boldsymbol{u}$ is the predictor variable. Eq. 1 can be derived from the weak phase approximation (Eq. S1). We confine $\boldsymbol{u}$ to the half-hexagon fitting region shown in Fig. 1b so that each $\boldsymbol{u}$ predicts a unique interferometry pattern. This region contains all possible shortest vectors from a lattice site in layer 1 to layer 2 or vice-versa, modulo inversion through the origin. This 180° phase ambiguity implies that AB and BA regions give identical interferometry patterns for on-zone-axis experiments. Subject to these constraints, the nonlinear regression finds the unique $\boldsymbol{u}$ that best predicts the intensity values for each pixel.

To account for the unknown prefactor coefficients $A_j$, a three-step fitting process is followed to obtain the optimized $A_j$ values (Supplementary Fig. 13). First, the $A_j$ values are estimated by manually defining an AB/BA-stacked region and averaging all virtual dark-field intensities within for each disk. Eq. 1 in the main text predicts that the AB/BA stacking order gives $\langle 1100 \rangle$ overlap regions of $0.25A_j$ and $\langle 2110 \rangle$ overlap regions of simply $A_j$ (see relative intensities in Supplementary Fig. 1d). This is only an initial estimate, chosen because AB/BA regions are the most readily spotted from superimposition of the virtual dark-field images. In the first regression, the $A_j$ calculated this way are held constant while the $\boldsymbol{u}$ is optimized separately for each pixel. Multiple local minima can arise on the optimization surface because of the trigonometric fitting function; therefore, twelve gradient-based optimization runs are initiated from different locations in the fitting region to ensure global convergence for each pixel. The initial estimate of the displacement map obtained this way is typically already quite good (Supplementary Fig. 13). In the second fit, the $A_j$ are allowed to optimize simultaneously with all ***u*** for each pixel, using the displacement map obtained in the first fit as a starting guess. Owing to the large scale of this regression, the multistart optimization strategy employed in the first fit cannot be used. Finally, the first pixel-by-pixel fit is performed again, but this time using the optimized values of the $A_j$ prefactors from the second fit. The displacement vectors from this third fit are lightly filtered to remove outliers on the basis of amplitude, via deviations from the median value in a 5 × 5 moving pixel window. Where possible, replacements are made from alternate displacement vector convergence locations (i.e. local minima) arising from the multistart displacement fit. Where no good multistart candidates exist, outlier displacement values are replaced by the median value in the moving window. The proportion of displacement vectors modified by outlier filtering is typically fewer than 5% for any given dataset. We use the results of the filtered third fit as the displacement map for each dataset.

### 3. Evaluation of Fitting Bias

Displacement histograms (Fig. 1b, bottom left) obtained through 4DSTEM interferometry frequently show geometric patterns, in which some displacement values cluster together and some displacement values are avoided. To investigate the origins of this, we performed a simulation mimicking the effects of a finite probe radius (Supplementary Fig. 28). The displacement half-hexagon fitting region was populated with a grid of points representing true probe positions. For each point, the interferometry patterns for all displacements within a 0.2 Å radius were calculated according to Equation 1 in the main text and averaged. The averaged pattern was then re-fit to a displacement vector using Equation 1. The discrepancy between the original probe position in displacement space and the final fitted displacement estimates the bias introduced from a finite probe width. Note that this a simplified model, not accounting for the radial intensity profile of the probe or the effects of reconstruction, which will produce a non-uniform density of points in displacement space (see Supplementary Fig. 14).

Supplementary Fig. 28 shows that probe averaging induces bias matching the clustering pattern in the displacement histogram (Fig. 1b, bottom left). The "avoided regions" with large biasing correspond to high symmetry interferometry patterns. For example, Equation 1 predicts an SP pattern will have 8 of 12 disks with zero intensity. Many of these disks will be nonzero for intermediate stacking orders surrounding pure SP stacking. Probe averaging thereby obtains an interferometry pattern that does not quite look like pure SP stacking, leading to the bias.

Despite these effects, the design of the data analysis procedure renders the strain maps robust against biasing artifacts. When the displacement histogram is phase unwrapped for strain mapping (see above), the avoided regions manifest themselves as losses in continuity at the boundaries between moiré unit cells. Application of the total generalized variation (TGV) filter removes these discontinuities, essentially eliminating the bias through interpolation. The strain filter parameters thus set the systematic uncertainty on the calculated strain values, as discussed in the Methods.

## 4. Heterostrain Calculations

We employ a previously published model[2] to compute heterostrain triangulation plots (Fig. 2b, Fig. 6a and Supplementary Figs. 11c, g). Briefly, it is assumed that one layer of the TBG sample remains unstrained while the other bears uniaxial tensile heterostrain, $\varepsilon_H$, at some angle relative to the lattice. This model provides three degrees of freedom for the distorted moiré geometry: the moiré angle $\theta_m$, the heterostrain magnitude $\varepsilon_H$, and the angle of heterostrain application $\theta_h$. By measuring the lengths of the three sides of each moiré triangle, these three variables can be fit uniquely and plotted for the triangulated moiré geometry.

## 5. Reconstruction-Induced Displacement

In the small twist angle regime, AB reconstruction in TBG may be simplistically modelled as rotation of large fixed plates in the opposite direction of the moiré angle. As all AB plates rotate in the same direction, the boundary of any two plates experiences a shearing mechanic due to the reconstruction. The displacement induced at such a boundary may be calculated geometrically using the moiré triangle side length $L = a/|\theta_m|$, where $a = 2.461$ Å is the lattice constant of graphene. The distance from the center of the triangular AB domain to the boundary point is $a/(2\sqrt{3}|\theta_m|)$. Under the small angle approximation, the displacement arising from rotation of one AB domain is $|s| = r|\theta_{AB}| = a|\theta_{AB}|/(2\sqrt{3}|\theta_m|)$, where here $\theta_{AB}$ is the reconstruction rotation in a single layer of graphene. The adjoining AB domain contributes an equal amount of displacement in the opposite direction. The total in-plane displacement in one layer of graphene at an AB boundary is therefore $|u_{bound}| = 2|s| = a|\theta_{AB}|/(\sqrt{3}|\theta_m|)$. The inset of Fig. 3h plots the induced displacement in a single layer of graphene according to this formula.

When traversing a soliton wall perpendicular to the direction of the wall, the total stacking order displacement change is $a/\sqrt{3}$ (Fig. 1e). If the rotation in the two graphene layers is equal and opposite and all of the stacking order change is produced by AB reconstruction rotation, then there must be a simple shear displacement of $|u_{bound}| = a/(2\sqrt{3})$ in both the top and bottom layer to satisfy the soliton wall boundary condition. Equating this with the displacement formula derived in the previous paragraph, we have $a|\theta_{AB}|/(\sqrt{3}|\theta_m|) = a/(2\sqrt{3})$. This simplifies to just $|\theta_{AB}| = |\theta_m|/2$, which is also the angle at which the AB reconstruction exactly cancels out the moiré rotation to form commensurate Bernal-stacked AB domains. So not only does AB reconstruction improve the interlayer stacking energy, but it also produces the correct boundary displacement for thin soliton walls. This shows that AB reconstruction can be viewed as a mechanism for generating soliton walls. Experimentally (Fig. 3h), as $\theta_m$ goes to zero, we indeed see $|\theta_{AB}|$ approaching $|\theta_m|/2$ and $|u_{bound}|$ approaching $a/(2\sqrt{3})$.

## 6. Uncertainty Quantification

When quantifying twist-angle disorder from the moiré geometry, it is important to ensure that the variation observed does not arise simply from the uncertainty in the AA geometry fit. To this end, we performed bootstrapping[3] on three AA regions for $\theta_m$ = 1.37. The standard error of the AA region ($x,y$) coordinates was 0.08 nm. Numerical error propagation simulations show this corresponds to a standard deviation of about 0.01° in the calculated twist angle distribution arising from fitting error. As the measured $\theta_m$ distribution standard deviations are well in excess of this value, around 0.3° (Fig. 2c), we conclude that the observed twist angle disorder is a real effect. Furthermore, spatially-localized twist angle disorder is visible from the displacement maps.

Analogous error propagation simulations show that AA registration uncertainty produces a heterostrain of 0.05% and a standard deviation of 0.026%. The non-zero value of heterostrain due to only AA registration uncertainty arises because the heterostrain triangulation model[2] is a biased estimator (i.e. it is impossible to have a negative heterostrain value under this model, only positive heterostrains that are oriented in different directions). Consequently, our measured heterostrain distributions could be systematically inflated, though we expect by no more than 0.05%, given the above calculations. Taken together, since the experimental spread in heterostrain is about three times greater than what could be explained by AA registration error alone, we can conclude this heterostrain disorder is likewise a real effect.

For homogenous sample strain trends, x-axis error bars are always given as standard deviations, reflecting the distribution of triangulated twist angles within each dataset. Y-axis error bars are 95% confidence intervals obtained from the variance across multiple domains measured within the image (for instance, multiple AA mask regions). For datasets acquired near a tear in one of the graphene layers, this approach cannot be followed because the twist angle is changing rapidly. Instead, each AA region is assigned an effective twist angle by averaging the triangulated moiré twist angles for all adjacent moiré unit cells. X-axis error bars are given as the standard deviation of these moiré angle values. Y-axis error bars are given as the standard deviation of the quantity of interest within the AA mask.

## 7. Intermediate Stacking Orders

To probe the impact of intermediate stacking orders, a second stacking area analysis was conducted using a five-category partition (Supplementary Fig. 17i). The AA region was divided into an inner AA region (displacement |**u**| < 0.35 Å) and an outer AA region (0.35 < |**u**| < 0.71 Å). In addition to the pure AB/BA and SP stacking orders, an "SP/AB transitional" stacking order was defined at exactly the average of the AB/BA and SP stacking order displacement vectors. All non-AA displacement vectors were assigned to the closest of these three stacking orders. We note that while there are several potential ways to define a five-category partition, this example serves as an illustration of the effect of intermediate stacking orders.

As in Figure 4a, the twist-angle-dependent stacking area % was analysed (Supplementary Fig. 17j). Between $\theta_m$ = 1.4° and 0.5°, decreasing $\theta_m$ leads to a decrease in both the inner and outer AA stacking area %, while SP, AB/BA, and AB/SP transitional stacking areas each increase slightly. Below the critical angle of $\theta_m$ = 0.5°, SP and AB/SP transitional stacking areas reverse trends and decrease in area % as the twist angle decreases, while AB/BA stacking increases sharply in area %. This behavior is the same as in the original partition in Figure 4a, in which SP stacking area rose from $\theta_m$ = 1.4° to 0.5° and decreased below $\theta_m$ = 0.5°, while AB/BA stacking rose modestly from $\theta_m$ = 1.4° to 0.5° and sharply below $\theta_m$ = 0.5°. This suggests that AA transitional stacking ("AA outer") diminishes as AA reconstruction takes place from $\theta_m$ = 1.4° to 0.5°, while AB/SP transitional stacking diminishes as AB reconstruction takes place below $\theta_m$ = 0.5°. These results are in agreement with intuitive expectations regarding reconstruction, and can be confirmed visually be examining the real-space stacking order assignment images (Supplementary Fig. 17e–h, bottom row). The same two regimes of reconstruction are visible regardless of the stacking partition employed.

The variation in local rotation and stacking order across an SP region was analysed for $\theta_m$ = 0.26° (Supplementary Figure 10c,e) and $\theta_m$ = 1.03° (Supplementary Figure 10d,f). The reconstruction rotation varies as a bell curve across the SP in both cases, but the change is much smaller for $\theta_m$ = 1.03 than $\theta_m$ = 0.26 as expected on the basis of the reconstruction regimes. To analyse the shift between the layers, we plotted the displacement change both parallel ($\Delta u_y$) and perpendicular ($\Delta u_x$) to a line traversing the SP (Supplementary Fig. 10b). For both values of $\theta_m$, $\Delta u_y$ is negligible, indicating that the displacement in SP regions is of the shear type as expected. Interestingly, $\Delta u_x$ displays a sigmoidal profile for $\theta_m$ = 0.26°, but a linear profile for $\theta_m$ = 1.03°. The former is expected for a shear soliton, while the latter is expected for a rigid moiré. This directly relates to the two regimes of reconstruction. Though reconstruction has taken place at $\theta_m$ = 1.03°, it dominantly occurs through rotation around the AA regions. These rotations are oppositely balanced at the center of the SP, leading to no net change of the rigid moiré displacement along the line chosen. For $\theta_m = 0.26°$, however, reconstruction dominantly occurs through AB rotation, which occurs perpendicular to the line chosen and leads to sigmoidal variation. For both values of $\theta_m$, the displacement and rotation profiles pass smoothly through intermediate stacking orders, indicating that perfect SP stacking only occurs along an infinitesimal line.

## 8. Reconstruction Model

To enable band structure calculations considering the effects of AA and AB reconstruction separately, a simple parameterized model was developed. AA reconstruction was modelled by a 2D Gaussian rotation field centred on each AA region. AB reconstruction was modelled by including a constant rotation field within each triangular AB domain. The AB domain edges were drawn at a buffer distance $b_{AB}$ from the lines connecting two AA regions, and the edges were then smoothed by a Gaussian filter. Consequently, the AB reconstruction rotation is constant within the centre of the AB domain and tapers off near the edges, matching our experimental observations. The reconstruction parameters used for the band structure calculations are given in Supporting Information Table 2. Note that the maximum applied rotation angle in the individual AA and AB domains ($\alpha_{AA}$ and $\alpha_{AB}$) are not mathematically the same as the fixed-body rotation of the sample in the AA and AB regions ($\theta_R^{AA}$ and $\theta_R^{AB}$), owing to the overlap between rotation fields centred on multiple AA and AB domains within this model. Reconstruction parameters were chosen to give a good match to the sample geometry and $\theta_R^{AA}$, $\theta_R^{AB}$ values within the constraints of the model.

## 9. Band Structures of Relaxed TBG From a Continuum Model

For evaluation of the interlayer electronic tunnelling functions, we use a tight-binding model extracted from DFT calculations[5]. We follow an electronic continuum model prescription[6], updated to allow for atomic relaxations. The relaxations are included by evaluating the Fourier transform of the interlayer tight-binding coupling at the relaxed configurations, $\tilde{t}(\boldsymbol{b} + \boldsymbol{u}(\boldsymbol{b}))$, instead of the bare configurations, $\tilde{t}(\boldsymbol{b})$ (see below).

The continuum model for the band structure of TBG[7] can be extended to include arbitrary relaxations[8,9]. The central idea is to extract the effective interlayer scattering terms between momenta $q_i$ and $q_j$, usually notated as $T_{ij}$, by Fourier transforming the interlayer orbital-to-orbital couplings, $t(b)$, where $b$ is the configuration (relative distance between the pairs of carbon atoms) and quickly falls to zero within 5Å. Generally, the $T$ matrices take the form

$$T_{ij} = \begin{pmatrix} \omega_0 e^{i\phi_{ij}^{11}} & \omega_1 e^{i\phi_{ij}^{12}} \\ \omega_1 e^{i\phi_{ij}^{21}} & \omega_0 e^{i\phi_{ij}^{22}} \end{pmatrix} \tag{S3}$$

where the phases $\phi_{ij}^{mn}$ depend on the position of the orbitals and the choice of origin, and the $\omega_0 \equiv \omega_{AA}$ and $\omega_1 \equiv \omega_{AB}$ correspond to the effective interlayer coupling strength between similar and dissimilar orbitals of opposite layers, respectively. Relaxations, defined by a vector field $u$ for interlayer relaxation, modify the relative displacement of each pair of atoms, and so we need to consider the transform of the object $t(b + u(b))$. For simplicity, we ignore out-of-plane corrugations of the two lattices, and consider only local rotations around AA and AB stacking sites as discussed in the Reconstruction Model section above. We also ignore in-plane corrections to the Hamiltonian caused by modifications to the (intralayer) couplings of monolayer graphene under shearing strain. These corrections require in-plane momentum scattering based on the Fourier coefficients of the relaxation pattern and, for the assumptions of Gaussian rotations, do not take simple analytic forms.

Fig. 5b–d shows the band structures under various relaxation assumptions. The effective interlayer coupling terms ($\omega_{AA}$, $\omega_{AB}$) are provided in Supplementary Table 3. Note that this model has not been explicitly symmetrized, and so some erroneous gaps of order 2 meV are present due to symmetry-breaking errors introduced during numerical interpolation of the interlayer tunnelling. These errors are most noticeable for the $\theta_m = 0.35°$ band structure (Fig. 5b). The rigid lattices have no moiré superlattice gaps, but including either the AA or AB relaxation assumption for $\theta_m = 1.15°$ opens up gaps above and below the flat-band manifold. In general, inclusion of either assumption reduces $\omega_{AA}$ and increases $\omega_{AB}$. At smaller angles AB rotation plays a larger role than AA rotation, while near the magic angle the opposite is true.

## 10. Computation of Interlayer Tunnelling Functions

The values given in Supporting Information Table 3 for ($\omega_{AA}$, $\omega_{AB}$) are only approximate measures of the effective interlayer electronic tunnelling. For a momentum basis centred at $K_0 = K_1$, the $K$-point of Layer 1's (bottom layer) Brillouin zone, there are three highest order scatterings to Layer 2, given by the three smallest values of $K_0 + G_2$, where $G_2$ is any reciprocal lattice vector of Layer 2. These couplings are given by $\tilde{t}(K_0 + G_2)$, and this is often the value taken to estimate $\omega_{AA}$ and $\omega_{AB}$. The general form of the tunneling is given by $\tilde{t}(K_0 + G_1 + G_2)$, where $G_1$ is any reciprocal lattice of Layer 1. As long as $G_1 + G_2 \approx 0$, the coupling stays near $K_0$, but is now sampled in a regular grid of the moiré reciprocal lattice. For rigid lattices, $\tilde{t}$ is smooth and non-zero in the vicinity of $K_0$, and so the approximation $\tilde{t}(K_0 + G_1 + G_2) \approx \tilde{t}(K_0) \equiv \omega_{AA} = \omega_{AB}$ is a fairly good choice[7]. However, for relaxed lattices, this assumption is not always well justified[8,9]. The variation in $\tilde{t}(k)$ can become quite severe near $K_0$, and can even be sampled right on a nodal point of the Fourier transform, as in the case of the fully relaxed AA coupling at $\theta_m = 0.5°$, shown in Fig. 5c. For this reason, analysis of the effective $\omega$ strengths at the magic-angle does not always generalize easily to relaxed TBG at smaller angles. The interlayer tunnelling functions for orbitals of similar type (i.e. A-to-A and B-to-B) for three different twist angles are displayed in Supplementary Figs. 21, 23, and 25, respectively. The interlayer tunnelling functions for orbitals of dissimilar type (i.e. A-to-B and B-to-A) are displayed in Supplementary Figs. 22, 24, and 26.

## 11. Definition of $\beta_{\omega_0}$ and $\beta_{\omega_1}$

In Fig. 5e of the main text, we presented variables $\beta_{\omega_0}$ and $\beta_{\omega_1}$ as the value of a relative "angle" between interlayer couplings $(\tilde{t}(k))$ for AA and AB rotations only versus the fully relaxed interlayer coupling. This is defined by introducing a generalized inner product between two different complex interlayer tunnelling functionals, $f(k)$ and $g(k)$:

$$\langle f, g \rangle = \int d\mathbf{k} f^*(\mathbf{k}) g(\mathbf{k}) \tag{S4}$$

The relative angle between two interlayer tunnelling functionals is then given by

$$\beta_{fg} = \cos^{-1}\left(\frac{Re(\langle f, g \rangle)}{\sqrt{\langle f, f \rangle \langle g, g \rangle}}\right) \tag{S5}$$

We define $\beta_{\omega_0} = \beta_{\tilde{t}^\mu_{AA} \tilde{t}^f_{AA}}$ and $\beta_{\omega_1} = \beta_{\tilde{t}^\mu_{AB} \tilde{t}^f_{AB}}$, with $\tilde{t}^\mu_{ij}$ the interlayer coupling between orbital $i$ and $j$ with relaxation indexed by $\mu$ for AA or AB rotation only, and $\tilde{t}^f_{ij}$ the coupling for the full relaxation model with both rotations.

## 12. Finite Element Simulations

The finite-element relaxation method uses a generalized stacking fault energy (GSFE) and elastic moduli for bilayer graphene, both of which are extracted from previous Density Functional Theory (DFT) calculations[10,11]. The elastic relaxation model consists of an elastic energy term, capturing the strain energy of each layer, and a GSFE term, capturing the variations in interlayer binding energy (see below). To allow for the evaluation of the relaxation for real space superlattices with heterostrain, we impose an initial spatially-dependent interlayer displacement $\mathbf{b}_0(\mathbf{r})$ that includes both rotation and shear. The total energy is then minimized by optimizing the interlayer relaxation field $\boldsymbol{u}(\boldsymbol{r})$. The initial constant shear in $\boldsymbol{b}_0$ is added to the gradients of $\boldsymbol{u}$ in the evaluation of the elastic energy. This allows for the optimization of the periodic function $\boldsymbol{u}$, instead of having to explicitly encode the twisted boundary conditions of the moiré superlattice.

The planar relaxation of twisted 2D bilayers can be well captured by finite-element approaches with parameters extracted from DFT[10]. The relative displacement between the layers is given by a spatially-varying interlayer displacement, which will be a sum of an initial displacement $b_0(r)$ and a relaxation field $u(r)$. The moiré supercell is defined by a pair of lattice vectors, which are columns of the $2 \times 2$ matrix

$$A_{sc} = \begin{pmatrix} m\sin(\beta/2) & -n\sin(\beta/2) \\ m\cos(\beta/2) & n\cos(\beta/2) \end{pmatrix} \quad (S6)$$

where $\beta$ is the interior angle of the moiré supercell and $m, n$ are the side lengths of the cell ($\beta = 60°$ and $m = n$ in the absence of heterostrain). The initial displacement between the layers is given by the vector field

$$b_0(r) = A_{uc} A_{sc}^{-1} r \quad (S7)$$

where $A_{uc}$ is the $2 \times 2$ matrix representing the unit cell of monolayer graphene

$$A_{uc} = \frac{l}{2}\begin{pmatrix} 3 & 3 \\ -\sqrt{3} & \sqrt{3} \end{pmatrix} \quad (S8)$$

with $l = 1.42$ Å, the nearest-neighbour bonding distance of graphene. The relative orientations of $A_{sc}$ and $A_{uc}$ ensure that the initial configuration is uniformly distributed over the supercell, and the shape of the supercell (e.g. the side lengths and angle $\beta$) determines the initial twist angle and heterostrain for the simulation. It also allows us to introduce a relaxation field that is periodic with respect to the moiré supercell as the twisted boundary condition is entirely captured by $b_0$, allowing for the derivatives of the relaxation field to be evaluated via its Fourier components, instead of finite-element derivative stencils.

We assume the two layers share the displacement equally, e.g. $u_1(r) = -u_2(r) = u(r)/2$. The elastic ("kinetic") energy, related to in-plane deformation of a single layer, is given by

$$E_{intra}(u) = \frac{1}{2} \int K \left( \frac{\partial u_x}{\partial x} + \frac{\partial u_y}{\partial y} \right)^2 + G \left[ \left( \frac{\partial u_x}{\partial x} - \frac{\partial u_y}{\partial y} \right)^2 + \left( \frac{\partial u_x}{\partial y} + \frac{\partial u_y}{\partial x} \right)^2 \right] dr \tag{S9}$$

with $[K, G] = [69.518, 47.352]$ eV per unit cell area of graphene[10].

To define the interlayer binding energy between the graphene layers, we employ a generalized stacking fault energy function ($V_{\text{GSFE}}(b)$) which represents the relative energy of each stacking configuration, as extracted from DFT calculations[5]. To respect the symmetries of graphene, we expand $V_{\text{GSFE}}$ in terms of its three lowest even Fourier components[10]:

$$\begin{aligned} V_{\text{GSFE}}(b) = \ & c_0 + c_1(\cos v + \cos w + \cos(v + w)) \\ & + c_2(\cos(v + 2w) + \cos(v - w) + \cos(2v + w)) \\ & + c_3(\cos(2v) + \cos(2w) + \cos(2v + 2w)) \end{aligned} \tag{S10}$$

where

$$\begin{pmatrix} v \\ w \end{pmatrix} = 2\pi A_{uc}^{-1} \begin{pmatrix} b_x \\ b_y \end{pmatrix} \tag{S11}$$

and $[c_0, c_1, c_2, c_3] = [6.832, 4.064, -0.374, -0.095]$ meV per unit cell area[11]. The total interlayer ('potential') energy is then given by

$$E_{inter}(b) = \int V_{\text{GSFE}}(b(r)) dr \tag{S12}$$

To find the relaxed geometry, we initialize the relaxation field $u(r) = 0$, and then minimize

$$E = 2E_{intra}\big((b_0 + u)/2\big) + E_{inter}(b_0 + u) \tag{S13}$$

Here we have used the assumption that the two layers have identical relaxations, and thus identical strain energy. During entry to $E_{intra}$, the gradients of $u$ are evaluated in the Fourier basis and then added to the gradients of $b_0$ (which are constant throughout the supercell). After optimizing $u$, the interlayer displacement is given by $u_{inter} = b_0 + u$, and the effective strain ($\gamma_{\max}$) is easily extracted.

**References**

(1) Vulović, M., et al. When to use the projection assumption and the weak-phase object approximation in phase contrast cryo-EM. *Ultramicroscopy* **136**, 61–66 (2014).

(2) Kerelsky, A. et al. Maximized electron interactions at the magic angle in twisted bilayer graphene. *Nature* **572,** 95–100 (2019).

(3) Kulesa, A.; Krzywinski, M.; Blainey, P.; Altman, N. Sampling distributions and the bootstrap. *Nat. Methods* **12**, 477–478 (2015).

(4) Zhang, K. & Tadmor, E. B. Structural and electron diffraction scaling of twisted graphene bilayers. *J. Mech. Phys. Solids* **112,** 225–238 (2018).

(5) Fang, S., & Kaxiras, E. Electronic structure theory of weakly interacting bilayers. *Phys. Rev. B* **93**, 235153 (2016).

(6) Massatt, D., Carr, S., Luskin, M., & Ortner, C. Incommensurate Heterostructures in Momentum Space. *Multiscale Model. Simul.* **16**, 429 (2018).

(7) Bistritzer, R. & MacDonald, A. H. Moiré bands in twisted double-layer graphene. *Proc. Natl. Acad. Sci. USA* **108,** 12233–12237 (2011).

(8) Carr, S., Fang, S., Zhu, Z., & Kaxiras, E. Exact continuum model for low energy electronic states of twisted bilayer graphene. *Phys. Rev. Research* 1, 013001 (2019).

(9) Guinea, F., & Walet, N. R. Continuum models for twisted bilayer graphene: Effect of lattice deformation and hopping parameters. *Phys. Rev. B* **99**, 205134 (2019).

(10) Carr, S. *et al.* Relaxation and domain formation in incommensurate two-dimensional heterostructures. *Phys. Rev. B* **98,** 224102 (2018).

(11) Zhou, S., Han, J., Dai, S., Sun, J., & Srolovitz, D. J. van der Waals bilayer energetics: Generalized stacking-fault energy of graphene, boron nitride, and graphene/boron nitride bilayers. *Phys. Rev. B* **92**, 155438 (2015).

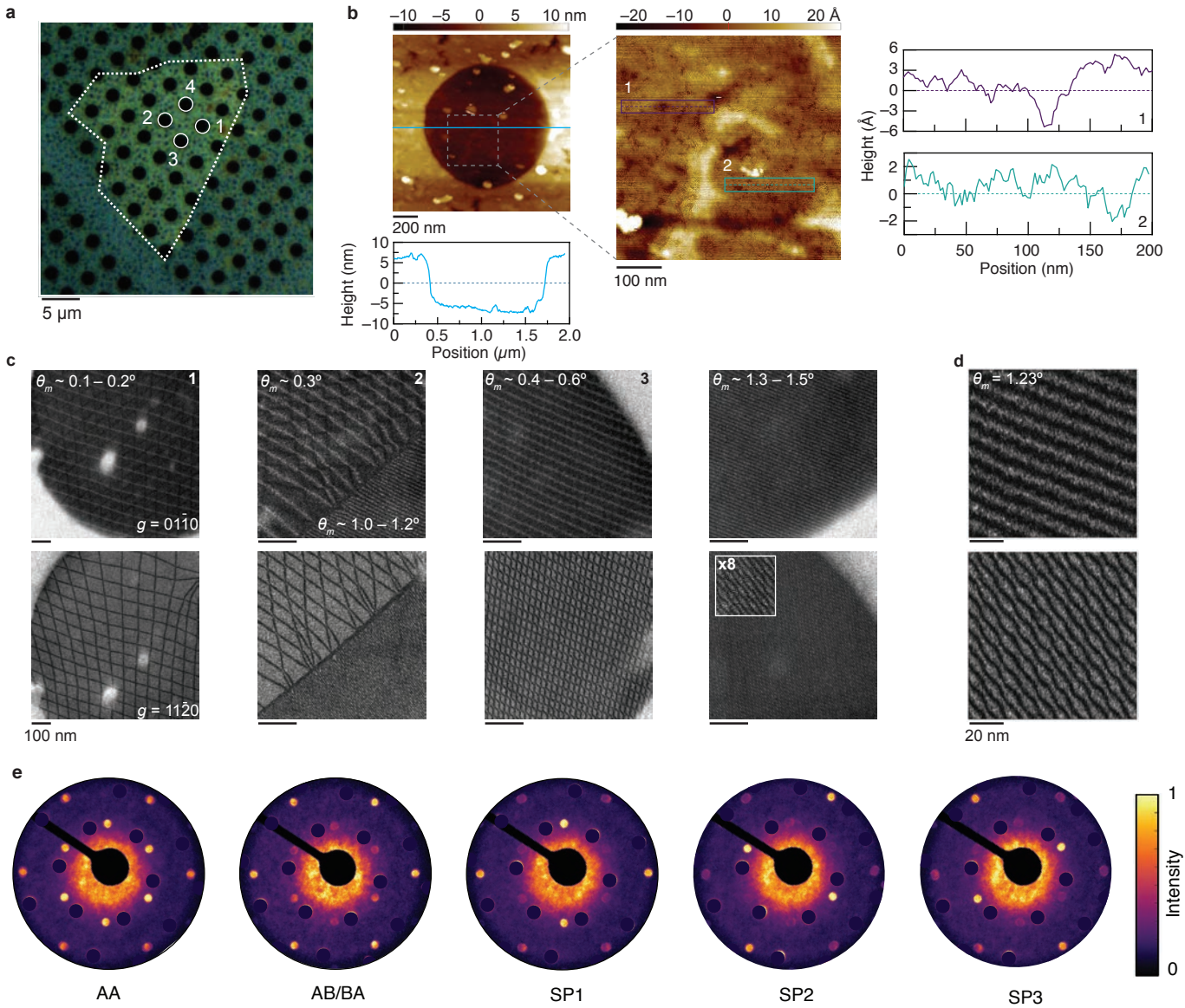

**Supplementary Fig. 1. Low magnification microscopy and dark-field imaging. a**, Optical micrograph of hBN/TBG heterostructure (outlined) on a holey silicon nitride membrane. **b**, Atomic force microsocopy of the hBN/TBG heterostructure on the silicon nitride membrane. **c**, TEM dark-field images obtained by selecting the ***g*** = 01$\bar{1}$0 and ***g*** = 11$\bar{2}$0 graphene diffraction peaks for regions with various twist angles. Locations used for imaging are labelled in **a**. Scale bars 100 nm. **d**, 'Virtual' dark field images reconstructed from 4D-STEM data of a TBG sample with $\theta_m$ = 1.23°. Scale bars: 20 nm. **e**, Averaged diffraction patterns for TBG stacking orders (hBN Bragg disks have been masked for clarity).

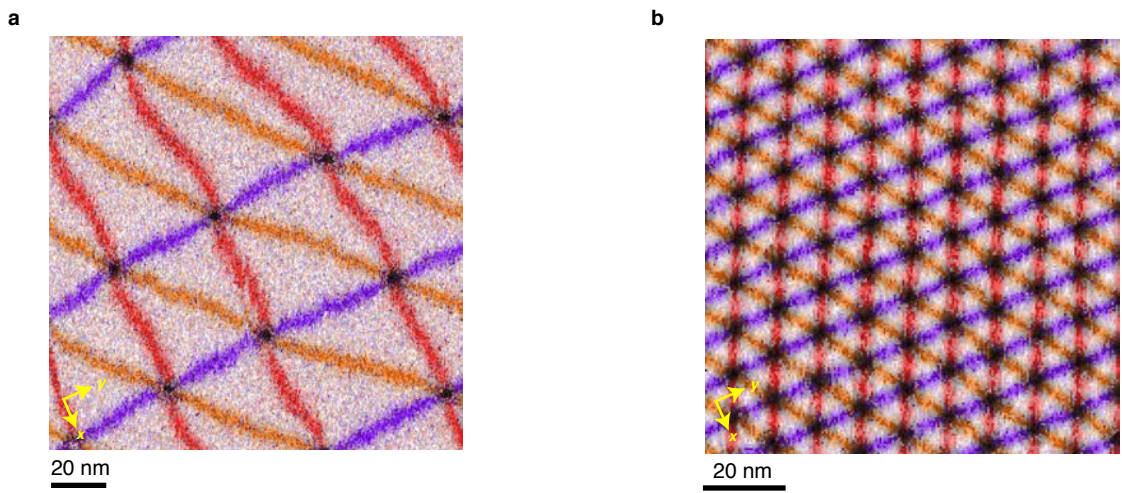

**Supplementary Fig. 2. Displacement field maps.** Maps of TBG possessing average $\theta_m$ of 0.26º (**a**) and 1.23º (**b**).

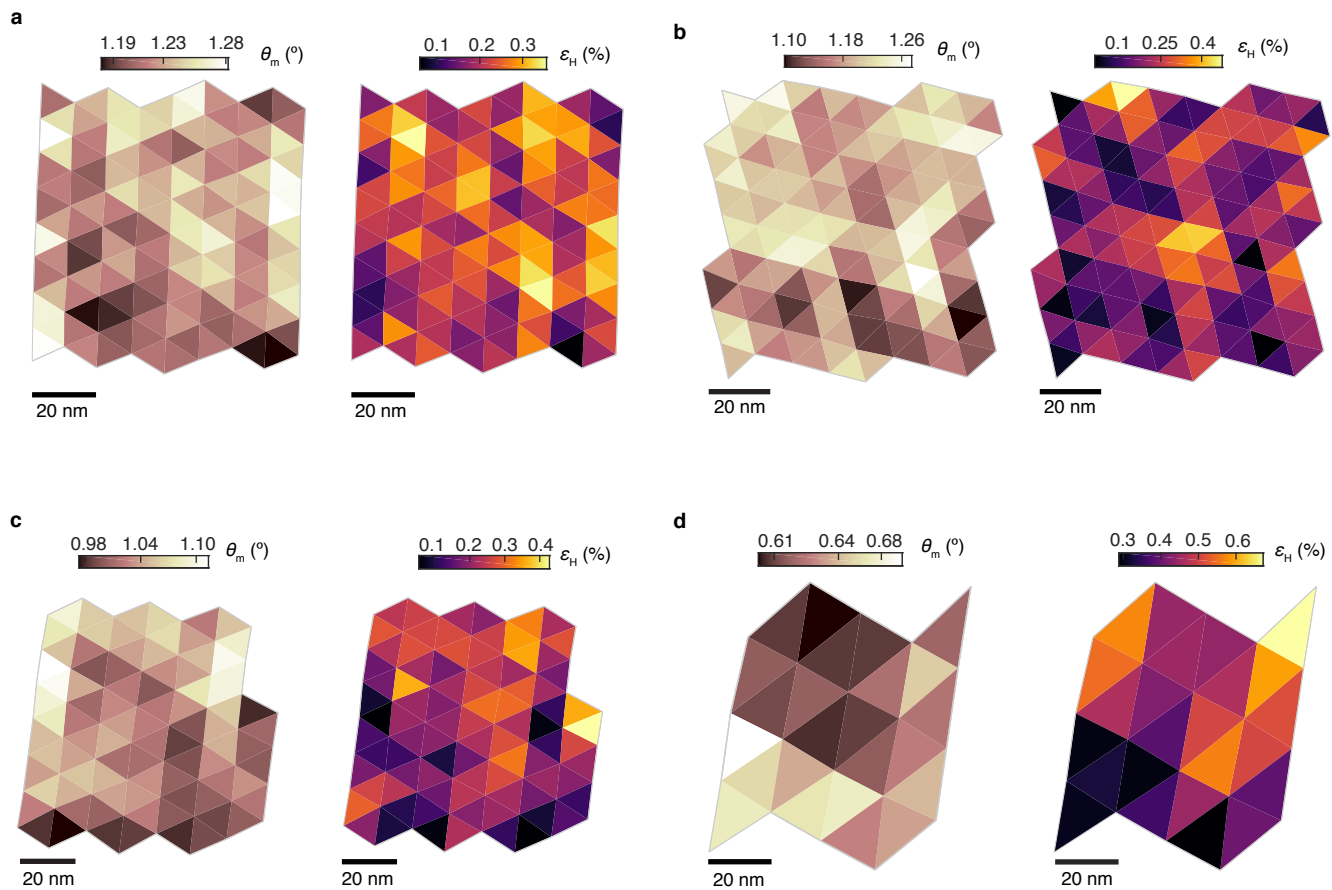

**Supplementary Fig. 3. Local twist angle and heterostrain triangulation maps.** Local twist angle (left) and uniaxial heterostrain (right) measurements for regions with average $\theta_m$ of 1.23º (**a**), 1.19º (**b**), 1.03º (**c**), and 0.63º (**d**).

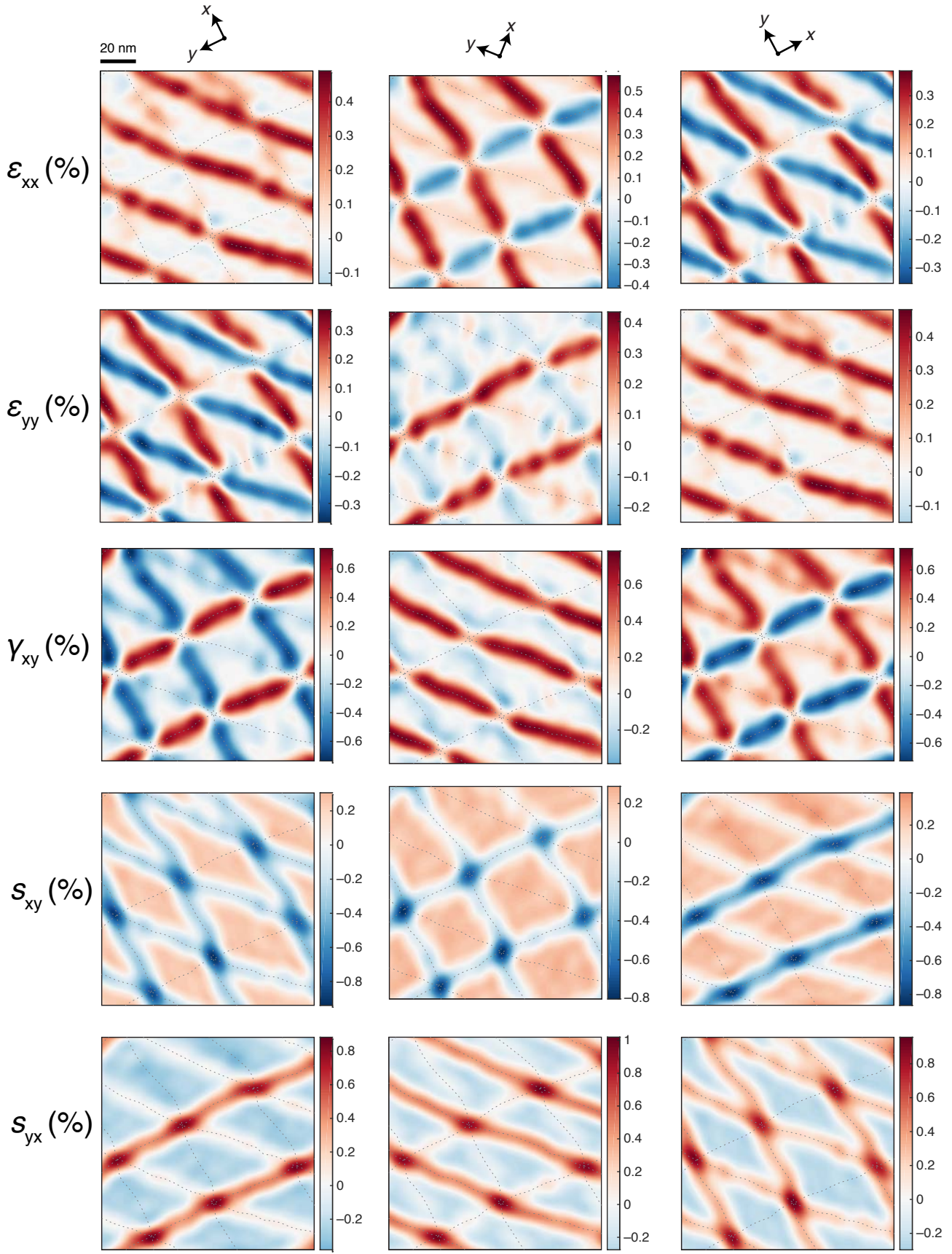

**Supplementary Fig. 4. Tensor rotations for 0.26° TBG.** Maps of normal strains, $\epsilon_{xx}$ and $\epsilon_{yy}$, engineering pure shear strain, $\gamma_{xy}$, and simple shear strains, $s_{xy}$ and $s_{yx}$, produced with tensor rotations wherein the x-axis is successively aligned perpendicular to the three SP directions. These maps arise from the same dataset as that displayed in Figs. 3a,c and Supplementary Figs. 2a and 15d.

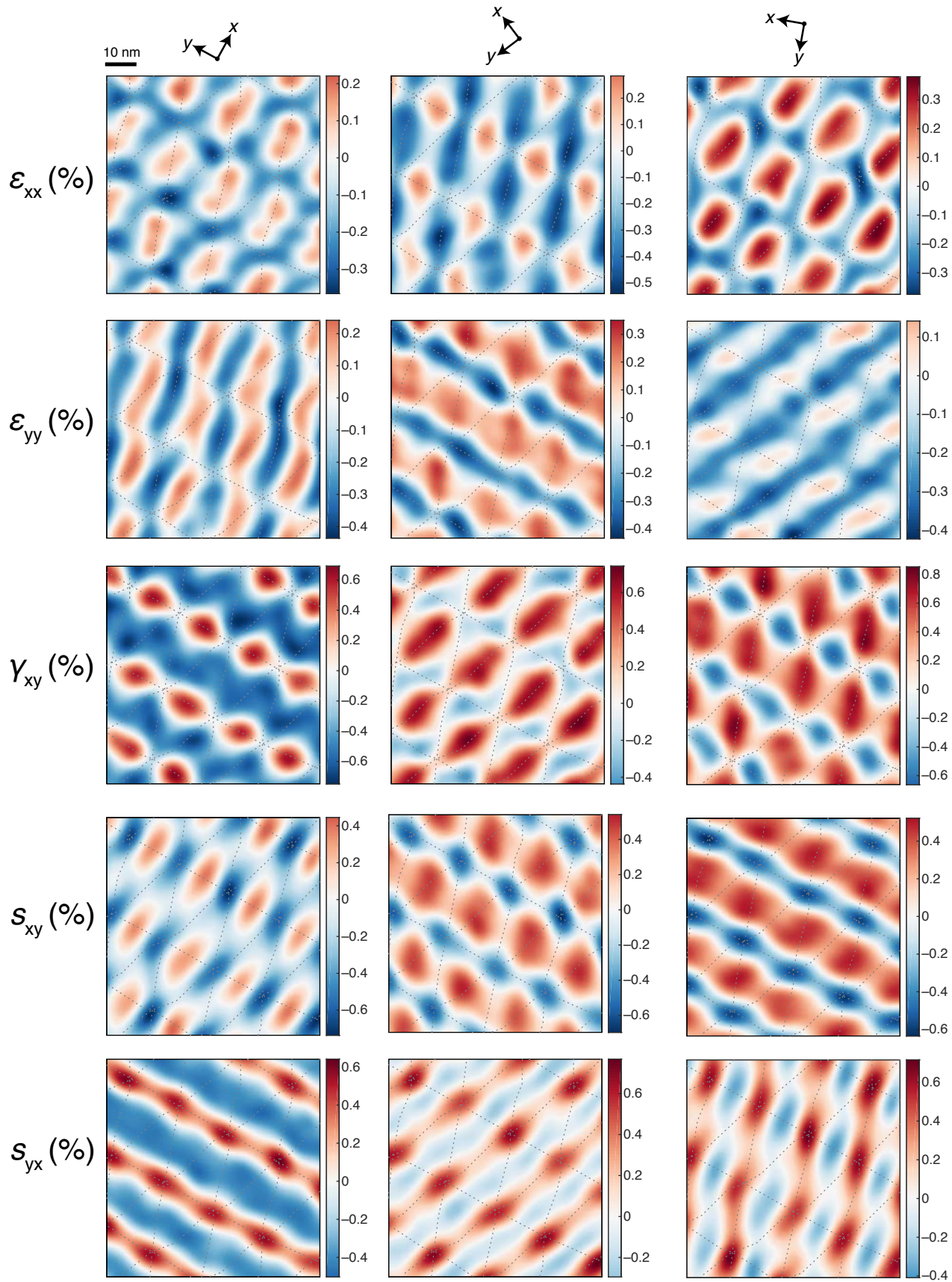

**Supplementary Fig. 5. Tensor rotations for 0.63° TBG.** Maps of normal strains, $\epsilon_{xx}$ and $\epsilon_{yy}$, engineering pure shear strain, $\gamma_{xy}$, and simple shear strains, $s_{xy}$ and $s_{yx}$, produced with tensor rotations wherein the x-axis is successively aligned perpendicular to the three SP directions. These maps arise from the same dataset that displayed in Fig. 1d, Fig. 6c, and Supplementary Fig. 15c.

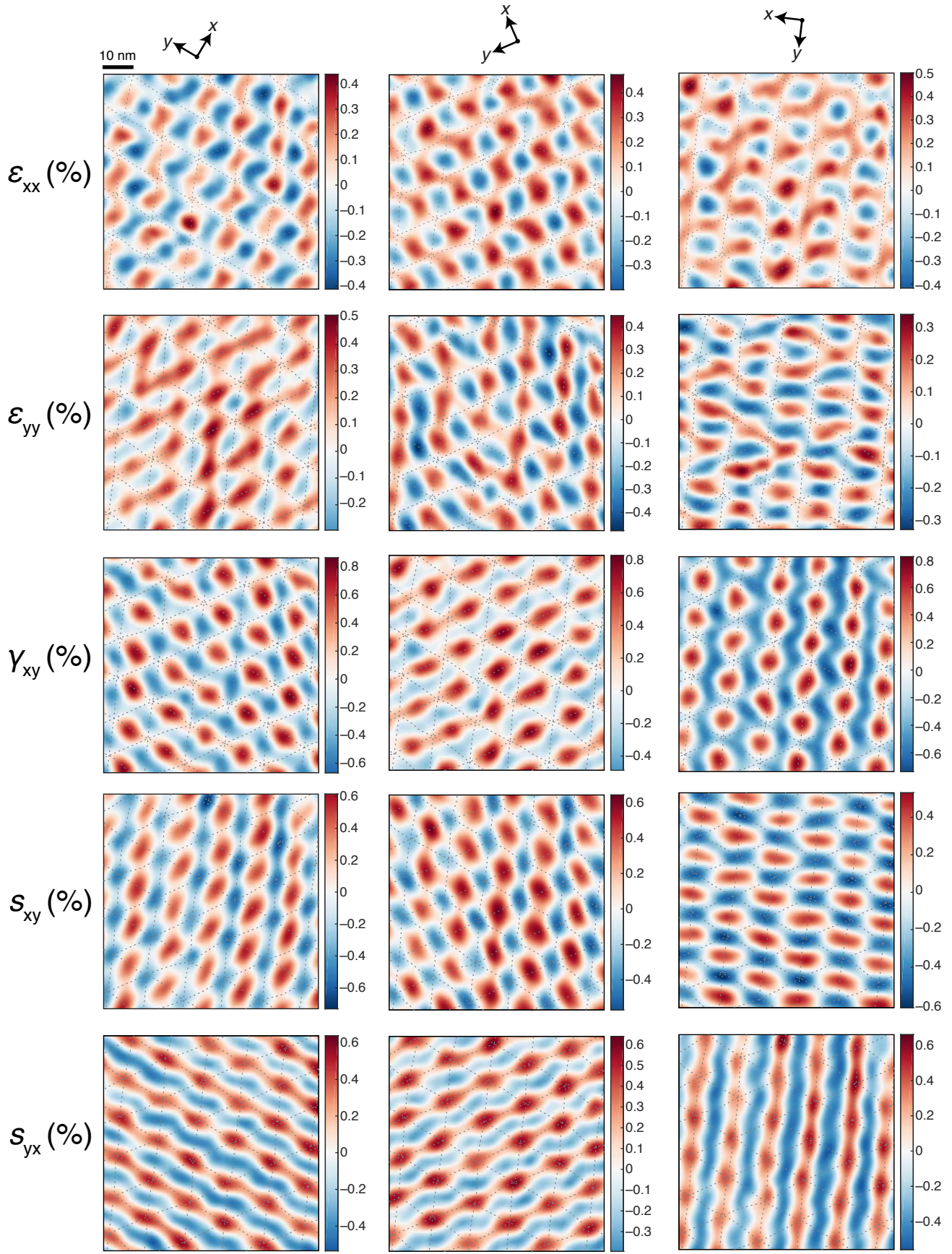

**Supplementary Fig. 6. Tensor rotations for 1.03° TBG.** Maps of normal strains, $\epsilon_{xx}$ and $\epsilon_{yy}$, engineering pure shear strain, $\gamma_{xy}$, and simple shear strains, $s_{xy}$ and $s_{yx}$, produced with tensor rotations wherein the x-axis is successively aligned perpendicular to the three SP directions. These maps arise from the same dataset as that displayed in Fig. 1e, Figs. 3b, d and Supplementary Fig. 15b.

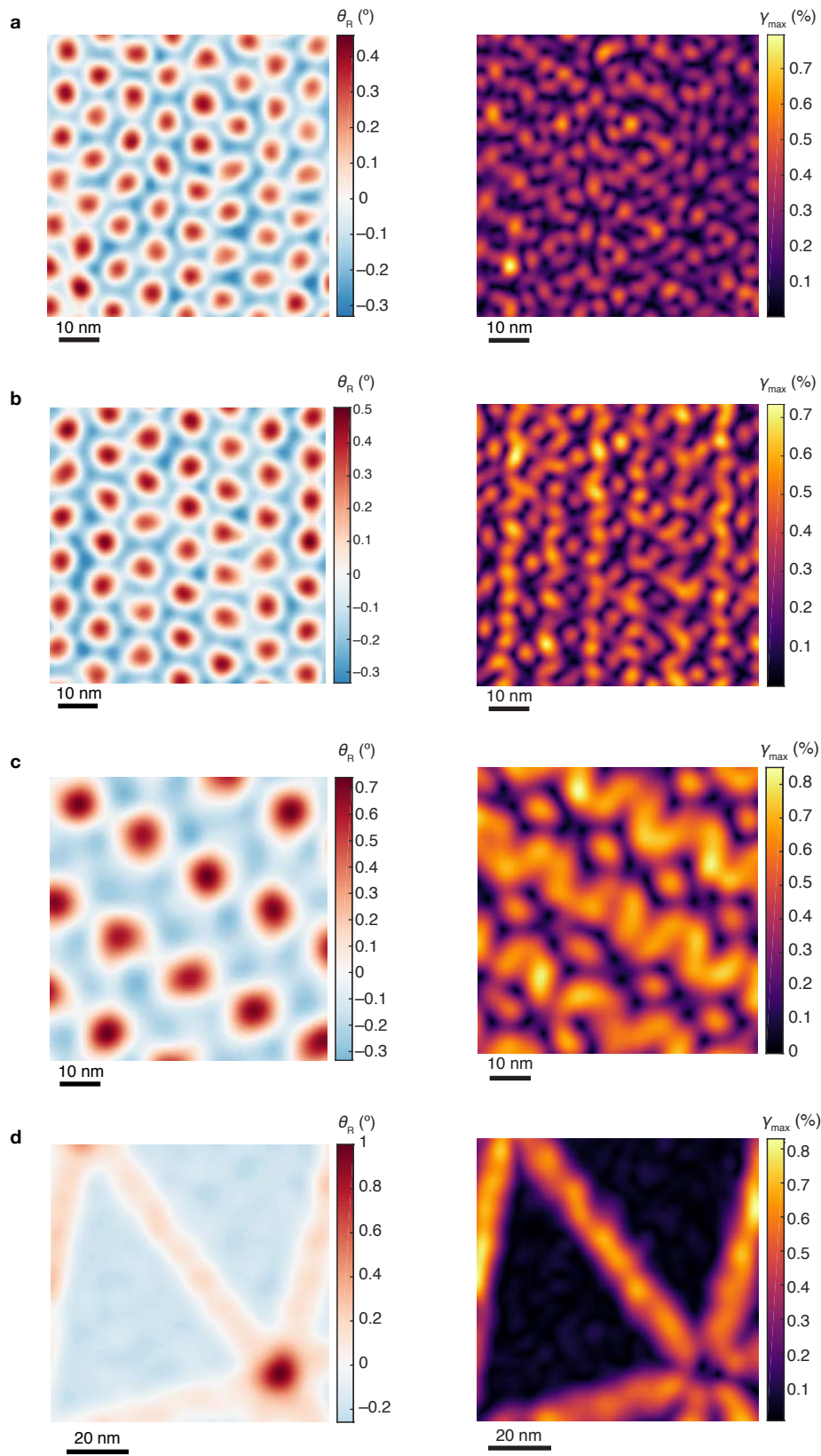

**Supplementary Fig. 7. Strain maps.** Interlayer reconstruction rotation (left) and maximum shear strain (right) for $\theta_m$ of 1.37° (**a**), 1.23 (**b**), 0.63 (**c**), and 0.16° (**d**). The maximum shear map in **c** is the same as that shown in Figure 6c of the main text, but is included here for completeness.

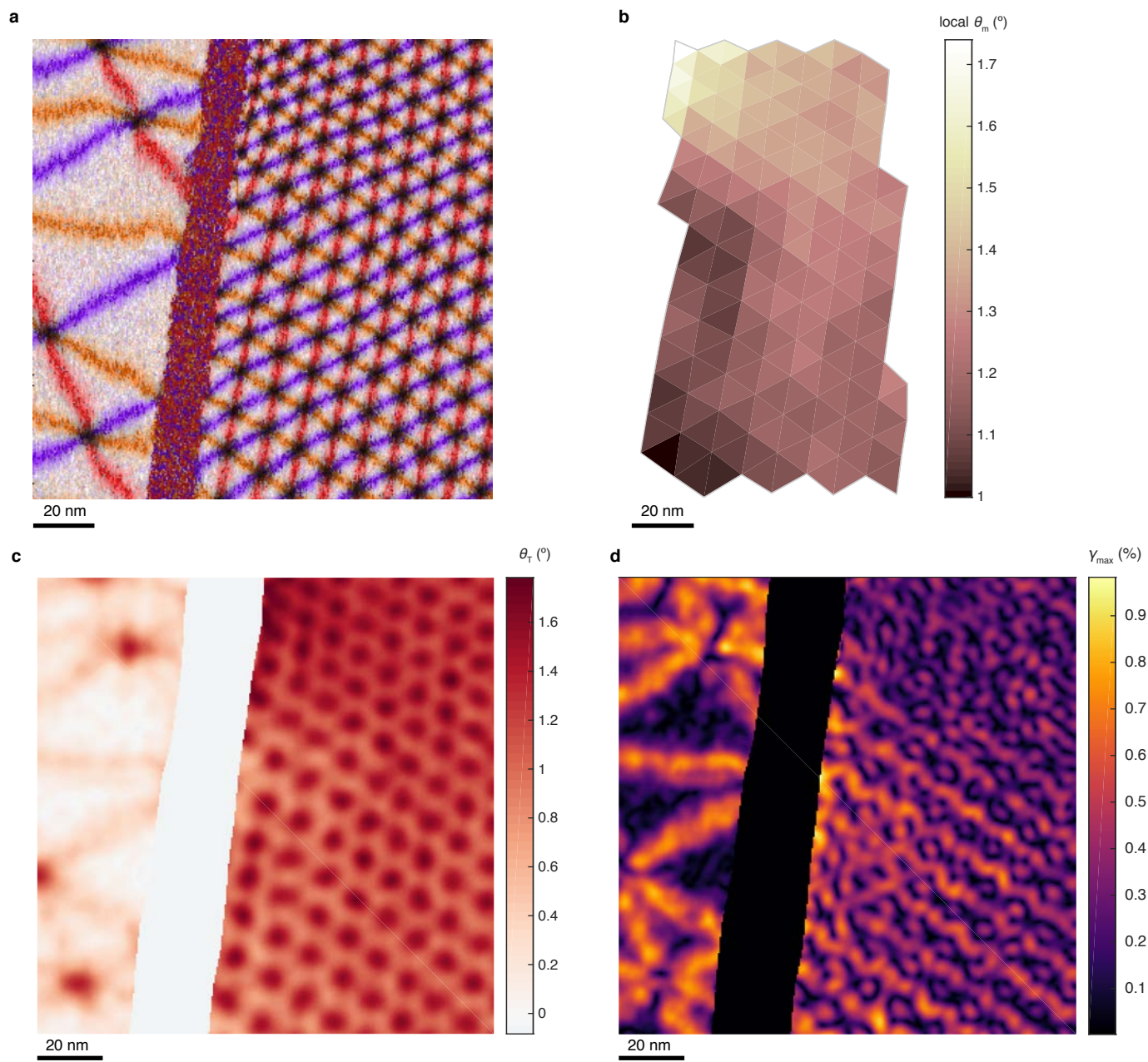

**Supplementary Fig. 8. Imaging near a tear in one graphene layer. a,** Displacement field map. **b**, Local twist angle triangulation map. **c**, Interlayer total rotation. **d**, maximum shear strain. Heterostrained regions (to the right of the tear and from middle–bottom) show strong 1D maximum shear features, consistent with the behaviour presented in Fig. 6 of the main text.

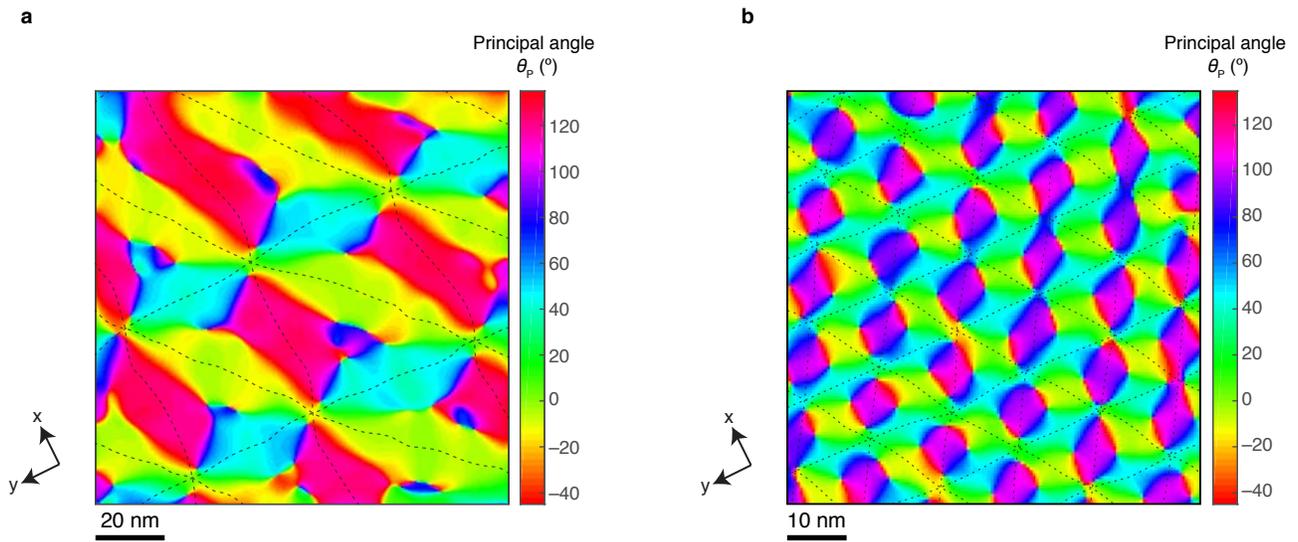

**Supplementary Fig. 9. Principal angle maps.** Orientation of the maximum principal strain component in TBG samples at twist angles of 0.26º (**a**) and 1.03º (**b**). These maps correspond to the datasets shown in Figure 3 of the main text. Overlaid dashed lines depict the moiré unit cell geometry from displacement maps, showing the strain direction is nearly uniform within each SP domain. Angles are computed relative to the positive x-axis displayed.

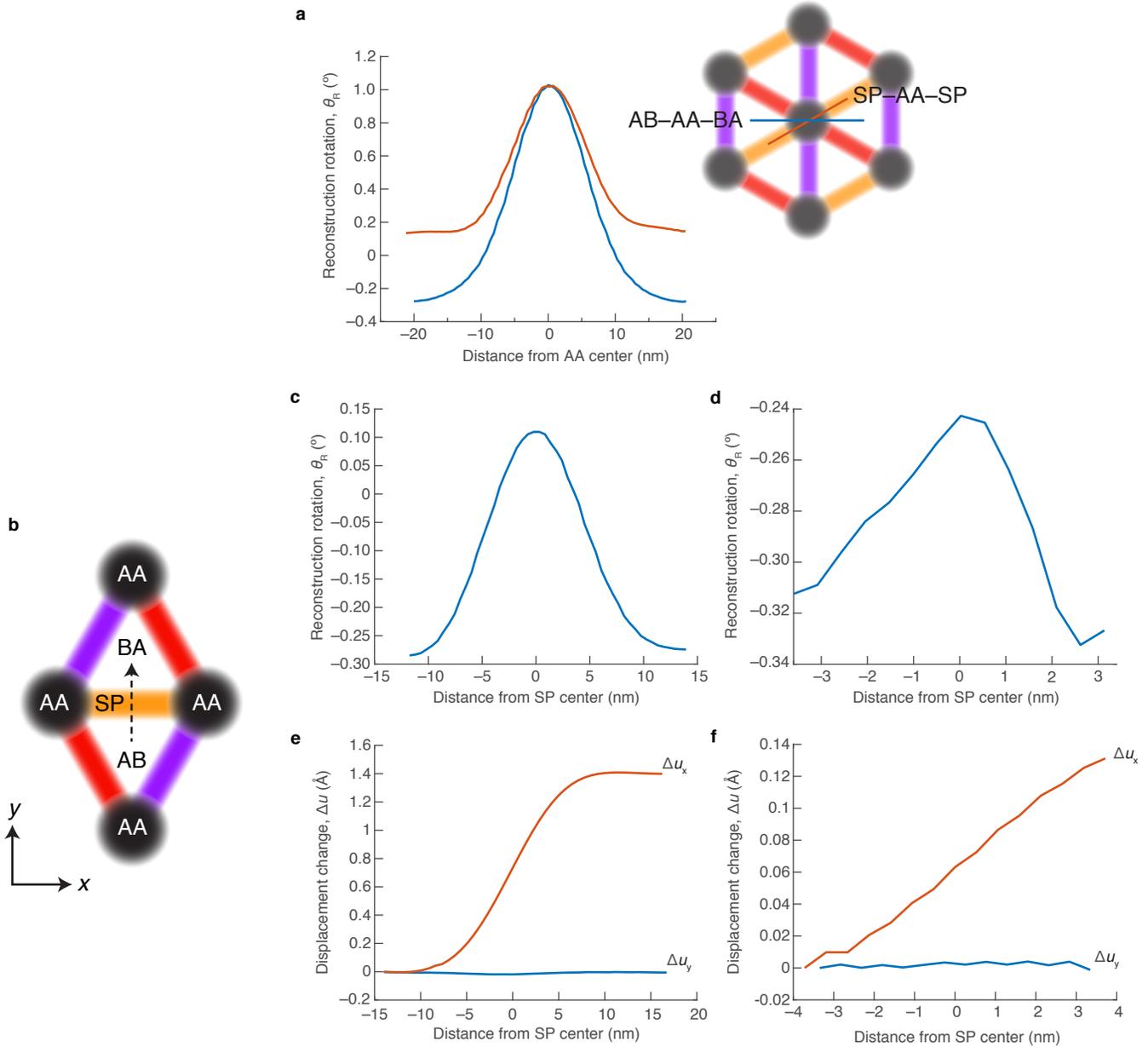

**Supplementary Fig. 10. Radial local rotation profile across AA domain. a,** Variation in local reconstruction rotation across an AA region in a TBG sample with twist angle 0.26° (Figure 3a) via two paths shown in the inset schematic. **b,** Illustration of path followed in traversing across an SP region in TBG. **c–f,** Variation in local reconstruction rotation across an SP region (**c,d**) and displacement components $\Delta u_x$ and $\Delta u_y$ across an SP region (**e,f**) in a TBG sample with twist angle 0.26° (**c,e**) and 1.03° (**d,f**) from the datasets shown in Figure 3a and 3b of the main text. Directions $x$ and $y$ are given by the axes of **b**. Sigmoidal shear displacement ($\Delta u_x$) is observed at 0.26° in **e** on account of AB/BA-dominated reconstruction producing soliton walls, while linear shear displacement is observed at 1.03° in **f** due to weak AB reconstruction at this angle. Instead, the linear displacement variation observed is due to the underlying moiré pattern.

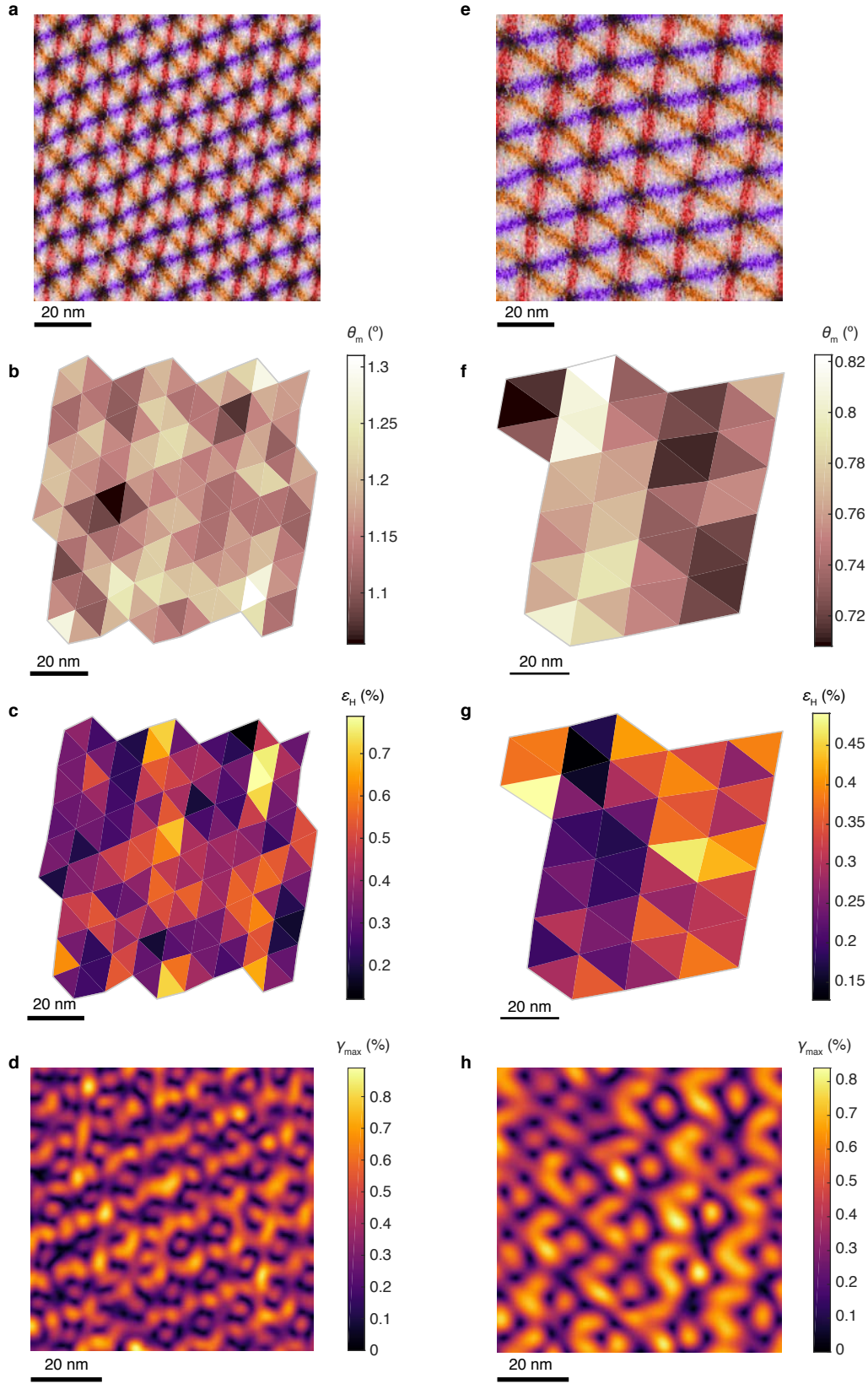

**Supplementary Fig. 11. Intralayer strain in uniaxially heterostrained TBG.** Displacement field (**a, e**), local twist triangulation (**b, f**), heterostrain triangulation (**c, g**), and maximum shear strain (**d, h**) for TBG at average twist angles of 1.17º (**a–d**) and 0.77º (**e–h**). Average heterostrain is 0.4% (**c**) and 0.3% (**g**). The maximum shear plots again display pronounced striped features in regions with higher local heterostrain, as observed in Figure 6. Increased heterostrain also correlates with increased local SP maximum shear strain loading, as observed in the transition from top-left to bottom-right in **g** and **h**.

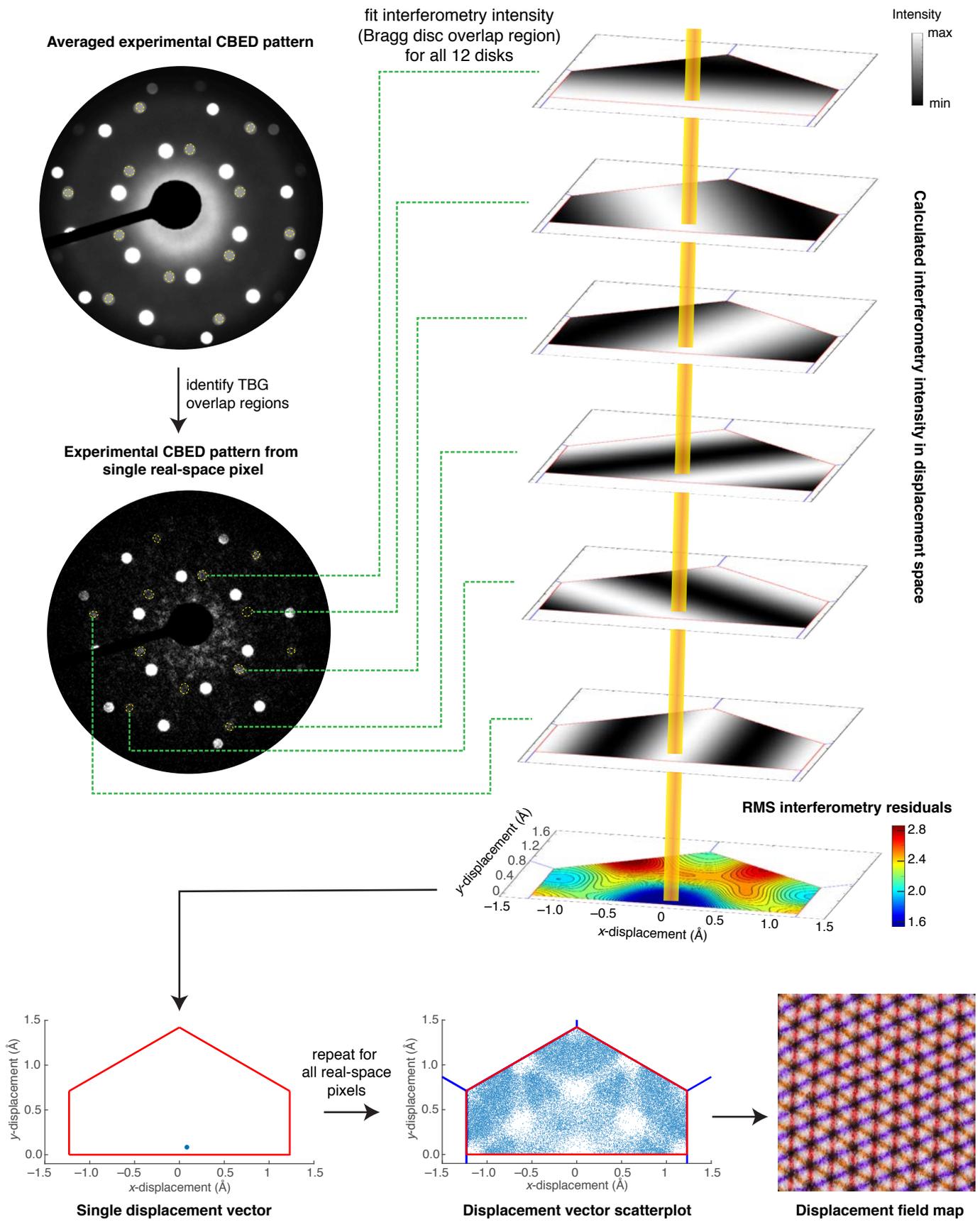

**Supplementary Fig. 12. 4D-STEM Bragg Interferometry.** Schematic of the fitting process showing how the TBG interferometry intensity in one 2D convergent beam electron diffraction (CBED) pattern is fit to extract a single displacement vector (here for an AA site) and then the entire 4D data set is assembled into a displacement field map for a 1.23º twist angle TBG.

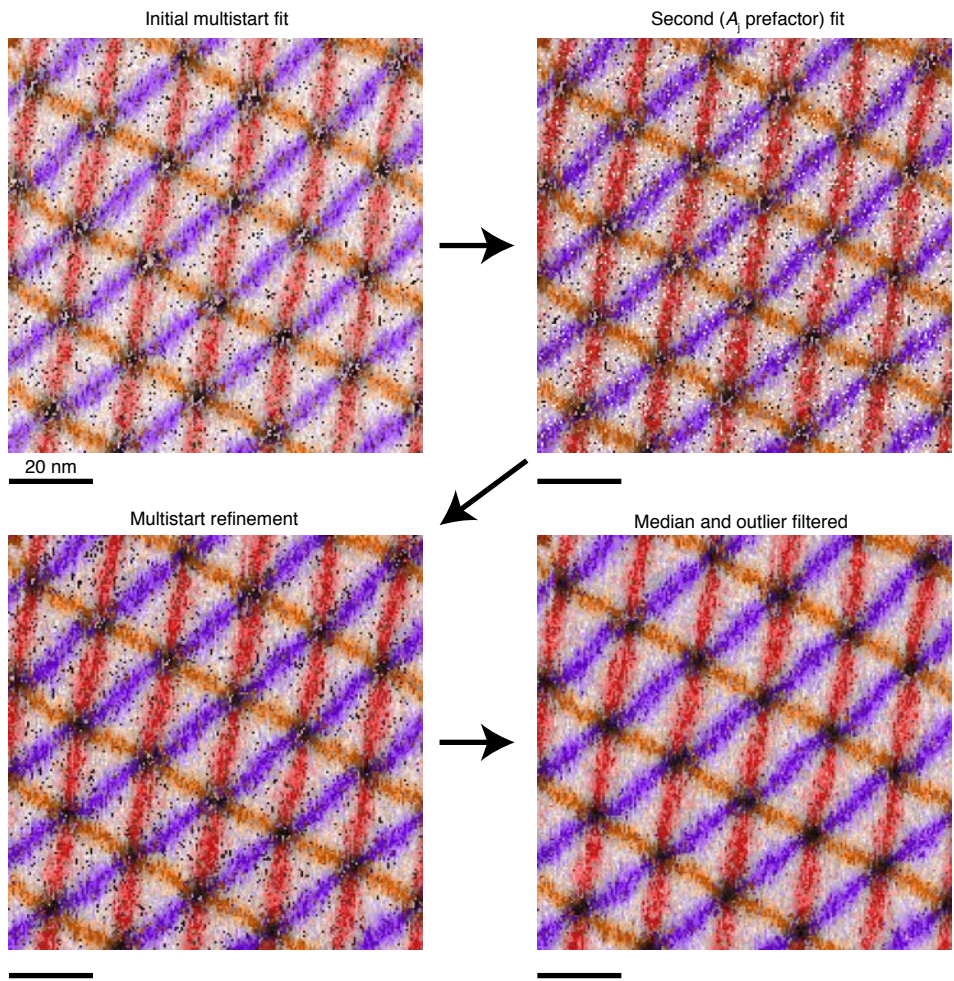

**Supplementary Fig. 13. Displacement field map refinement.** Sequence of fitting, refinement, and filtering used to reconstruct a displacement field map for a 0.63º twist angle TBG.

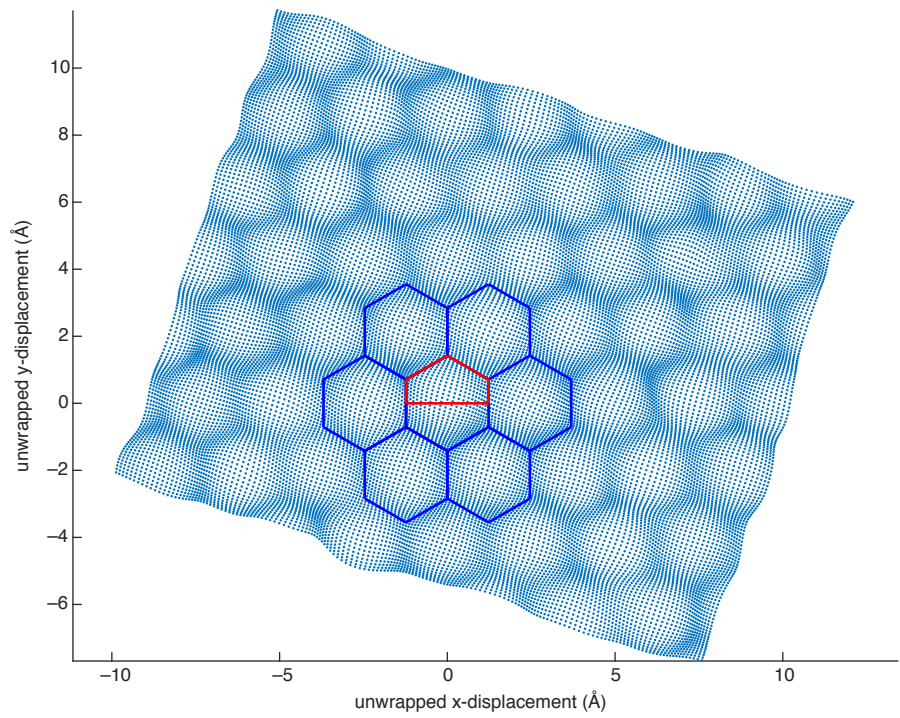

**Supplementary Fig. 14. Unwrapped displacement vector scatter plot.** Phase-unwrapping and TGV filtering the displacement vector field converts the displacement half-hexagon into a continuous vector field between adjacent moiré domains that is amenable to differentiation.

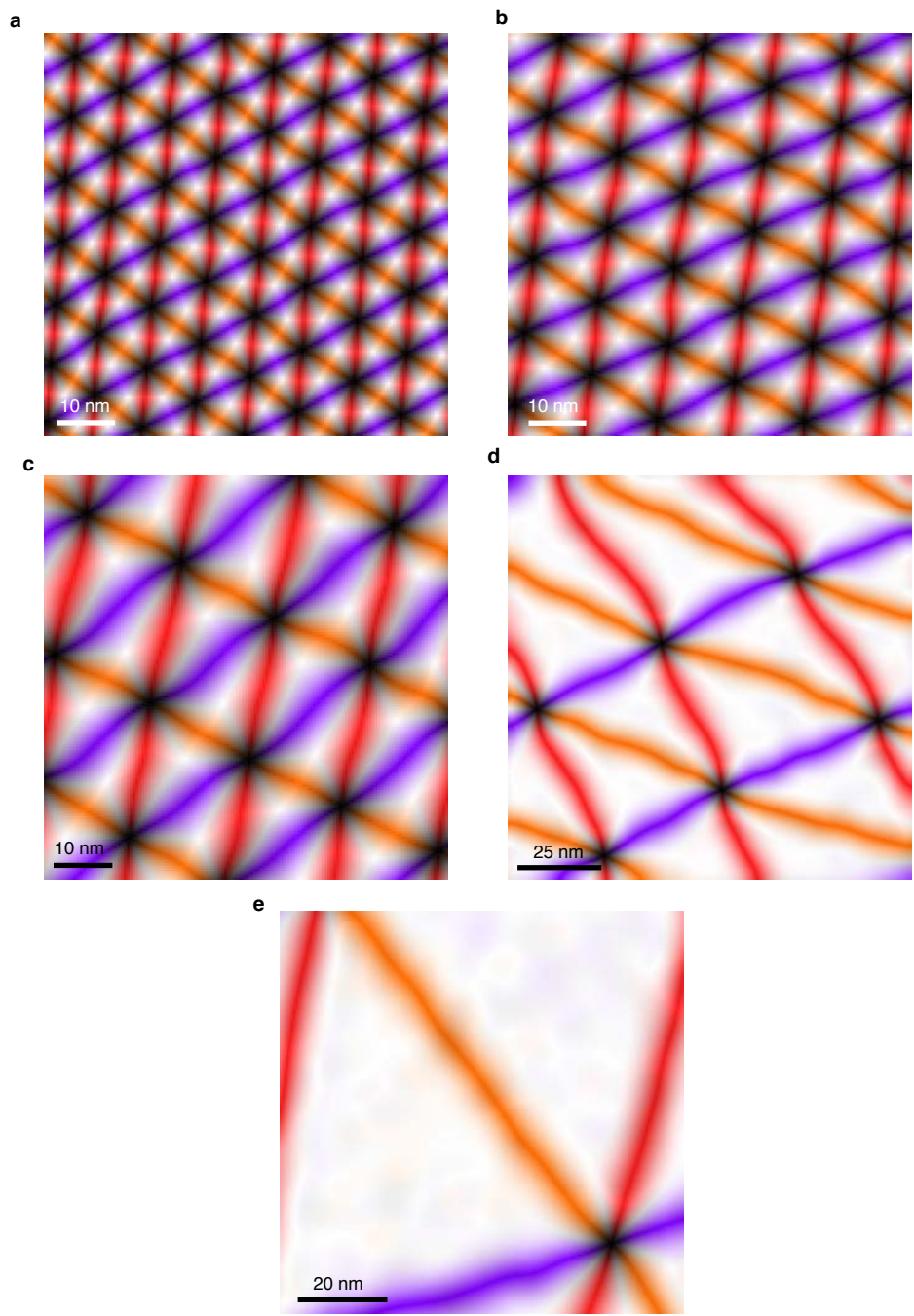

**Supplementary Fig. 15. Denoised displacement field maps.** Phase-unwrapped and TGV denoised displacement maps for the datasets shown in Figs. 1c–f and Supplementary Fig. 2 with average $\theta_m$ of 1.37° (**a**), 1.03° (**b**), 0.63° (**c**), 0.26° (**d**), and 0.16° (**e**). A border of ≤30 nm is lost in the unwrapping–denoising process compared to the original displacement field maps.

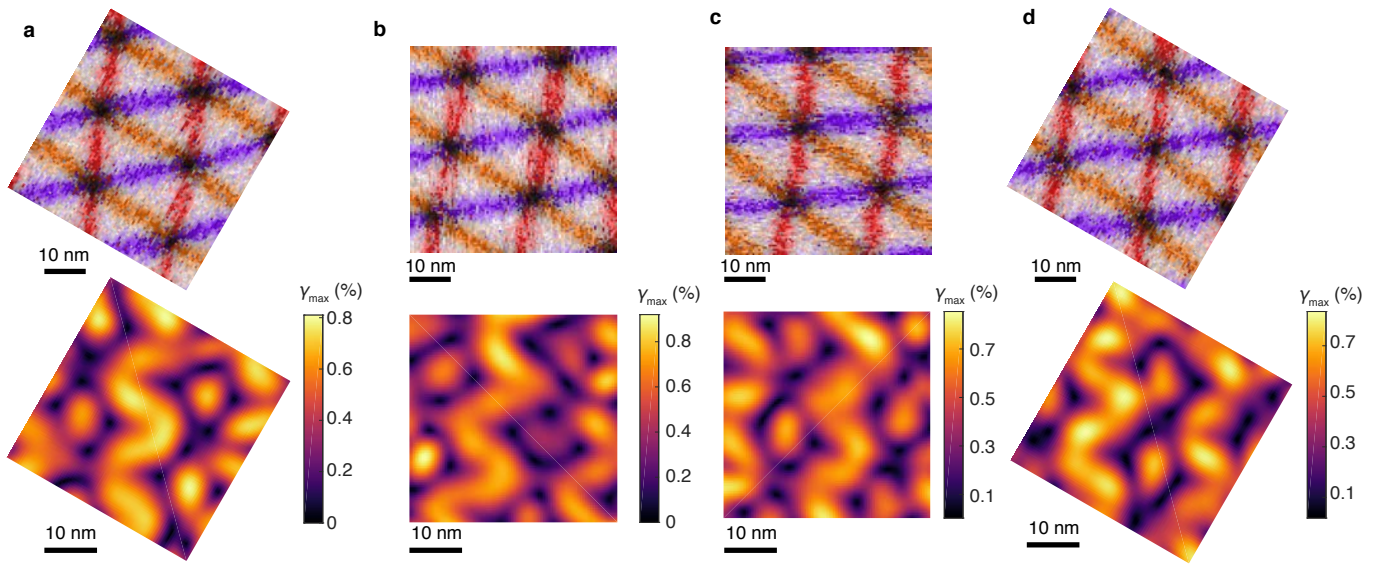

**Supplementary Fig. 16. Scan direction and sample drift verification.** Displacement field (top) and maximum shear strain (bottom) maps. Replicate pairs of images at different STEM scan directions in twist angle homogeneous regions with $\theta_m = 0.64°$ (**a, b**) and $\theta_m = 0.65°$ (**c, d**). STEM scan directions are 210° (**a**), 180° (**b**), 270° (**c**), and 210° (**d**). These regions are in close proximity (within ~50 nm of each other) and exhibit heterostrain in dark-field TEM images (Panel 3 in Supplementary Fig. 1b). The maps have been counterrotated by the scan direction. Since the displacement field and strain maps are aligned in both sets of images (**a** aligns with **b** and **c** aligns with **d**), we conclude that drift is negligible and cannot be responsible for the observed heterostrain. Analysis of SP intersection angles (see Methods and Supplementary Table 1) and the dark-field TEM images provides additional corroboration.

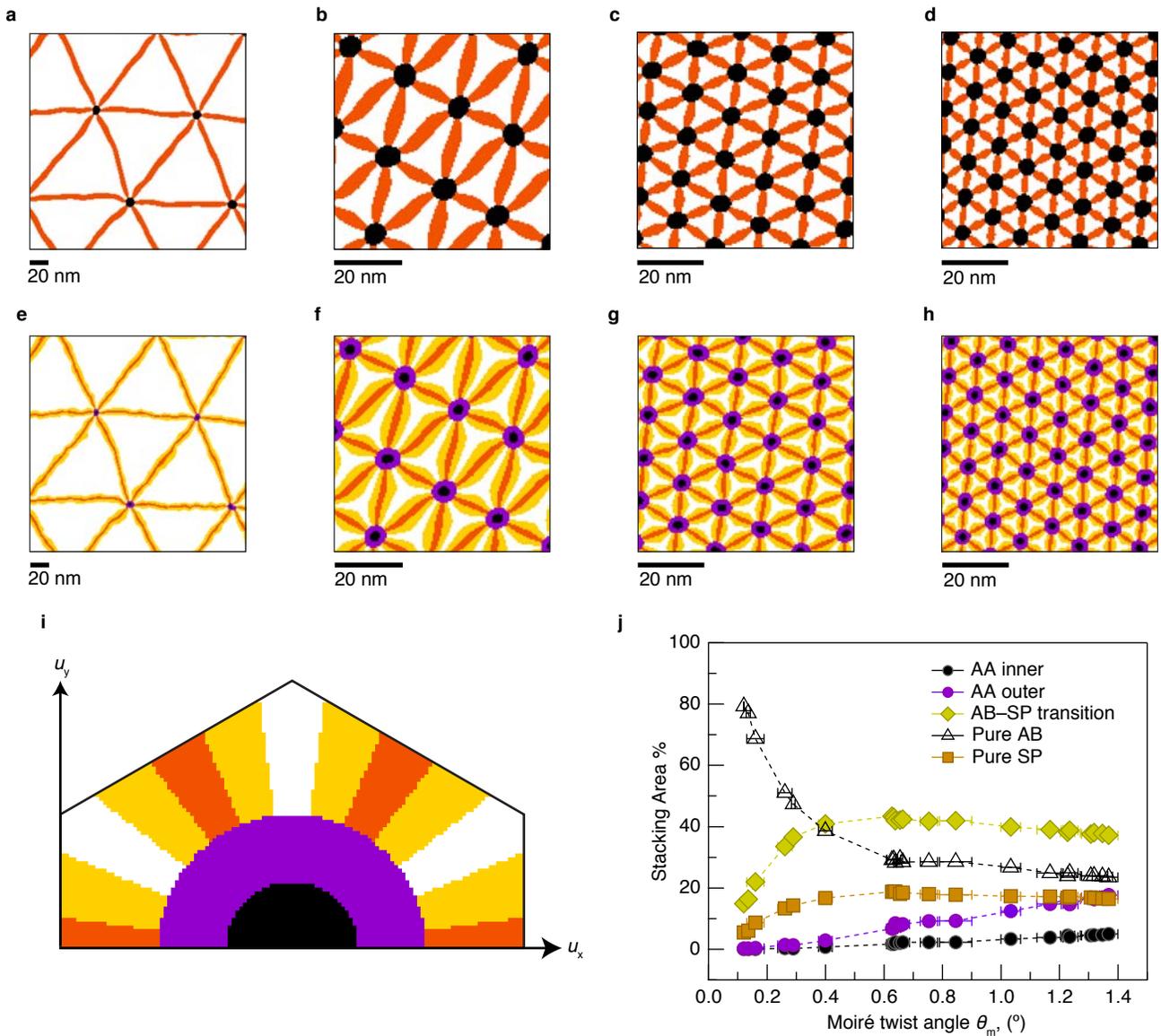

**Supplementary Fig. 17. Stacking order assignments. a–h,** Stacking assignments for displacement field plots of TBG at representative twist angles of 0.14º (**a**, **e**), 0.63º (**b**, **f**), 1.03º (**c**, **g**), and 1.37º (**d**, **h**). White = AB/BA, black = AA, orange = SP. Displacements in **a–d** are assigned according to the three-region partition displayed in the inset of Fig. 4a of the main text: White = AB/BA, black = AA, orange = SP. Displacements in **e–h** are assigned according to the five-region partition shown in **i**: Black = inner AA, Purple = outer AA, Orange = SP, White = AB/BA, and Yellow = SP/AB transitional. **j,** Stacking area % trends computed on the basis of the five-region partition.

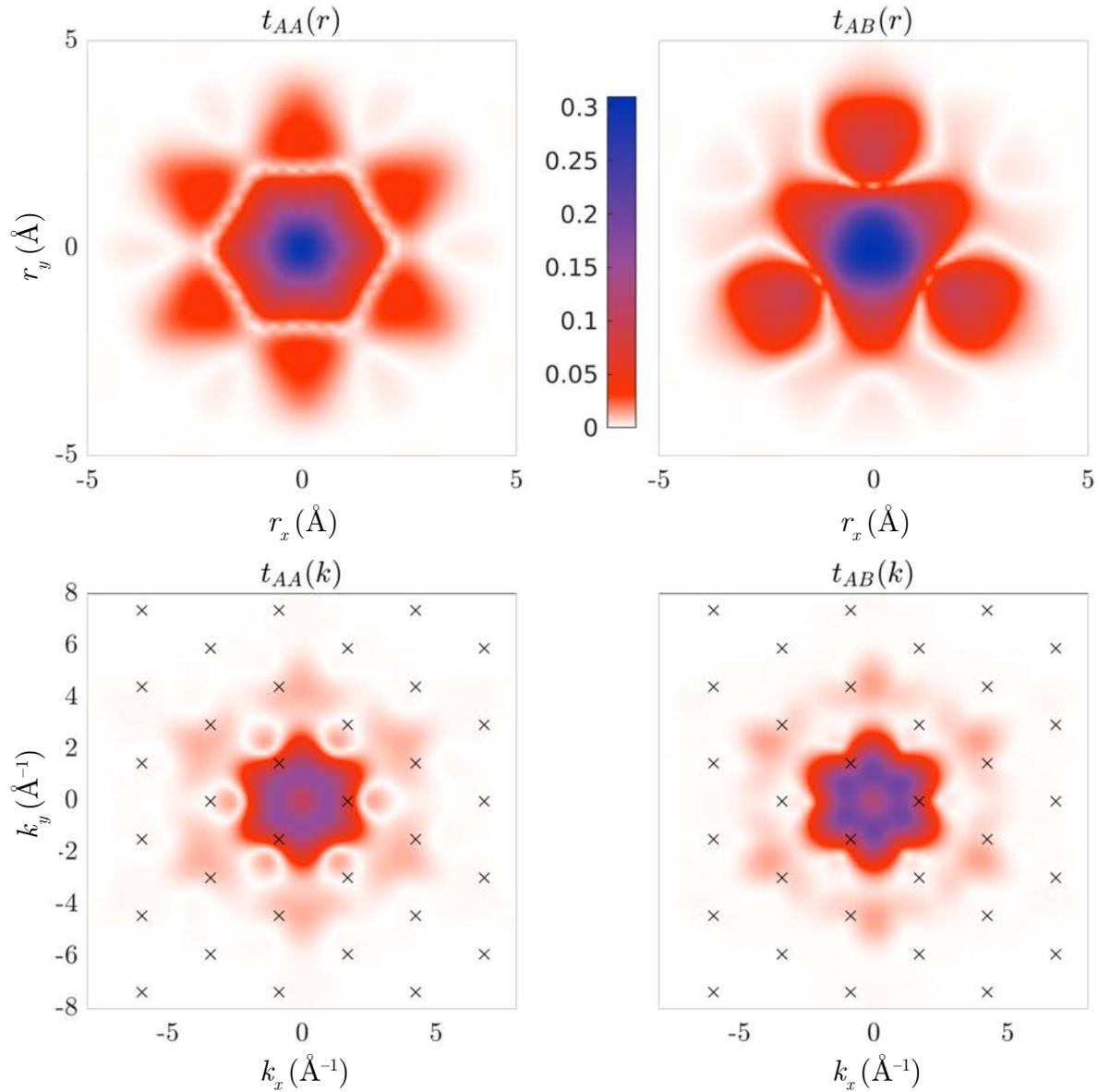

**Supplementary Fig. 18. Relaxed interlayer couplings for $\theta_m = 1.15°$ TBG.** Absolute values of the interlayer coupling for the fully relaxed TBG model at $\theta_m = 1.15°$ (i.e., both AA and AB centred rotations included). Top panels present the real space interlayer coupling after full relaxation, while bottom panels display their Fourier transform. The black "×" marks indicate the relevant scattering momenta, $K_0 + G$, where $K_0$ is the momentum corresponding to the valley we are expanding around and G is the reciprocal lattice of untwisted graphene. Left panels show couplings between similar orbitals (A–A, or $\omega_0$), while right panels show couplings between dissimilar orbitals (A–B, or $\omega_1$).

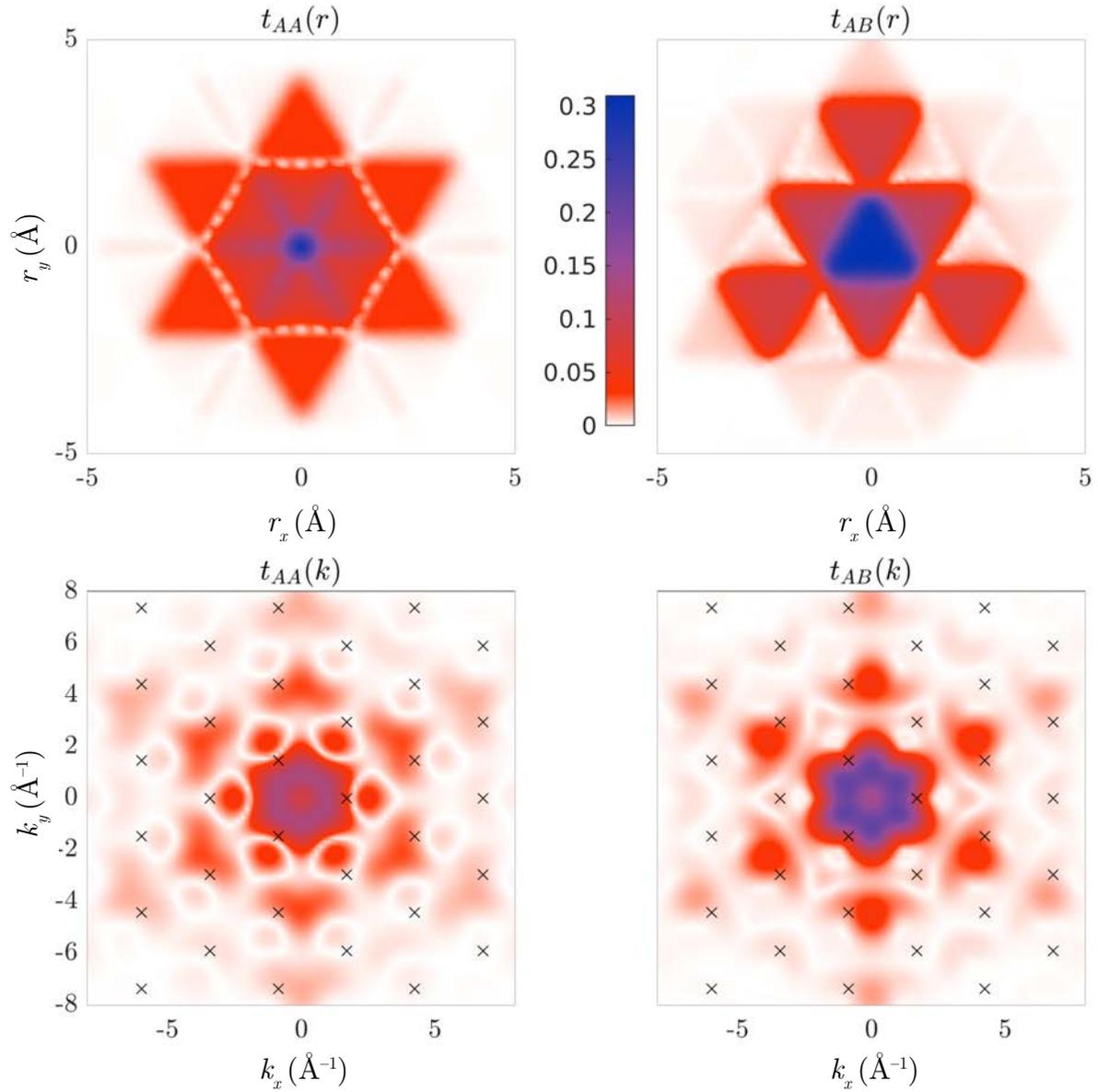

**Supplementary Fig. 19. Relaxed interlayer couplings for $\theta_m = 0.5°$ TBG.** Absolute values of the interlayer coupling for the fully relaxed TBG model at $\theta_m = 0.5°$ (i.e., both AA and AB centred rotations included). Top panels present the real space interlayer coupling after full relaxation, while bottom panels display their Fourier transform. The black "×" marks indicate the relevant scattering momenta, $K_0 + G$, where $K_0$ is the momentum corresponding to the valley we are expanding around and G is the reciprocal lattice of untwisted graphene. Left panels show couplings between similar orbitals (A–A, or $\omega_0$), while right panels show couplings between dissimilar orbitals (A–B, or $\omega_1$).

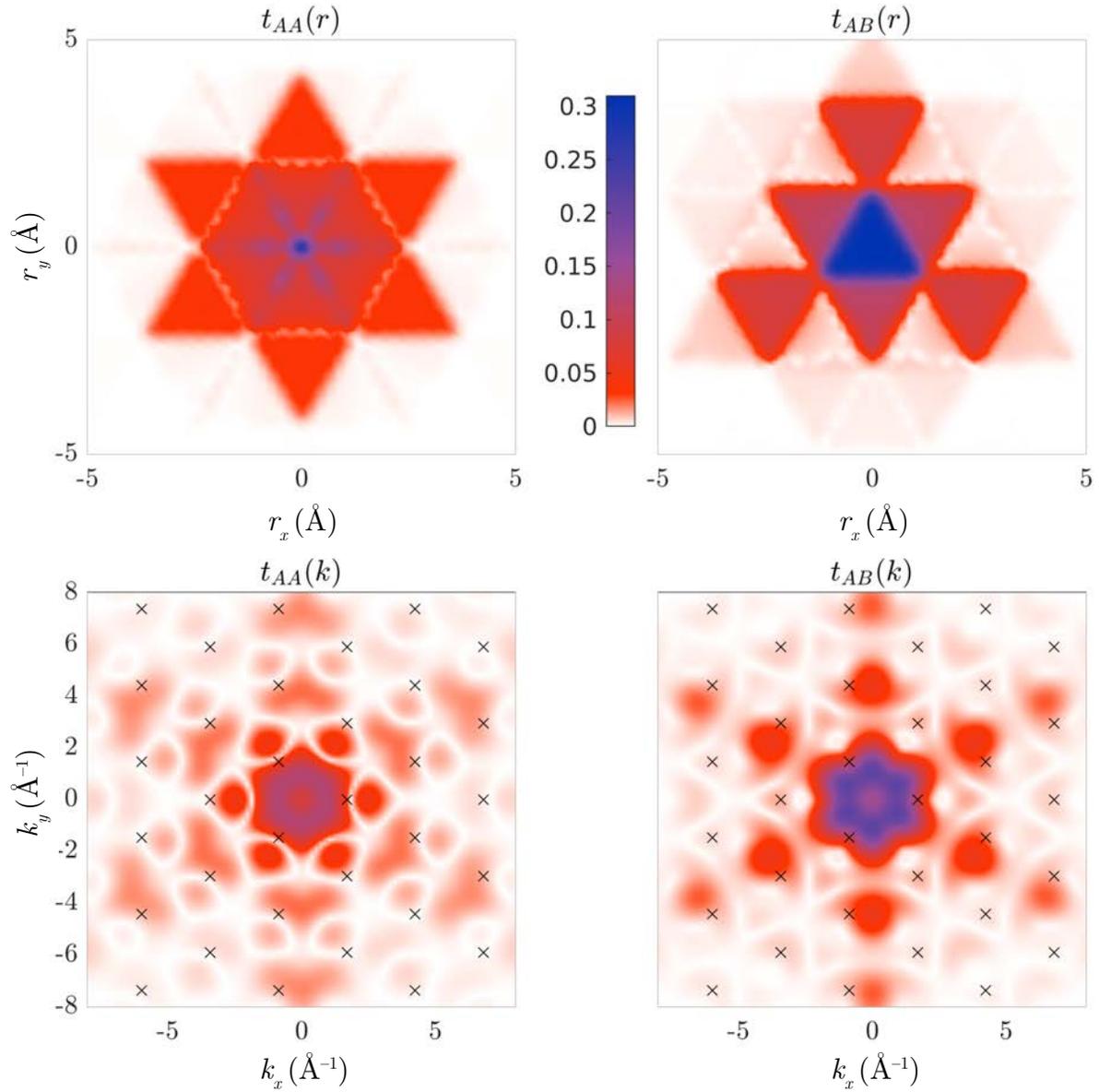

**Supplementary Fig. 20. Relaxed interlayer couplings for $\theta_m = 0.35°$ TBG.** Absolute values of the interlayer coupling for the fully relaxed TBG model at $\theta_m = 0.35°$ (i.e., both AA and AB centred rotations included). Top panels present the real space interlayer coupling after full relaxation, while bottom panels display their Fourier transform. The black "×" marks indicate the relevant scattering momenta, $K_0 + G$, where $K_0$ is the momentum corresponding to the valley we are expanding around and G is the reciprocal lattice of untwisted graphene. Left panels show couplings between similar orbitals (A–A, or $\omega_0$), while right panels show couplings between dissimilar orbitals (A–B, or $\omega_1$).

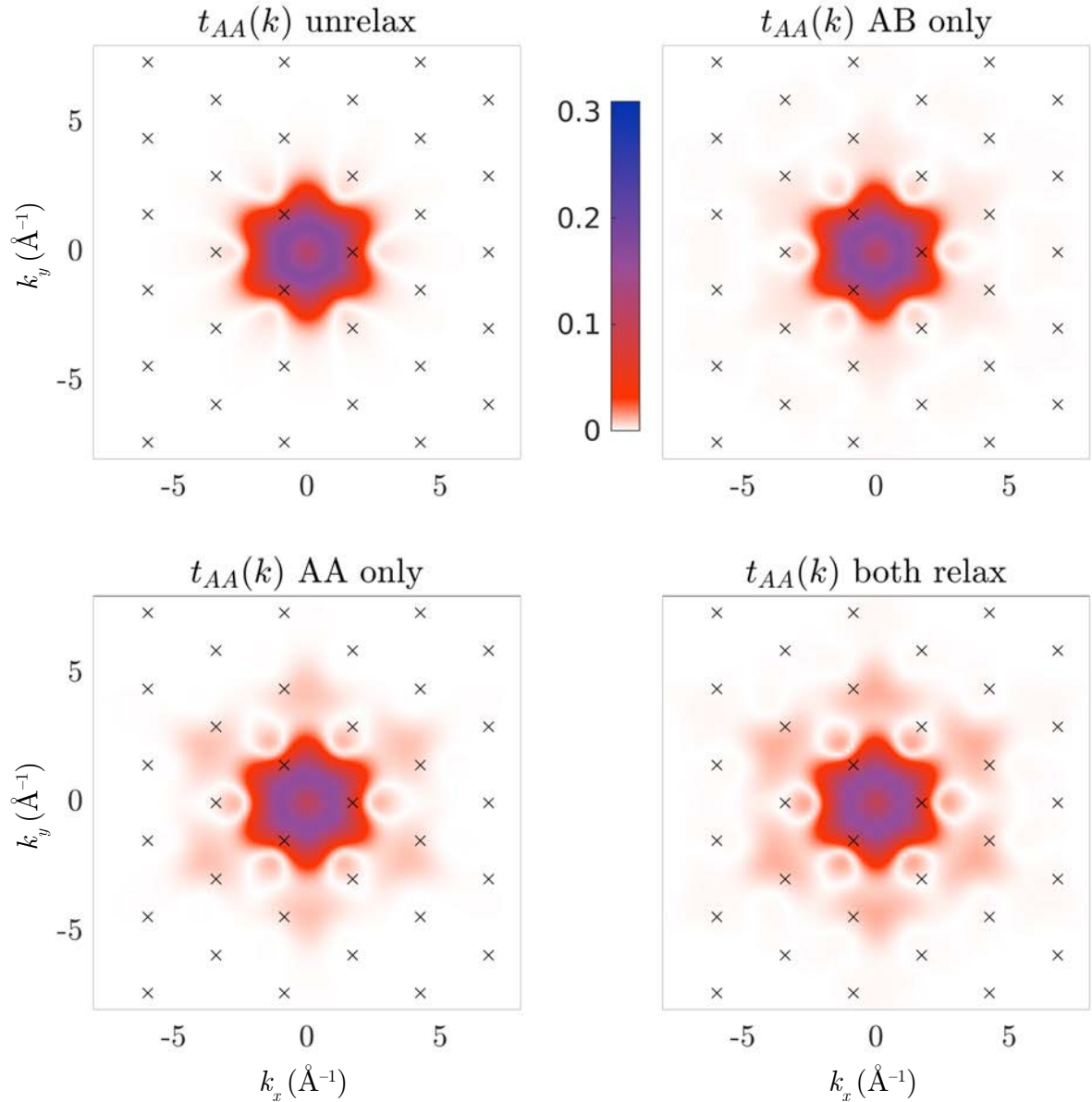

**Supplementary Fig. 21. Reconstruction-dependent interlayer couplings for orbitals of similar type at $\theta_m = 1.15°$ TBG.** Absolute value of interlayer A-to-A and B-to-B ($\omega_0$) scattering between the layers (in momentum space) for TBG at $\theta_m = 1.15°$ for all four possible relaxation assumptions. The black "×" marks indicate the momenta which all relevant scatterings are near, $K_0 + G$.

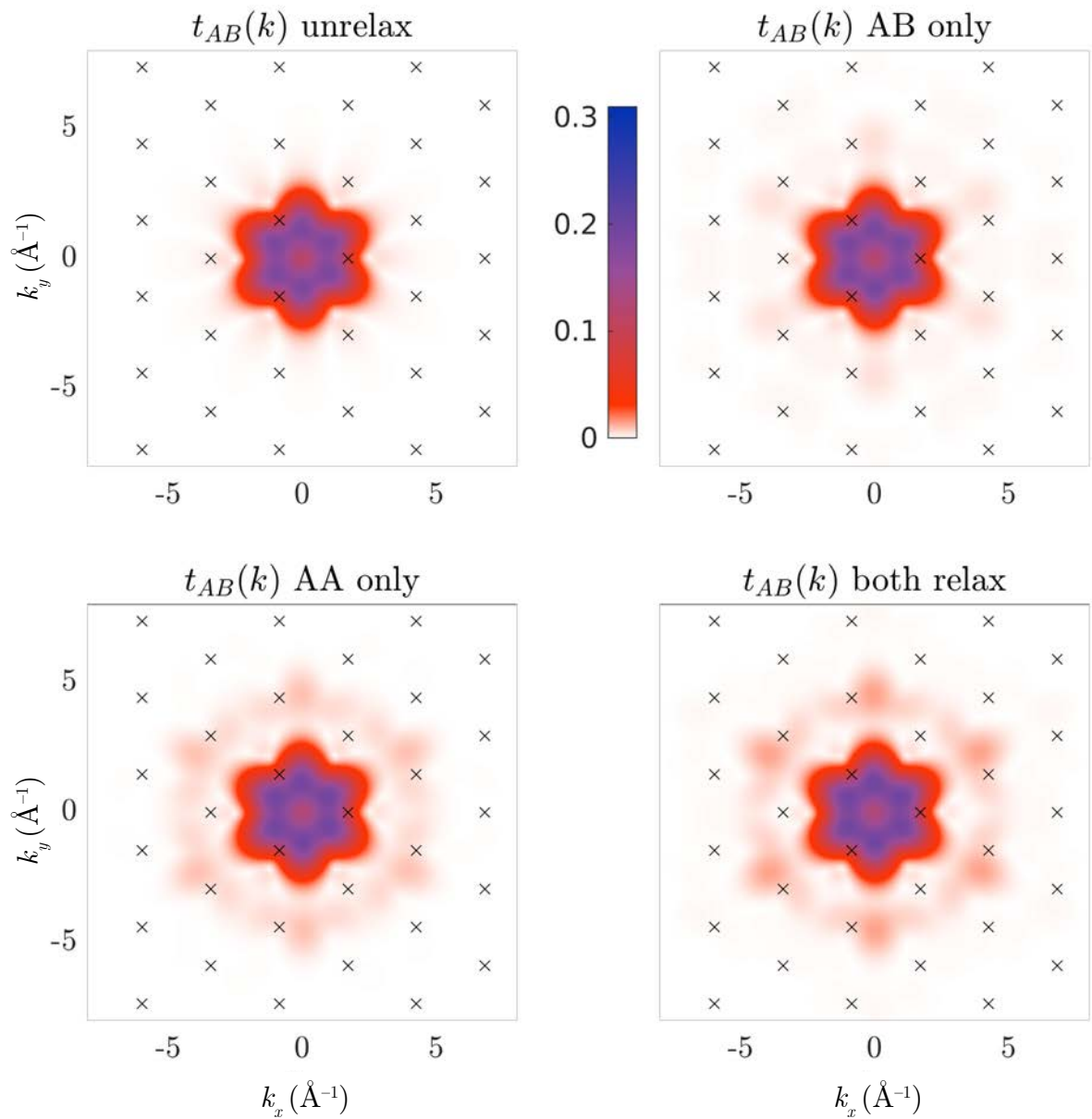

**Supplementary Fig. 22. Reconstruction-dependent interlayer couplings for orbitals of dissimilar type at $\theta_m$ = 1.15º TBG.** Absolute value of interlayer A-to-B and B-to-A ($\omega_1$) scattering between the layers (in momentum space) for TBG at $\theta_m$ = 1.15º for all four possible relaxation assumptions. The black "×" marks indicate the momenta which all relevant scatterings are near, $K_0 + G$.

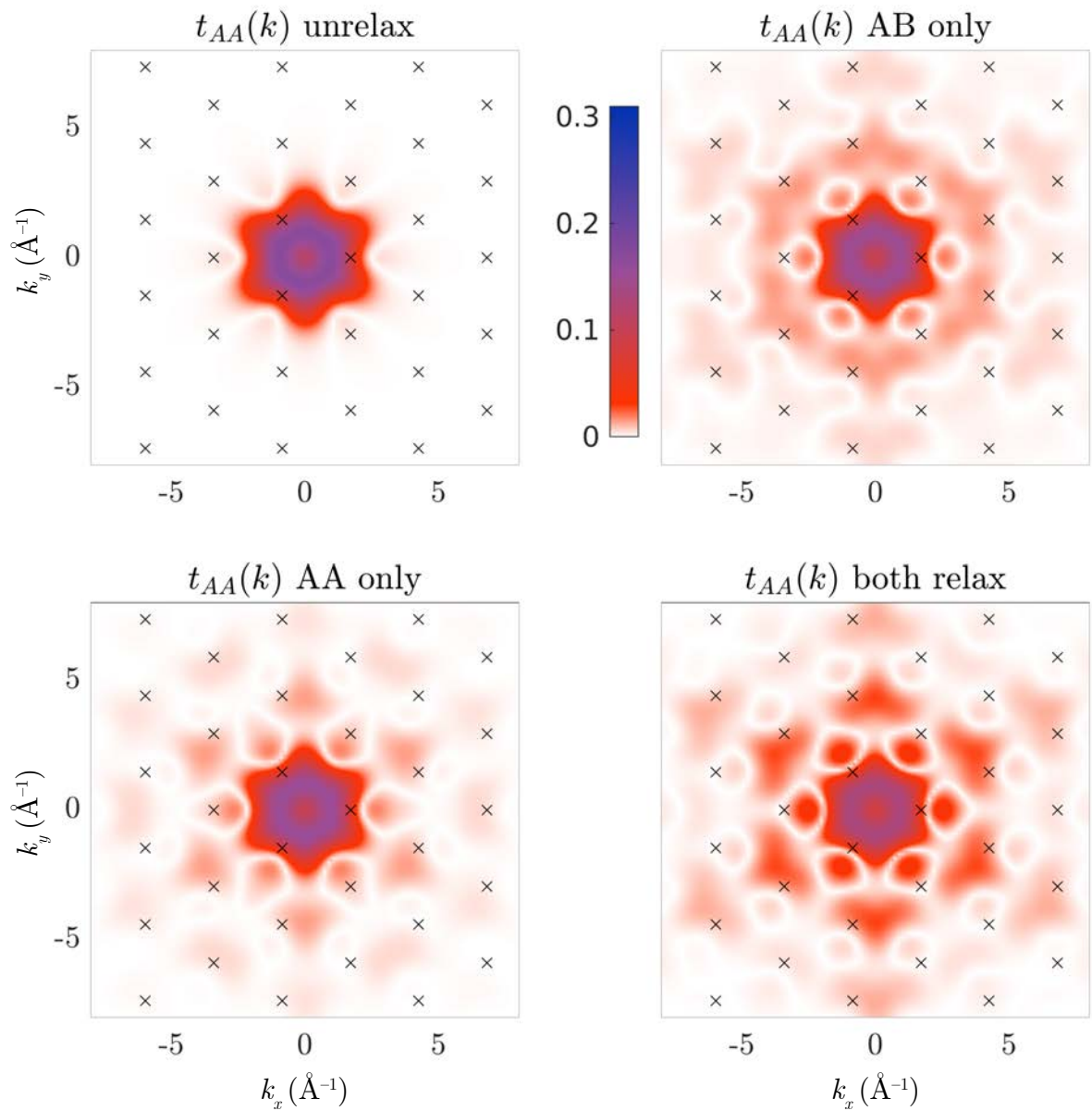

**Supplementary Fig. 23. Reconstruction-dependent interlayer couplings for orbitals of similar type at $\theta_m$ = 0.5º TBG.** Absolute value of interlayer A-to-A and B-to-B ($\omega_0$) scattering between the layers (in momentum space) for TBG at $\theta_m$ = 0.5º for all four possible relaxation assumptions. The black "×" marks indicate the momenta which all relevant scatterings are near, $K_0 + G$.

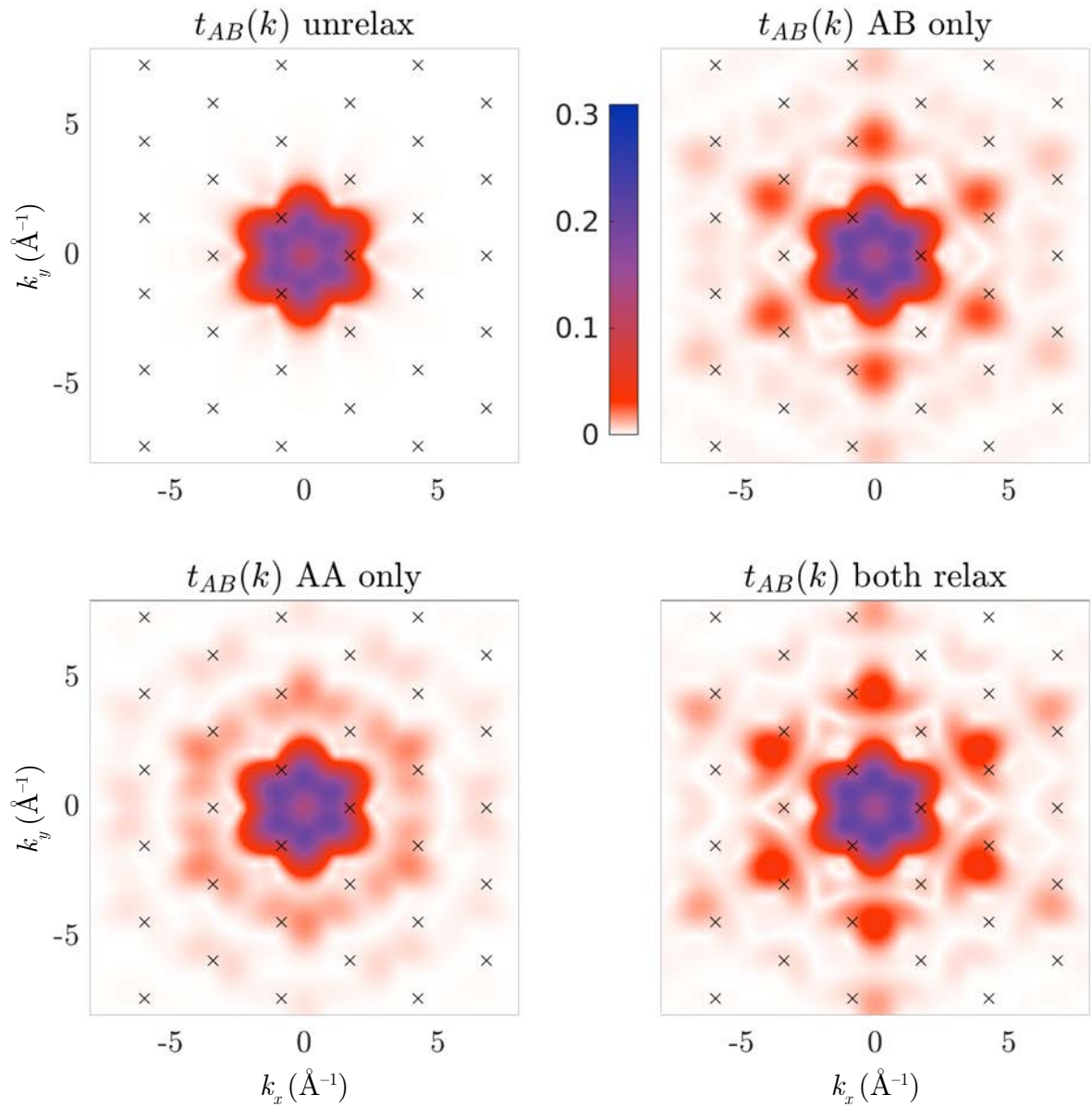

**Supplementary Fig. 24. Reconstruction-dependent interlayer couplings for orbitals of dissimilar type at $\theta_m$ = 0.5º TBG.** Absolute value of interlayer A-to-B and B-to-A ($\omega_1$) scattering between the layers (in momentum space) for TBG at $\theta_m$ = 0.5º for all four possible relaxation assumptions. The black "×" marks indicate the momenta which all relevant scatterings are near, $K_0 + G$.

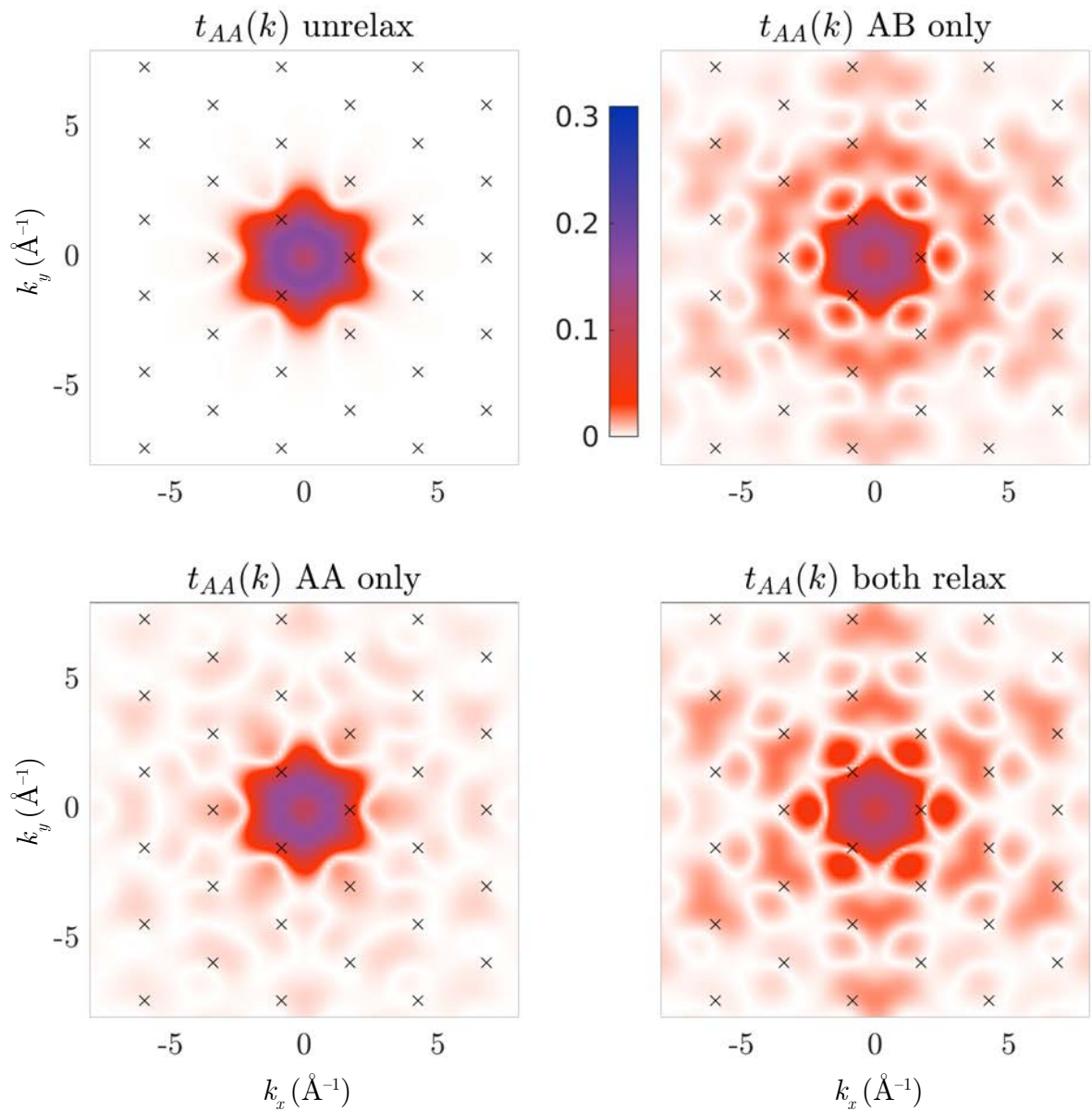

**Supplementary Fig. 25. Reconstruction-dependent interlayer couplings for orbitals of similar type at $\theta_m$ = 0.35º TBG.** Absolute value of interlayer A-to-A and B-to-B ($\omega_0$) scattering between the layers (in momentum space) for TBG at $\theta_m$ = 0.35º for all four possible relaxation assumptions. The black "×" marks indicate the momenta which all relevant scatterings are near, $K_0 + G$.

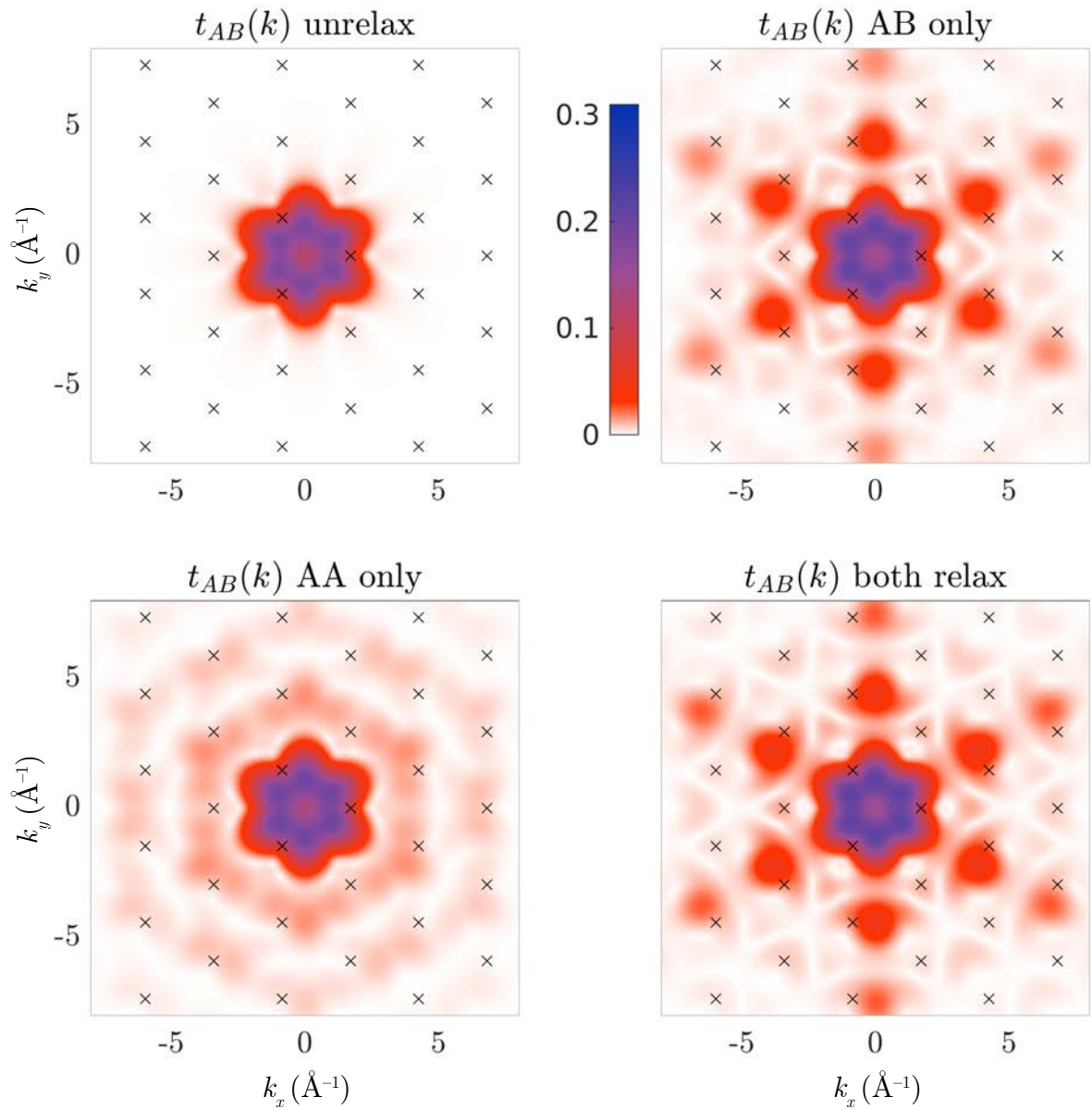

**Supplementary Fig. 26. Reconstruction-dependent interlayer couplings for orbitals of dissimilar type at $\theta_m$ = 0.35° TBG.** Absolute value of interlayer A-to-B and B-to-A ($\omega_1$) scattering between the layers (in momentum space) for TBG at $\theta_m$ = 0.35° for all four possible relaxation assumptions. The black "×" marks indicate the momenta which all relevant scatterings are near, $K_0 + G$.

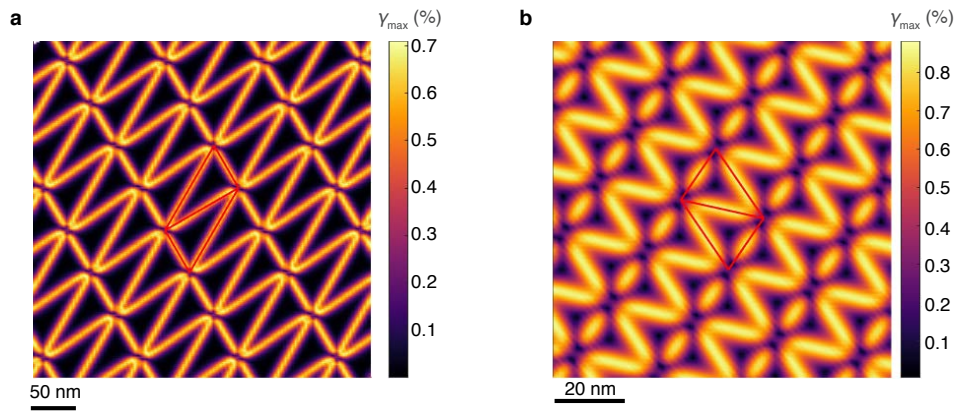

**Supplementary Fig. 27. FEM heterostrain calculations.** Finite-element-relaxed maximum shear strain ($\gamma_{max}$) maps for (**a**) $\theta_m = 0.14°$, $\varepsilon_H = 0.31\%$, (b) $\theta_m = 0.63°$, $\varepsilon_H = 0.45\%$.

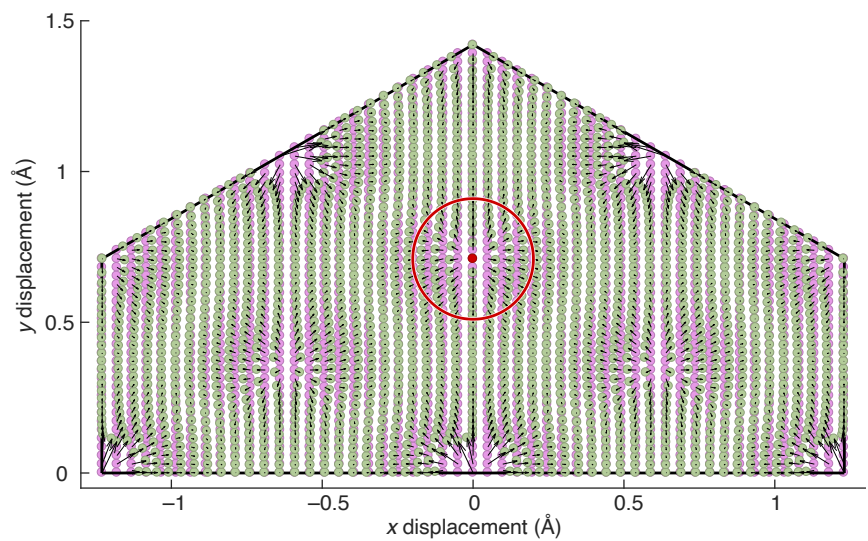

**Supplementary Fig. 28. Probe Averaging Simulation.** Simulation of bias induced by the finite probe width according to the methodology described in Supplementary Information Section 3. Purple markers represent beam positions in displacement space. Predicted interferometry patterns are averaged in a 0.2 Å radius circle around each point; averaging radius shown with red circle for one example probe position. Fitted displacement values are shown in green markers, with a black arrow connecting the fitted displacement value to the true probe position.

|  | Region 1 ($\theta_m = 0.65°$) | | Region 2 ($\theta_m = 0.64°$) | |
| --- | --- | --- | --- | --- |
| Angle between SPs (°) | Scan direction = 270° | Scan direction = 210° | Scan direction = 210° | Scan direction = 180° |
| Purple–Red | 71.8 | 73.5 | 68.5 | 67.6 |
| Red–Orange | 61.9 | 60.2 | 66.1 | 68.2 |
| Orange–Purple | 46.3 | 46.3 | 45.4 | 44.2 |

**Supplementary Table 1.** Effect of STEM scan direction rotation on observed moiré distortions (see also Supplementary Fig. 16).

| $\theta_m$ (°) | $\alpha_{AA}$ (°) | $\alpha_{AB}$ (°) | $\sigma_{AA}$ (Å) | $\sigma_{AB}$ (Å) | $b_{AB}$ (Å) |
| --- | --- | --- | --- | --- | --- |
| 1.15 | 0.65 | −0.20 | 45 | 15 | 10 |
| 0.50 | 0.65 | −0.35 | 50 | 25 | 15 |
| 0.30 | 0.85 | −0.30 | 50 | 25 | 15 |

**Supplementary Table 2.** Parameters used in simplified reconstruction model. Here $\alpha_{AA}$ and $\alpha_{AB}$ give the applied rotation field centred on each individual AA or AB domain, $\sigma_{AA}$ and $\sigma_{AB}$ give the Gaussian standard deviations for AA rotation decay and the AB smoothing kernel, and $b_{AB}$ gives the buffer distance for defining the AB rotation field area.

|  | No reconstruction | | AA rotation only | | AB rotation only | | Both rotations | |
| --- | --- | --- | --- | --- | --- | --- | --- | --- |
| $\theta_m$ (°) | $\omega_{AA}$ | $\omega_{AB}$ | $\omega_{AA}$ | $\omega_{AB}$ | $\omega_{AA}$ | $\omega_{AB}$ | $\omega_{AA}$ | $\omega_{AB}$ |
| 0.35 | 88 | 88 | 73 | 93 | 45 | 95 | 27 | 95 |
| 0.50 | 88 | 88 | 67 | 95 | 63 | 95 | 39 | 97 |
| 1.15 | 88 | 88 | 78 | 92 | 84 | 90 | 74 | 94 |

**Supplementary Table 3.** Calculated interlayer coupling terms (in meV) for TBG at three values of $\theta_m$ under four relaxation assumptions: no reconstruction, AA rotation alone, AB rotation alone, and full reconstruction (both rotations).